\documentclass[10pt,a4paper,twoside]{report}

\usepackage[utf8]{inputenc}   

\usepackage[english]{babel} 

\newcommand{\acknowledgments}{@undefined} 
\addto\captionsenglish{\renewcommand{\acknowledgments}{Acknowledgments}}

\addto\captionsportuguese{\renewcommand{\acknowledgments}{Agradecimentos}}
\addto\captionsportuguese{\renewcommand{\nomname}{Lista de S\'{i}mbolos}} 

\newcommand{\coverThesis}{@undefined} 
\newcommand{\coverSupervisors}{@undefined} 
\newcommand{\coverExaminationCommittee}{@undefined} 
\newcommand{\coverChairperson}{@undefined} 
\newcommand{\coverSupervisor}{@undefined} 
\newcommand{\coverMemberCommittee}{@undefined} 
\addto\captionsenglish{\renewcommand{\coverThesis}{Thesis to obtain the Master of Science Degree in}}
\addto\captionsenglish{\renewcommand{\coverSupervisors}{Supervisor(s)}}
\addto\captionsenglish{\renewcommand{\coverExaminationCommittee}{Examination Committee}}
\addto\captionsenglish{\renewcommand{\coverChairperson}{Chairperson}}
\addto\captionsenglish{\renewcommand{\coverSupervisor}{Supervisor}}
\addto\captionsenglish{\renewcommand{\coverMemberCommittee}{Member of the Committee}}
\addto\captionsportuguese{\renewcommand{\coverThesis}{Disserta\c{c}\~{a}o para obten\c{c}\~{a}o do Grau de Mestre em}}
\addto\captionsportuguese{\renewcommand{\coverSupervisors}{Orientador(es)}}
\addto\captionsportuguese{\renewcommand{\coverExaminationCommittee}{J\'{u}ri}}
\addto\captionsportuguese{\renewcommand{\coverChairperson}{Presidente}}
\addto\captionsportuguese{\renewcommand{\coverSupervisor}{Orientador}}
\addto\captionsportuguese{\renewcommand{\coverMemberCommittee}{Vogal}}

\def\FontLb{
  \usefont{T1}{phv}{b}{n}\fontsize{16pt}{16pt}\selectfont}

\def\FontMb{
  \usefont{T1}{phv}{b}{n}\fontsize{14pt}{14pt}\selectfont}
\def\FontSn{
  \usefont{T1}{phv}{m}{n}\fontsize{12pt}{12pt}\selectfont}

\usepackage{geometry}	
\usepackage{setspace}

\usepackage{graphicx}

\usepackage{amsmath}  
\usepackage{amsthm}   
\usepackage{amsfonts} %

\usepackage{subfigure}

\usepackage{subfigmat}

\usepackage{dcolumn}
\newcolumntype{d}{D{.}{.}{-1}} 
\newcolumntype{e}{D{E}{E}{-1}} 

\usepackage{nomencl}
\makenomenclature
\RequirePackage{ifthen} 
\ifthenelse{\equal{\languagename}{english}}%
    { 
    \renewcommand{\nomgroup}[1]{%
      \ifthenelse{\equal{#1}{R}}{%
        \item[\textbf{Roman symbols}]}{%
        \ifthenelse{\equal{#1}{G}}{%
          \item[\textbf{Greek symbols}]}{%
          \ifthenelse{\equal{#1}{S}}{%
            \item[\textbf{Subscripts}]}{%
            \ifthenelse{\equal{#1}{T}}{%
              \item[\textbf{Superscripts}]}{}}}}}%
    }{
    \renewcommand{\nomgroup}[1]{%
      \ifthenelse{\equal{#1}{R}}{%
        \item[\textbf{Simbolos romanos}]}{%
        \ifthenelse{\equal{#1}{G}}{%
          \item[\textbf{Simbolos gregos}]}{%
          \ifthenelse{\equal{#1}{S}}{%
            \item[\textbf{Subscritos}]}{%
            \ifthenelse{\equal{#1}{T}}{%
              \item[\textbf{Sobrescritos}]}{}}}}}%
    }%

\usepackage[number=none]{glossary}
\setglossary{gloskip={}}
\makeglossary

\usepackage{rotating}

\usepackage[pdftex]{hyperref} 
\hypersetup{colorlinks,       
            linkcolor=black,  
            anchorcolor=black,
            citecolor=black,  
            filecolor=black,  
            menucolor=black,  
            pagecolor=black,  
            urlcolor=black,   
	          bookmarks=true,         
	          bookmarksopen=false,    
	          bookmarksnumbered=true, 
	          pdftitle={Thesis},
            pdfauthor={Andre C. Marta},
            pdfsubject={Thesis Title},
            pdfkeywords={Thesis Keywords},
            pdfstartview=FitV,
            pdfdisplaydoctitle=true}

\usepackage[figure,table]{hypcap}

\usepackage[numbers]{natbib} 

\usepackage{notoccite}

\usepackage{multirow}

\usepackage{booktabs}

\usepackage{pdfpages}

\usepackage{amsthm}
\usepackage{amssymb}

\usepackage{float}

\usepackage[normalem]{ulem}

\newtheorem{dfn}{Definition}[section]

\newtheorem{nota}{Notation}[section]
\newtheorem{thm}{Theorem}[section]
\newtheorem{lmm}{Lemma}[section]
\newtheorem{ex}{Example}[section]
\newtheorem{rmk}{Remark}
\newtheorem{alg}{Algorithm}

\begin{document}

\pagestyle{plain}

\pagenumbering{roman}


\thispagestyle {empty}

\includegraphics[bb=9.5cm 11cm 0cm 0cm,scale=0.29]{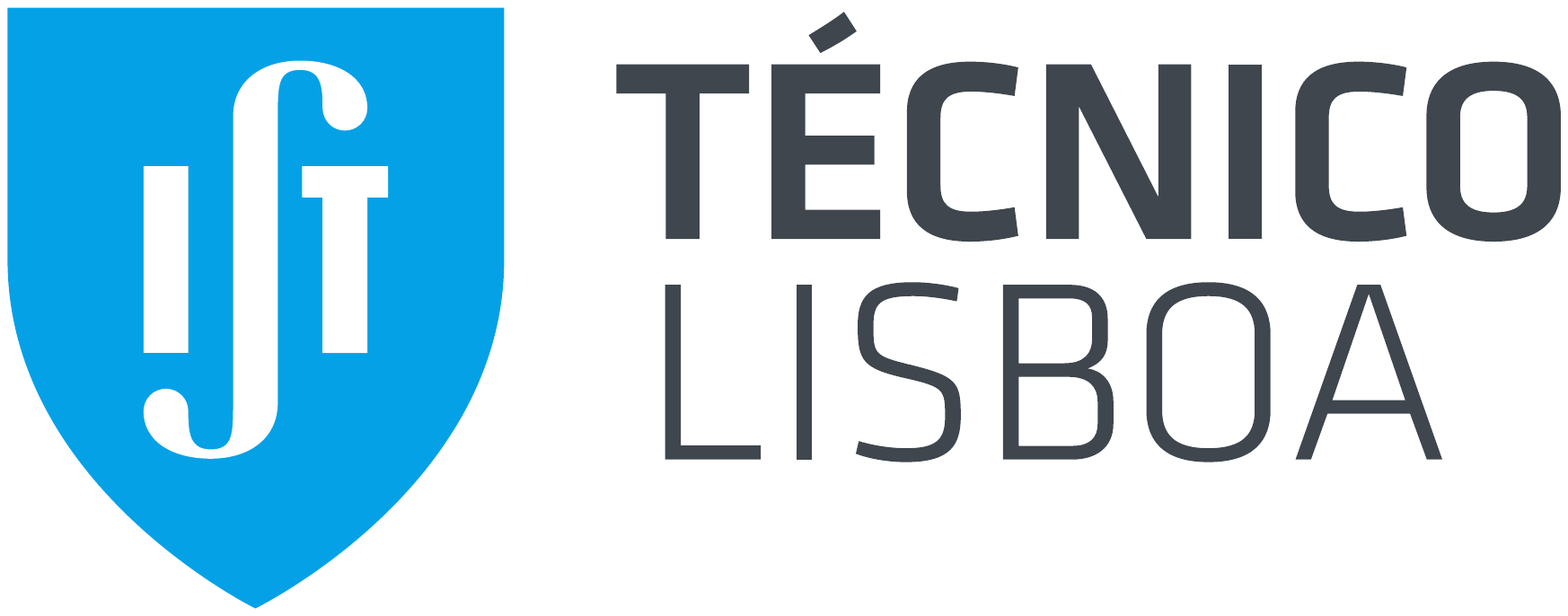}

\begin{center}
%
\vspace{2.5cm}
\includegraphics[height=40mm]{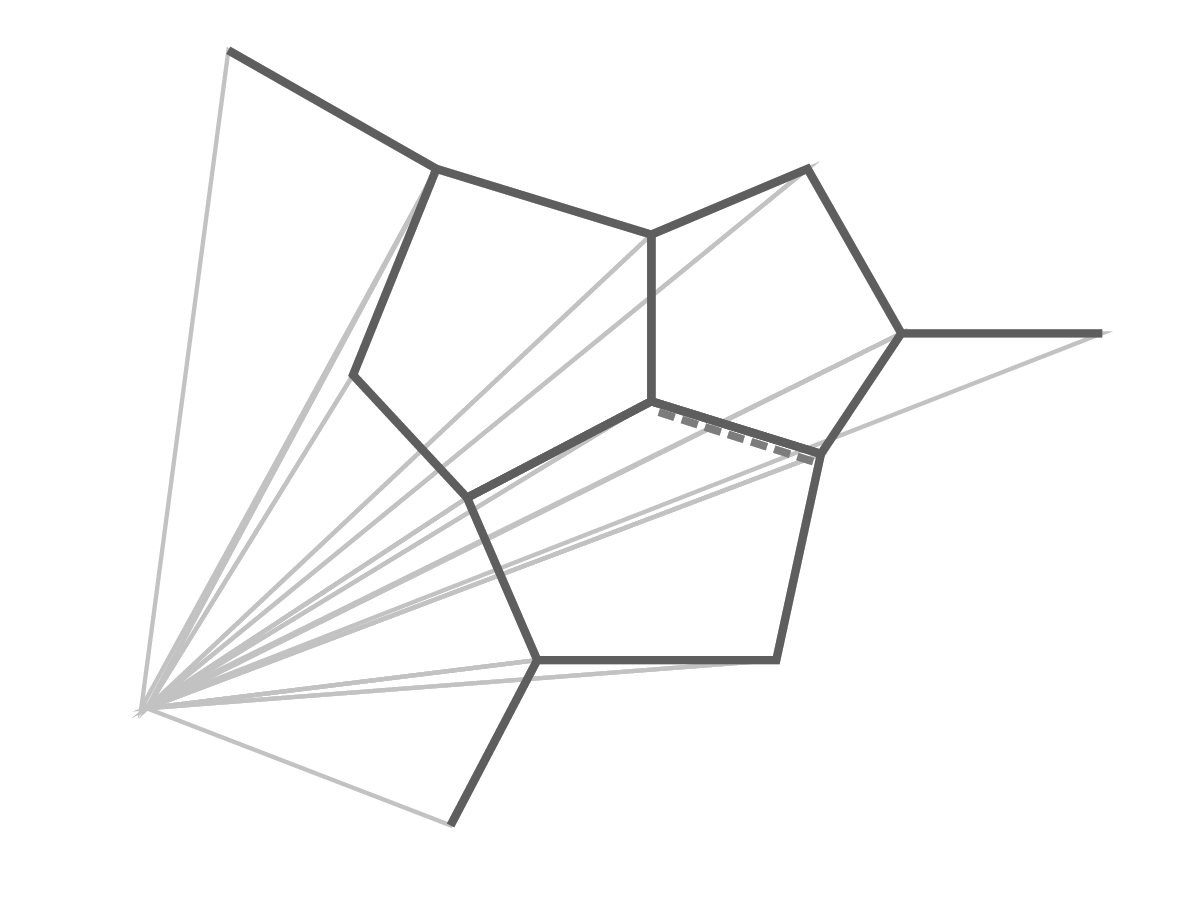}

\vspace{1.0cm}
{\FontLb An analysis of the Geodesic Distance and other comparative metrics for tree-like structures} \\ 
\vspace{2.6cm}
{\FontMb Bernardo Lopo Tavares Fernandes} \\ 
\vspace{2.0cm}
{\FontSn \coverThesis} \\
\vspace{0.3cm}
{\FontLb Mathematics and Applications} \\ 
\vspace{1.0cm}
{\FontSn %
\begin{tabular}{ll}
 \coverSupervisors: & Prof. Alexandre Paulo Lourenço Francisco \\ 
                    & Prof. Pedro Alves Martins Rodrigues    
\end{tabular} } \\
\vspace{1.0cm}
{\FontMb \coverExaminationCommittee} \\
\vspace{0.3cm}
{\FontSn %
\begin{tabular}{c}
\coverChairperson:     Prof. Maria Cristina De Sales Viana Serôdio Sernadas          \\ 
\coverSupervisor:      Prof. Alexandre Paulo Lourenço Francisco \\ 
\coverMemberCommittee: Prof. Francisco Miguel Dionísio           
\end{tabular} } \\
\vspace{1.5cm}
{\FontMb November 2018} \\ 
\end{center}

\cleardoublepage


\null\vskip5cm%
\begin{flushright}
     Dedicated to Luna, Igor and\\
     all those living between\\
     dream and reality.
\end{flushright}
\vfill\newpage

\cleardoublepage


\section*{\acknowledgments}

\addcontentsline{toc}{section}{\acknowledgments}

I am very grateful to professor Alexandre Francisco and professor Pedro Martins Rodrigues for all patience and availability to answer my questions and address my curiosity in regards to comparison metrics, such as the courses they taught during my academic journey in Técnico Lisboa.\par

I owe my thanks to the university and the Mathematics Department for providing a calm workspace to learn and work. I'm also grateful that this degree helped me develop persistence, systematic approach and analytic perspective to challenges, this was useful not only as a student but also for my personal life.\par

To my family and friends.\par

\cleardoublepage



\section*{Abstract}

\addcontentsline{toc}{section}{Abstract}

Graphs are interesting structures: extremely useful to depict real-life problems, extremely easy to understand given a sketch, extremely complicated to represent formally, extremely complicated to compare.
Phylogeny is the study of the relations between biological entities. From it, the interest in comparing tree graphs grew more than in other fields of science. Since there is no definitive way to compare them, multiple distances were formalized over the years since the early sixties, when the first effective numerical method to compare dendrograms was described.
This work consists of formalizing, completing (with original work) and give a universal notation to analyze and compare the discriminatory power and time complexity of computing the thirteen here formalized metrics. We also present a new way to represent tree graphs, reach deeper in the details of the Geodesic Distance and discuss its worst-case time complexity in a suggested implementation. 
Our contribution ends up as a clean, valuable resource for anyone looking for an introduction to comparative metrics for tree graphs.

\vfill

\textbf{\Large Keywords:} Geodesic Distance, Comparative Metrics, Graph Theory, Complexity, Phylogeny.

\cleardoublepage


%
\tableofcontents
\cleardoublepage 

%
\phantomsection
\addcontentsline{toc}{section}{\listtablename}
\listoftables
\cleardoublepage 

%
\phantomsection
\addcontentsline{toc}{section}{\listfigurename}
\listoffigures
\cleardoublepage 

%
%
\phantomsection
\addcontentsline{toc}{section}{\nomname}
\printnomenclature
\cleardoublepage

%
%
%
%
%
%


\glossary{name={\textbf{Phylogenetics}},description={Study of relations between biological entities}}

\glossary{name={\textbf{Distance, Metric}},description={Definition \ref{dfn:metric} in page~\pageref{dfn:metric}}}

\glossary{name={\textbf{Graph}},description={Definition \ref{dfn:basic1} in page \pageref{dfn:basic1}}}

\glossary{name={\textbf{Big-Oh notation, $O(f(n))$}},description={Definition \ref{dfn:BON} in page~\pageref{dfn:BON}}}

\glossary{name={\textbf{Tree}},description={Definition \ref{dfn:basic1} in page \pageref{dfn:basic1}}}

\glossary{name={\textbf{Path}},description={Definition \ref{dfn:basic1} in page \pageref{dfn:basic1}}}

\glossary{name={\textbf{Cycle} (graph theory)},description={Definition \ref{dfn:basic1} in page \pageref{dfn:basic1}}}

\glossary{name={\textbf{Directed/Undirected Graph}},description={Definition \ref{dfn:basic1} in page \pageref{dfn:basic1}}}

\glossary{name={\textbf{Neighboor}},description={Definition \ref{dfn:basic1} in page \pageref{dfn:basic1}}}

\glossary{name={\textbf{Degree} (vertex)},description={Definition \ref{dfn:basic1} in page \pageref{dfn:basic1}}}

\glossary{name={\textbf{Label} (vertex)},description={check after Definition \ref{dfn:basic1} in page \pageref{dfn:basic1}}}

\glossary{name={\textbf{Weight} (edge)},description={check after Definition \ref{dfn:basic1} in page \pageref{dfn:basic1}}}

\glossary{name={\textbf{Leaf} (vertex)},description={Definition \ref{dfn:basic} in page \pageref{dfn:basic}}}

\glossary{name={\textbf{Rooted/Unrooted} tree (vertex)},description={Definition \ref{dfn:basic} in page \pageref{dfn:basic}}}

\glossary{name={\textbf{Depth} (vertex, tree)},description={Definition \ref{dfn:basic} in page \pageref{dfn:basic}}}

\glossary{name={\textbf{Ancestor (or direct ancestor)} (vertex)},description={Definition \ref{dfn:basic} in page \pageref{dfn:basic}}}

\glossary{name={\textbf{Descendant (or direct descendant)} (vertex)},description={Definition \ref{dfn:basic} in page \pageref{dfn:basic}}}

\glossary{name={$\gamma_S$},description={Definition \ref{dfn:gammas} in page \pageref{dfn:gammas}}}

\glossary{name={$\gamma^w_S$},description={Definition \ref{dfn:wlt} in page \pageref{dfn:wlt}}}

\glossary{name={\textbf{Identical} (trees)},description={Definition \ref{dfn:ide} in page \pageref{dfn:ide}}}

\glossary{name={\textbf{Set of Labels $S$}},description={Check Remark \ref{rmk:slab} in page \pageref{rmk:slab}. Also, for the Geodesic Distance, $S=\{0,1,2,...,n\}$ (bold at the end of section \ref{section:MTMethod})}}

\glossary{name={\textbf{Contraction of Bourque}},description={Definition \ref{dfn:alphaop} in page \pageref{dfn:alphaop}}}

\glossary{name={\textbf{Powerset}},description={Given a set $A$, the powerset of $A$, $\mathcal{P}(A)$ is the set of all subsets of $A$.}}

\glossary{name={\textbf{Cluster} and \textbf{Cluster Representation}},description={Notation \ref{nota:day} in page \pageref{nota:day}}}

\glossary{name={\textbf{Labeled Tree}},description={Definition \ref{dfn:lt} in page \pageref{dfn:lt}}}

\glossary{name={\textbf{Clade}},description={Definition \ref{dfn:basic} in page \pageref{dfn:basic}}}

\glossary{name={\textbf{Binary tree}},description={Definition \ref{dfn:basic} in page~\pageref{dfn:basic}}}

\glossary{name={\textbf{Forest} (trees)},description={Definition \ref{dfn:basic} in page~\pageref{dfn:basic}}}

\glossary{name={\textbf{Dendogram}},description={Definition \ref{dfn:basic} in page~\pageref{dfn:basic}}}

\glossary{name={\textbf{Strict Consensus Method, Strict Consensus Tree}},description={Definition \ref{dfn:scmsct} in page~\pageref{dfn:scmsct}}}

\glossary{name={\textbf{Weighted tree}},description={Definition \ref{dfn:wlt} in page~\pageref{dfn:wlt}}}

\glossary{name={\textbf{Partitioning Function}},description={Notation \ref{dfn:partfunc} in page~\pageref{dfn:partfunc}}}

\glossary{name={\textbf{Matched} (edges)},description={Notation \ref{dfn:medg} in page~\pageref{dfn:medg}}}

\glossary{name={\textbf{Connected Components} (graph theory)},description={Let $T=(V,E)$ be a graph such that $V=\{V_1,V_2,...,V_k\}$ and there's no path between any two sets $V_i$ and $V_j$ for all $i\neq j$. $T$ in a unconnected graph with connected components $\{V_1,V_2,...,V_k\}$. }}

\glossary{name={\textbf{Quartet}},description={Let $T=(V,E)$ be a tree. Quartet $Q\subsetneq V$ such that $Q$ has size $4$ and all elements of are leaves of $T$.}}

\glossary{name={\textbf{Triplet}},description={Let $T=(V,E)$ be a tree. Quartet $H\subsetneq V$ such that $H$ has size $3$ and all elements of are leaves of $T$.}}

\glossary{name={\textbf{Unresolved} Quartet/Triplet},description={Figure \ref{fig:QTTopo} in page \pageref{fig:QTTopo}}}

\glossary{name={\textbf{Internal Edge}},description={An edge $uv$ such that neither $u$ or $v$ are leaves.}}

\glossary{name={\textbf{Orthant}},description={Definition \ref{dfn:stsn} in page~\pageref{dfn:stsn}}}

\glossary{name={\textbf{Space of trees of $n$ labels,} $\mathcal{T}_n$},description={Definition \ref{dfn:stsn} in page~\pageref{dfn:stsn}}}

\glossary{name={\textbf{Split}, ($L$-Split, $X$-Split)},description={Definition \ref{dfn:split} in page~\pageref{dfn:split}}}

\glossary{name={\textbf{Compatible} split/split set},description={Definition \ref{dfn:cscss} in page~\pageref{dfn:cscss}}}

\glossary{name={\textbf{Compatible Edges}},description={Definition \ref{dfn:ces} in page~\pageref{dfn:ces}}}

\glossary{name={\textbf{Split Equivalence Theorem}},description={Theorem \ref{thm:set} in page~\pageref{thm:set}}}

\glossary{name={\textbf{Minimum rooted subtree} (geodesic distance)},description={Definition \ref{dfn:mrs} in page~\pageref{dfn:mrs}}}

\glossary{name={\textbf{Path Space and Path Space Geodesic}},description={Definition \ref{dfn:psg} in page~\pageref{dfn:psg}}}

\glossary{name={\textbf{Proper Path and Proper Path Space}},description={Definition \ref{dfn:pps} in page~\pageref{dfn:pps}}}

\glossary{name={\textbf{Cone Path}},description={after Theorem \ref{thm:p3} in page~\pageref{thm:p3}}}

\glossary{name={\textbf{GTP Algorithm}},description={Algorithm \ref{alg:GTP} in page~\pageref{alg:GTP}, Section \ref{section:GTP}}}

\glossary{name={\textbf{Split Node}},description={Definition \ref{dfn:iodr} in page~\pageref{dfn:iodr}}}

\glossary{name={\textbf{In-Degree, Out-Degree}},description={Definition \ref{dfn:iodr} in page~\pageref{dfn:iodr}}}

\glossary{name={\textbf{Reticulation node}},description={Definition \ref{dfn:iodr} in page~\pageref{dfn:iodr}}}

\glossary{name={\textbf{In-Degree, Out-Degree}},description={Definition \ref{dfn:iodr} in page~\pageref{dfn:iodr}}}

\glossary{name={\textbf{Lexicographical Order}},description={Definition \ref{dfn:lex} in page~\pageref{dfn:lex}}}

\glossary{name={\textbf{Incompatibility Graph}},description={Definition \ref{dfn:igraph} in page~\pageref{dfn:igraph}}}

\glossary{name={\textbf{Extension Problem}},description={Definition \ref{dfn:extp} in page~\pageref{dfn:extp} and Definition \ref{dfn:extp2} in page~\pageref{dfn:extp2}}}

\glossary{name={\textbf{Independent Set, Maximum Weight Independent Set}},description={Definition \ref{dfn:mwis} in page~\pageref{dfn:mwis}}}

\glossary{name={\textbf{Vertex Cover, Minimum Weight Vertex Cover}},description={Definition \ref{dfn:mwvc} in page~\pageref{dfn:mwvc}}}

\glossary{name={\textbf{Max flow problem}},description={Definition \ref{dfn:mfp} in page~\pageref{dfn:mfp}}}

\glossary{name={\textbf{Min cut problem}},description={Definition \ref{dfn:mcp} in page~\pageref{dfn:mcp}}}

\glossary{name={\textbf{Max-Flow Min-cut problem}},description={Definition \ref{thm:mfmct} in page~\pageref{thm:mfmct}}}

\glossary{name={\textbf{Flow Equivalent Graph}},description={Definition \ref{dfn:feg} in page~\pageref{dfn:feg}}}

\glossary{name={\textbf{Edmonds-Karp algorithm}},description={Subsection \ref{subsection:EKAlg} in page~\pageref{subsection:EKAlg}}}

\glossary{name={\textbf{Residual Graph}},description={Definition \ref{dfn:resg} in page~\pageref{dfn:resg}}}

\glossary{name={\textbf{Common Edges} (geodesic distance)},description={Let $T_1,T_2\in\gamma^w_S$ with edge sets $E_1$ and $E_2$ respect., two edges $e_1\in E_1$ and $e_2\in E_2$ are common if they are \textbf{matched}}}


\phantomsection
\addcontentsline{toc}{section}{\glossaryname}
\printglossary
\cleardoublepage

%
\setcounter{page}{1}
\pagenumbering{arabic}



\chapter{Introduction}
\label{chapter:introduction}

\section{Motivation}
\label{section:motivation}

Phylogeny is the study of the relations between biological entities. From it, the need to compare tree-like graphs has risen and several metrics were established and researched, but since there is no definitive way to compare them, its discussion is still open nowadays. All of them emphasize different features of the structures and, of course, the efficiency of these computations also varies.

\section{Topic Overview}
\label{section:ioverview}

The topic of this work is comparison metrics. Given two classifications, the challenge is to generate some parameter that expresses how similar these classifications are.
A \textbf{classification} is some \textit{data structure} that expresses relations between the data. For instance:

One of the big applications of comparison metrics is \textbf{phylogenetics}. \textbf{Phylogenetics} is the study of relations between biological entities: may it be genes, species or individuals. If we consider Darwin's theory of evolution, we can think of every species as having a direct ancestor (or multiple direct ancestors, but for sake of this example, lets assume there's only one) and multiple successors, and then, it's only natural to suggest that the whole map of species evolution can be described by a \textit{tree-like graph}. We say that one specific tree hypothesis for the arrangement of species is a \textbf{classification}.

One of the first \textit{data structures} considered for the first known problem of classification were the dendrograms. \textbf{Dendrograms} are tree graphs where the data is concentrated in the leaves, while the rest expresses the relation between it. This was a way to relate data by hierarchical clustering: close data is on the same cluster (cluster as a \textit{bulk} of closely related data), and how close they are on the tree dictates how alike the information stored in the leaves is. The first effective numerical method to compare dendrograms was developed by Sokal and Rohlf in 1962, is known as the \textbf{cophenetic correlation} and it will be defined in the appropriate section. At the time, it was a method created to compare dendrograms generated from numerical taxonomic research.

However, since these methods were born out of a necessity to compare \textit{tree-like structures}, a lot of other metrics were proposed, some of them formalized for quirky data structures and to emphasize different properties (based on topology, edge weight, and other parameters). This happened since there was no definitive method to compare trees, from a discriminative point of view and also from lack of efficiency: in computer science, graphs are a complicated structure to work with, and with the growth of the field of application and amount of available data over time, it's no surprise the need for an efficient and personalized way to deal with these problems rises.

In recent years this field of study has widened and our knowledge of this problem deepened since more people started working on it. Probably the most relevant \textbf{metric} (or \textbf{distance}) is \textit{Robinson Foulds} since it can generate a parameter in linear time in the number of vertices of our trees (which is fairly good), but, as stated before, this does not mean that this metric suits all problems, hence the need to create others.

\section{Objectives}
\label{section:objectives}

In our work, we have as objective providing an introduction to comparative metrics for tree graphs structures and dive into the specifics of computing the geodesic distance, provide an implementation and analyze its complexity with detail.

\section{Thesis Outline}
\label{section:outline}

In this thesis we will focus on two main tasks: First, expose an overview of all the available metrics and formalize them mathematically, separating them from the methods to compute them. In this first chapter we approach \textit{Robinson Foulds}, a distance built from the minimum amount of contractions and decontraction of edges between two trees, \textit{Robinson Foulds Lenght}, a variation of \textit{Robinson Foulds} for weighted trees, \textit{Quartets} and \textit{Triplets}, which takes in consideration the similarity between the subtrees containing sets of four and tree leafs respectively and \textit{Triplets Lenghts}, a variant of the latter that takes in consideration edge lengths, the \textit{Geodesic Distance}, which formalizes a space for trees with leaf label set of size $n$ and defines itself as the regular euclidean distance between two trees in this space, the \textit{Maximum Agreement Subtree}, that is determined from the set of leaves of the maximum subtree between the trees we want to compute the distance, \textit{Align} that uses the partitions induced in the two trees by their respective edge sets, \textit{Cophenetic correlation coefficient}, the first method to compare trees that uses a rank established by the scientist of the least deepest vertex in the subtrees for each two leaves, \textit{Node distance}, which is a variant of the latter formalized to improve on its limitations on discriminatory power, \textit{Similarity based on probabily}, characterised by its probabilist approach, the \textit{Hybridization Number}, for acyclic directed graphs and finally the \textit{Subtree prune and regraft}, built with the same heuristic of \textit{Robinson Foulds} but considering the prune and regraft operation, that consists in separating subtrees and join them in other vertex to reach from one tree to the other. In all these we will try to formalized them as mathematically possible and discuss its discriminatory power; Secondly, we will discuss in further detail the implementation of the \textit{GTP Algorithm}, starting with the specification of a different way how to represent weighted trees built from defining an order for the partitions of size two for the set of labels $S$, that actually correspond to the vector in the space formalized for the \textit{geodesic distance}, and then discussing the details behind the computation of the \textit{GTP Algorithm}, such as presenting a worst-case time complexity analysis to a proposed implementation.
 We close this document by briefly presenting the conclusions of our work, such as proposals for future work. You can also check the implementation of all code written for this thesis in the annexes.

 \section{Original Work}
 \label{section:originalwork}

 As mentioned previously, most bibliography written about these metrics is vague, sometimes incomplete and all of them have a different notation for the same concepts. Here we give a uniform formulation for all the metrics, such as providing definitions for multiple distances that sometimes were merely vaguely outlined (for example, the space outlined for the geodesic distance is a two page description in the original paper, no definition is given for it). Plus, the following is also original work:

 \begin{itemize}
 	\item Theorem \ref{thm:rfldif}, together with exposing the problems with the formulation of the \textit{Robinson Foulds Length};
 	\item Theorem \ref{thm:orig1} and Definition \ref{dfn:TC}.
 	\item Chapter \ref{chapter:GeodesicAnalysis} is all original work, even though the \textit{GTP Algorithm} the implementation relies on is not.
 	\item Table \ref{tab:dcomp} and Table \ref{tab:dprob} compiles succinctly our contribution in Chapter \ref{chapter:background}.
 \end{itemize}

\cleardoublepage


\chapter{Background}
\label{chapter:background}

The work in this chapter is mainly expositive and a lifting from a collection of papers and articles. You can find the references in the appropriate section.


Our main goal is to go over these metrics, formalizing them and discussing its relevant aspects such as the advantages, disadvantages and what differences it from the others.

\raggedbottom

\section{Methodology and relevant aspects}
\label{section:methodology}

As stated before, no metric should be considered as default for all problems. Depending on the problem at hand, choosing a metric over the other can be an advantage given the goal we want to achieve.
However, this idea revolves around two important concepts. Since graphs are difficult to handle from a computation point of view, it is an advantage to know how fast are we able to compute the metric. On the other hand, since the output parameter describes how close two trees are the metric might benefit certain properties over others. These two can be referred to as \textbf{efficiency} and \textbf{discriminatory power}, respectively.

\textbf{Efficiency} can be seen from a \textbf{complexity} point of view, but complexity varies with the implementation of the algorithms. Although most of the times the description for the metrics might describe an algorithm, it might exist an equivalent algorithm implementation that is not the literal \textit{translation} from that description but outputs the same values for the same inputs, although has a lower complexity.

The \textbf{Discriminatory Power} will depend on the metric. For instance, some metrics might benefit the tree topology over edge length (or weight) and some might have associated errors or output the same value for some types of trees. Being aware of these properties it is important to recognize the best metric to solve a problem at hand, however (and unlike efficiency) it is not a parameter that we can quantify, so we will go over the discriminatory power of each metric on the appropriate section.

The following work is structured according to the analyzed metrics. First, we will go over the most commonly used metrics and then the others. The latter have a less common use since they may be formalized for a different type of data structures (e.g.: \textit{Hybridization number}) or aren't that practical to compute in present days (e.g.: \textit{Subtree Prune and Regraft}). In each subsection our main focus (besides defining the metrics) will be their \textbf{efficiency} and \textbf{discriminatory power}.

\section{Basic definitions}
\label{section:basicdef}

Next, we'll go over some definitions needed to define the metrics later on.

\begin{dfn}\label{dfn:basic1}{\textbf{Graph theory basic definitions}} \\
Let $G=(V,E)$ where $V$ is a set of vertices (or nodes) and $E$ a set of edges. An edge is a pair $(v_1,v_2)$ of vertices from $V$ (for a lighter notation, one may write $v_1v_2$ instead of $(v_1,v_2)$). $G$ is a \textbf{graph}.\par
A \textbf{path} is a subset $P\subseteq E$ of size $k$ that can be ordered in a way that for the $i$-th element of $P$ $(v_{i,1},v_{i,2})$: $v_{i,1}=v_{i-1,2}$, $v_{i,2}=v_{i+1,1}$. We say that a path is a \textbf{cycle} if, on top of being a \textit{path}, $v_{0,1}=v_{k,2}$. We name a connected graph without cycles a \textbf{tree graph} (or just \textbf{tree}). The length of a \textit{path} $P$ with no cycles is $k=|P|$.\par
In \textbf{undirected graphs} the edges $(a,b)$ and $(b,a)$ are equal. We will work with undirected graphs unless is differently stated\par
Let $u,v\in V$. We say that $v$ is \textbf{neighbor} of $u$ if $(u,v)\in E$ and we write $u\sim v$. In an \textit{undirected graph} this relation is reflexive. The \textbf{degree} of a vertex is the number of neighbors it has.
\end{dfn}

For the next set of concepts, it is important we assume that graph vertices can have \textbf{labels}, this means that exists a function $label:V\longrightarrow \mathrm{STR}\cup\{'\mathrm{NULL}'\}$ that for each vertex returns a \textit{string} (that we call \textit{label}) or $\mathrm{NULL}$. The same way we define a concept of edge \textbf{weight} (or \textbf{length}), as a function $weight:E\longrightarrow\mathbb{R}$. That's actually needed to be considered for some problems and could represent, building over the \textit{phylogeny} application, for example, how many years are between species. In both cases, they are just ways to hold information in these data structures, if needed. 

\begin{dfn}{\textbf{Tree specific concepts}}\\
\label{dfn:basic}
Let $T=(V,E)$ be a \textit{tree graph}.
\begin{itemize}
  \item A \textbf{leaf} is a vertex with degree $1$.
  \item If exists one (and only one) vertex $v\in V$ such that $label(v)='\mathrm{root}'$ then we say that $v$ is the \textbf{root vertex} and that $T$ is a \textbf{rooted tree}. If a \textit{root vertex} does not exist, then $T$ is an \textbf{unrooted tree}. Assuming $T$ is \textit{rooted} we can now define a new set of concepts:
      \begin{itemize}
        \item The \textbf{depth of a vertex} $v$ is the number of edges on the path (that on a tree is singular) from the \textit{root} vertex to $v$. The \textbf{depth of a tree} is the maximum depth between all nodes. (We will refer to the depth of a vertex $v$ as $depth(v)$, and the depth of a tree $T$ as $depth(T)$)
        \item Let $u, v\in V$. We say that $u$ is an \textbf{direct ancestor} of $v$ if $(u,v)\in E$ and $depth(u)<depth(v)$. In this case, $v$ is also a \textbf{direct descendant} of $u$. Also, we generally say that $u$ is an \textbf{ancestor} of $v$ if there is a path of direct ancestors from $v$ to $u$ (similarly we define the same general concept for \textbf{descendant}).
        \item A \textbf{clade} consists of a vertex and all its lineal descendants.
      \end{itemize}

  \item A \textbf{dendrogram} is a tree where only leaves (and, in case of a rooted tree, the root) have labels.

  \item A \textbf{binary tree} is a tree in which every vertex has degree at most $3$.

  \item A \textbf{forest} is a collection $F=\{T_1,T_2,...,T_k\}$ where for every $i$ $T_i$ is a tree.

\end{itemize}
\end{dfn}

\begin{dfn} 
\label{dfn:metric}
Let $d:X\times X \longrightarrow \mathbb{R}^+_0$ be an injective function. We say that $d$ is a \textbf{metric} over $X$ (and $d$ is called the \textbf{distance function}) if: \textbf{(1)} $d(x,y)=0$ if and only if $x=y$; \textbf{(2)} $d$ is symmetrical, that is $d(x,y)=d(y,x)$; \textbf{(3)} $d$ satisfies the triangular inequality, that is for all $x,y,z\in X$ $d(x,z)\leq d(x,y)+d(y,z)$.

\end{dfn}

In regard to efficiency, we now define a notation that will be useful to talk about program complexity.

\begin{dfn}\label{dfn:BON}{\textit{Big-O Notation}}\\
If $f$ and $g$ are two functions from $\mathbb{N}$ to $\mathbb{N}$, then we: \textbf{(1)} say that $f=O(g)$ if there exists a constant $c$ such that $f(n)\leq c \cdot g(n)$ for every sufficiently large $n$, \textbf{(2)} say that $f=\Omega(g)$ if $g=O(f)$, \textbf{(3)} say that $f=\Theta(g)$ if $f=O(g)$ and $g=O(f)$, \textbf{(4)} say that $f=o(g)$ if for every $\epsilon > 0$, $f(n)\leq \epsilon \cdot g(n)$ for every sufficiently large $n$, and \textbf{(5)} say that $f=\omega(g)$ if $g=o(f)$.\\
To emphasize the input parameter, we often write $f(n) = O(g(n))$ instead of $f = O(g)$, and use similar notation for $o$, $\Omega$, $\omega$, $\Theta$.
\end{dfn}

When one refers \textit{complexity} it is common to use \textit{Big-O Notation}, but there are other notations. This will be important to understand how the computation time varies with the size of the input. So if we say that an implementation runs in $O(n)$ time it means time grows linearly with input growth. As expected, \textit{how fast} a program runs will depend on its complexity. Given two functions $f$ and $g$, $f=O(p_f(n))$ and $g=O(p_g(n))$ the function with higher complexity is determined by which of $p_f(n)$ and $p_g(n)$ as a higher growth rate. For example, if $p_f(n)=e^n$ and $p_g(n)=n^5+n^3+10$, then $f$ has a higher complexity.


\section{Usual Approaches}
\label{section:uapproach}

In this section, the reader will find the most used metrics in comparing classifications. \textbf{Reading this first subsection about \textit{Robinson Foulds} and \textit{Robinson Foulds Length} is strongly advised even if it is not your point of interest since might introduce concepts or notation that will be important later on for other metrics}.

\subsection{Robinson Foulds, Robinson Foulds Length}
\label{subsection:rf}

Arguably one of the most used metrics for comparing classifications, \textbf{Robinson Foulds} was the result of the continued work of David F. Robinson and Leslie R. Foulds (published in 1981 on the Mathematical Biosciences journal) to compare \textit{phylogenetic trees}.



Originally, this metric was defined for \textit{binary dendrograms} and the motivation behind this later formalized definition comes from the attempt to know how far was, given two trees, one from the other, considering a specific operation that consisted in \textit{gluing} adjacent vertices and erasing the edge between or (for the inverse operation) splitting one vertex in two new vertices connected by a new edge.

To talk about the background of the \textit{Robinson Foulds distance} we need some extra notation:

\begin{dfn} \label{dfn:lt}
Given a tree $T=(V,E)$ we define $S=\{x\in \mathrm{STR}: \exists v\in V\ s.t.\ label(v)=x\}$ (that is $S=label(V)$) as the \textbf{set of labels} of $T$. A \textbf{labeled tree} consists of a \textit{4-tuple} $T_l=(V,E,label,S)$ where $(V,E)$ is a tree, $label$ a labeling function $label: V\longrightarrow S$ and $S$ the corresponding set of labels.
\end{dfn}

\begin{dfn}\label{dfn:gammas}The set of all labeled trees with $S$ as the set of labels is defined as:
\begin{equation}
\gamma_S = \{(V,E,label,S): label:S\longrightarrow V\}
\end{equation}
For sake of simplicity, we will omit the $label$ function from the elements of $\gamma_S$.
\end{dfn}

One should realize that, given this definition, \textbf{dendrograms} are in fact trees $T\in\gamma_S$ such that the function $label$ is injective and $label(V_T)=\{v\in V_T: degree(v)=1\}$.

\begin{dfn}
\label{dfn:ide}
 We say that two trees are \textbf{identical} if there is a bijective map between them that preserves labeling, meaning that, for two identical labeled trees $A,B\in\gamma_S$ exists $h:V_A\longrightarrow V_B$ bijective, such that $x,y\in V_A$ and $xy\in E_A$ if and only if $h(x)h(y)\in E_B$ and $label(x)=label(h(x))\wedge label(y)=label(h(y))$.
\end{dfn}

\begin{rmk}
\label{rmk:slab}
In all extension of our work we assume that $S$ is the set of labels of the leaves (and usually consists in all natural numbers until some $k\in\mathbb{N}$), meaning that leaves of trees in $\gamma_S$ must be exactly $|S|$ and its labels will be non repeated labels from $S$. The label 'root' is not in $S$. 
\end{rmk}

\begin{dfn}
(\textbf{Operation} $\alpha$) \label{dfn:alphaop} \\
Let $T=(V,E,S)\in\gamma_S$, $|V|=m$ and $v_iv_j\in E$. Then, $\alpha : U_S \longrightarrow \gamma_S$ is a function such that $U_S=\{(T,e):T\in\gamma_S;e\in E_T\}$ and $\alpha(T,v_i v_j)=(V',E',S)$ where:
\begin{itemize}
  \item $V'=(V\cup\{v_{m+1}\})\backslash\{v_i,v_j\}$;
  \item $E'=[(E\backslash E^i)\backslash E^j]\cup \{v_{m+1}v_h: v_hv_i\in E^i\ or\ v_hv_j\in E^j, h\neq i\ or\ h\neq j\}$, where $E^k=\{v_kv_q:v_kv_q\in E, v_q\in V\}$ is the set of edges incident with $v_k$.
  \item $label(v_{m+1})=label(v_i)\cup label(v_j)$
\end{itemize}
\end{dfn}

The reader should understand that this operation does nothing more than to collapsing edges and vertices on their ends into a new vertex $v_{m+1}$. One should note as well that, in this case, the co-domain of the $label$ function is the \textit{powerset} of $STR$, or information on the labels would be lost between $\alpha$ operations.

We will not, similarly to the reference that establishes it, formalize a definition for the operation $\alpha^{-1}$, but one should have a straightforward idea of how it works, being less straightforward only in regards of label and neighborhood attribution between the new vertices. Since the objective is to transform, with applying the minimum amount of $\alpha$ and $\alpha^{-1}$ operations, one tree into another, one should choose the distribution of neighbors and labels according to whats best to reach that end, when applying $\alpha^{-1}$.

The operations $\alpha$ and $\alpha^{-1}$ are also called as \textbf{contraction} and \textbf{decontraction of Bourque} respectively.

This leaves us with the original definition stated:

\begin{dfn}{\textbf{Robinson Foulds distance}}\\
\label{dfn:rf2}
Let $S$ be a set of labels and $A,B\in\gamma_S$. The Robinson Foulds distance between $A$ and $B$, $d'_{RF}(A,B)$, is defined as the minimum number of contractions and decontractions of Bourque necessary to apply on $A$ to get $B$.
\end{dfn}

One should note that, given three rooted trees $A,B,C\in\gamma_S$:

\begin{itemize}
  \item $d'_{RF}(A,B)=0$ then $A$ is identical to $B$;
  \item $d'_{RF}(A,B)>0$ then $A$ is not identical to $B$;
  \item $d'_{RF}(A,B)=d'_{RF}(B,A)$;
  \item $d'_{RF}(A,C)\leq d'_{RF}(A,B) + d'_{RF}(B,C)$.
\end{itemize}

All these items should cause no trouble for the reader to prove as true considering the definition so, in fact, $d'_{RF}$ is a well defined metric.

Another thing to keep in mind is how the operation $\alpha^{-1}$ is not defined formally. However, instead of having to consider decontractions we can instead define the \textit{Robinson Foulds distance} as the minimum $k$ such that there are sequences of trees $A=T_0,T_1,...,T_k=B$ with $T_i\in\gamma_S$ and $e_i\in E_{T_i}$ in which for all $0\leq i < k$ either $T_{i+1}=\alpha(T_i,e_i)$ or $T_i=\alpha(T_{i+1},e_{i+1})$. We will, however, to be coherent with the references, still use the contractions and decontractions of Bourque throughout our work.

The following definition (which is the most usual to state in literature) was proven to be equal to the distance in Definition \ref{dfn:rf2}, even though there is some clear abuse of notation since $C_A$ and $C_B$ aren't built over the same set of vertices and edges (but instead over $V_A$, $V_B$ and $E_A, E_B$ respectively):

\begin{dfn} \label{dfn:rf1} {\textbf{Robinson Foulds distance}}\\
Let $A$ and $B$ be two rooted trees with the same number of leaves and $C_A$ and $C_B$ the set of all clades for $A$ and $B$ respectively. The \textbf{Robinson Foulds} distance $d_{RF}$ is defined as 
\begin{equation}
d_{RF}(A,B)=|C_A\backslash C_B|+|C_B\backslash C_A|
\end{equation}
\end{dfn}

The proof that states that $d'_{RF}$ is actually the same as $d_{RF}$ from Definition \ref{dfn:rf1} can be found in \cite{bib:10}, taking in consideration that in that article the conclusion is reached not in terms of \textit{clades} but of \textit{edges}, but they are actually the same since there is a one-to-one correspondence between edges and clades in these structures (removing an edge from a tree would lead to an unconnected graph with two connected components. The one-to-one correspondence is given by associating that edge with the connected component that contains the deepest vertex of the two vertices the edge was connecting). The main idea to understand the equality lies on the existence of a third \textit{midway} tree $C\in\gamma_S$ between the sequence of applying the $\alpha$ and $\alpha^{-1}$ operations that contains the clades that are both in $A$ and $B$ (for $A,B\in\gamma_S$), and one should account $1$ for each collapsed and generated edge in this process. We'll talk about this \textit{midway} tree later (Definition \ref{dfn:scmsct}).

When it comes to implementation of the algorithm to calculate this metric, most authors refer it as fairly simple, but wasn't until William H. E. Day formalized an algorithm in 1985 that showed that to compute this was actually a linear time problem.

Given the result reached in \cite{bib:10}, the problem shifted from counting the number of $\alpha$ and $\alpha^{-1}$ operations between the two trees (which could be seen as an actual challenge to compute) to counting clades. In William H. E. Day article \cite{bib:11} he actually solves the problem for a group of similar problems in the field of study, which include the implementation of the Robinson Foulds distance as well. With little expression manipulation one can conclude that $d_{RF}$ is actually also equal to, given two trees $A$ and $B$
\begin{equation}
d_{RF}(A,B)=|C_A\backslash C_B|+|C_B\backslash C_A|=|C_A|+|C_B|-2|C_A \cap C_B|
\end{equation}
where $C_X$ is the set of clades of the tree $X$. Since the number of clades of a generic tree is easily calculated in linear time (given the one-to-one correspondence we approached earlier), the problem is reducted to calculate the clades of $A$ that are also clades of $B$.

However, Day refers to \textit{clades} indirectly, since he works with \textbf{clusters} through the whole article, which can be seen as the sets of labels on the clade's leaves. Therefore we can also formalize a \textbf{cluster representation} for trees.

\begin{nota}
\label{nota:day}
  In William Day's work, for every tree $T$ there is a cluster representation given by a set of sets of labels. Each set of labels is, in fact, the set of labels of the leaves for every clade of the tree. This cluster representation is denoted as $T'$.

\end{nota}

To calculate the $d_{RF}$ distance, Day actually formalizes a new structure which he calls as \textit{Strict Consensus Tree}:

\begin{dfn} \label{dfn:scmsct} (\textbf{Strict Consensus Method}, William Day (1985))
Let $S$ be a set of labels and $C: (\gamma_S)^k \longrightarrow \gamma_S$ a function such that, for $T_1, T_2, ..., T_k \in \gamma_S$, 
\begin{equation}
(C(T_1,T_2,...,T_k))'=\bigcap_{1\leq i \leq k} (T_i)'
\end{equation}
In this case, we say that $C$ is a \textbf{strict consensus method} and $C(T_1,T_2,...,T_k)$ the \textbf{strict consensus tree} between $T_1, T_2, ..., T_k$.
\end{dfn}

\textbf{Let us add to Day's definition that not only $C(T_1,T_2,...,T_n)$ has to satisfy its condition, but also it is the smaller tree (in terms of vertex count) to satisfy it.}

This means that, looking back on our Robinson Foulds distance, the $|C_A \cap C_B|$ term of the expression can be rewritten as $|(C(A,B))'|$. The algorithm defined in Day's paper is, in fact, an algorithm to calculate the strict consensus tree between $T_1,T_2,...,T_k\in\gamma_S$ with $|S|=n$ and the conclusion is that this algorithm is capable of doing it in $O(kn)$ time. Implementation, complexity reasoning and respective empirical verification are available in the article \cite{bib:11}.

Even though these results and proofs were published, as referred earlier, in 1981, David Robinson and Leslie Foulds gave us a blink of it in 1978 in \textit{Lecture Notes in Mathematics, vol. 748} \cite{bib:12}. However, this 1978 text was actually a revision of a previously submitted work that never got published: the unpublished work specified the \textit{Robinson Foulds} metric that we just discussed, and the published work specified that this unpublished metric was actually a particular case of a new metric there proposed, particular case in which every edge of the classifications fed to this new distance had weight $1$. Later on we will understand that this was not the case since $d_{RFL}$ has problems in its formulation.

\begin{dfn}
\label{dfn:wlt}
A \textbf{weighted labeled tree} consists of a \textit{4-tuple} $T_{wl}=(V,E,S,w)$ where $(V,E,S)$ is a labeled tree and $w$ the corresponding \textit{weight function} $w:E\longrightarrow\mathbb{R}^+_0$.
The set of all weighted labeled trees with $w$ as weight function and $S$ as the set of labels is denoted by $\gamma^w_S$.  We say that two trees are \textbf{weight-identical} if there is a bijective map between them that preserves labeling and weight of the edges, meaning that, for two weight-identical trees $A,B\in\gamma^w_S$ exists $h:V_A\longrightarrow V_B$ bijective, such that $x,y\in V_A$ and $xy\in E_A$ if and only if $h(x)h(y)\in E_B$, $label(x)=label(h(x))\wedge label(y)=label(h(y))$ and $w(xy)=w(h(xy))$.
\end{dfn}

Once we widen the \textit{Robinson Foulds} metric, we also realize that this new distance would need to account not only for the difference in tree topology but for edge's lengths as well. With this in mind, and not wanting to increase the complexity of the problem, \cite{bib:11} formalizes \textbf{Robinson Foulds Length} after presenting the following definitions:

\begin{dfn}\label{dfn:partfunc} {\textbf{Partitioning Function}}\\
Let $A\in\gamma^w_S$ such that $A=(V,E,S,w)$ and $Z_S$ the set of all proper partitions of $S$ into two subsets. Let $f:E\longrightarrow Z_S$ such that, for every edge $e\in E$, $f(e)$ returns the set of $Z_S$ that corresponds to the partition of $S$ given by (according to \textit{Day}'s notation, Notation \ref{nota:day}) the clusters of both connected components of $(V,E\backslash \{e\}, S, w)$. We say that $f$ is the \textbf{partitioning function} of $A$.
\end{dfn}

\begin{figure}[!htb]
  \centering
  \includegraphics[width=0.61803398875\textwidth]{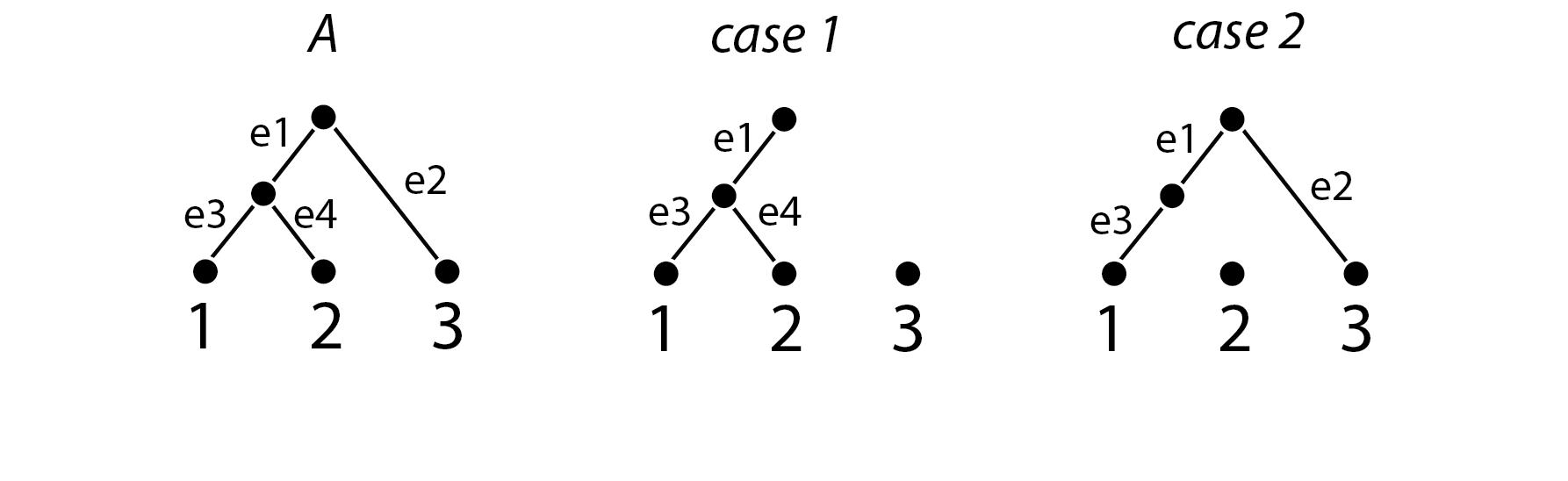}
  \caption[Connected components upon removal of edges]{Depiction of the various connected components of the graph on the left (with leaf set $S=\{``1",``2",``3"\}$) uppon removal of $e_2$ and $e_4$.}
  \label{fig:PF}
\end{figure}

\begin{ex}
  Consider the tree $A\in \gamma^w_{\{``1",``2",`3"\}}$, depicted in Figure \ref{fig:PF}, and with partitioning function $f_A$. In case $1$, removing the edge $e_2$ would lead to the depicted connected components, concluding that $f_A(e_2)=\{\{``1",``2"\},\{``3"\}\}$. In case $2$, the same reasoning will lead us to conclude that $f_A(e_4)=\{\{``2"\},\{``1",``3"\}\}$.
\end{ex}

\begin{dfn} \label{dfn:medg}
  Let $A,B\in\gamma^w_S$ with edge sets $E_A$ and $E_B$ and partitioning functions $f_A$ and $f_B$. The edges $e_A\in A$ and $e_B\in B$ are \textbf{matched} if and only if \begin{equation}f_A(e_A)=f_B(e_B)\end{equation}

\end{dfn}

This last definition can help us build concepts for matching functions from $E_1$ to $E_2$, and actually, one of those is needed for the definition of the \textit{Robinson Foulds Length} distance. Consider that $h_{(A,B)}:E_A\longrightarrow E_B$ is a function that, given $e_A\in E_A$, if exists $e_B\in B$ such that $f_A(e_A)=f_B(e_B)$ then $h_{(A,B)}(e_A)$ is defined and equals $e_B$, undefined otherwise.

\begin{dfn}{\textbf{Robinson Foulds Length distance}}\\
Let $A,B\in\gamma^w_S$, $E_A$, $E_B$, $E_{C(A,B)}$ the respective sets of edges, $f_A$ and $f_B$ the respective partitioning functions and $h_{(A,B)}$ the matching function from $A$ to $B$. The \textbf{Robinson Foulds Length distance} is defined as

\begin{equation} d_{RFL}(A,B)= \Bigg(\sum _{e\in (E_A\backslash E'_A)} w(e) \Bigg) + \Bigg(\sum _{e\in (E_B\backslash E'_B)} w(e)\Bigg) + \Bigg(\sum _{e\in E'_A} | w(e) - w(h_{(A,B)}(e))|\Bigg) \end{equation}

where:

$$ E'_A=\{e_A:e_A\in E_A, \exists e_B\in E_B\ s.t.\ f_A(e_A)=f_B(e_B) \}$$
$$ E'_B=\{e_B:e_B\in E_B, \exists e_A\in E_A\ s.t.\ f_B(e_B)=f_A(e_A) \}$$

\end{dfn}


After defining this new metric there's a few important things to note: first, how this distance behaves in a usual scenario considering its application, secondly, the relation between $d_{RF}$ and $d_{RFL}$, the relation with the sets $E'_A$ and $E'_B$ with the strict consensus tree of $A$ and $B$ and how that translates into an algorithm implementation for $d_{RFL}$.

Regarding the first topic, one should understand that, for \textit{identical} trees $A,B\in\gamma^w_S$, since exists a bijective function $h_V:V_A\longrightarrow V_B$, the matching function $h:E_A\longrightarrow E_B$ can actually be given (informally) by $h(v_1v_2)=h_V(v_1)h_V(v_2)$. This implies that the matching function $h_{(A,B)}$ is \textbf{bijective} and $(E_A=E'_A)\ \wedge\ (E_B=E'_B)$ and, as a consequence (reinforcing that this only holds for $A,B\in\gamma^w_S$ identical) \begin{equation} d_{RFL}(A,B)= \sum _{e\in E'_A} | w(e) - w(h_{(A,B)}(e))|^k,\ for\ k=1 \end{equation}
The fact that usually, given the field of application, trees are \textit{roughly identical} influenced that most literature that covers \textit{Robinson Foulds Length} only considers this part of the distance function to discriminate distance between trees. That can be seen, for example, in \cite{bib:1} which also refers to variations of $d_{RFL}$ by raising every part of the sum by a power of some $k\in\mathbb{N}$, which is the case o Kuhner and Felsenstein (1994).

\begin{figure}[!htb]
  \centering
  \includegraphics[width=0.61803398875\textwidth]{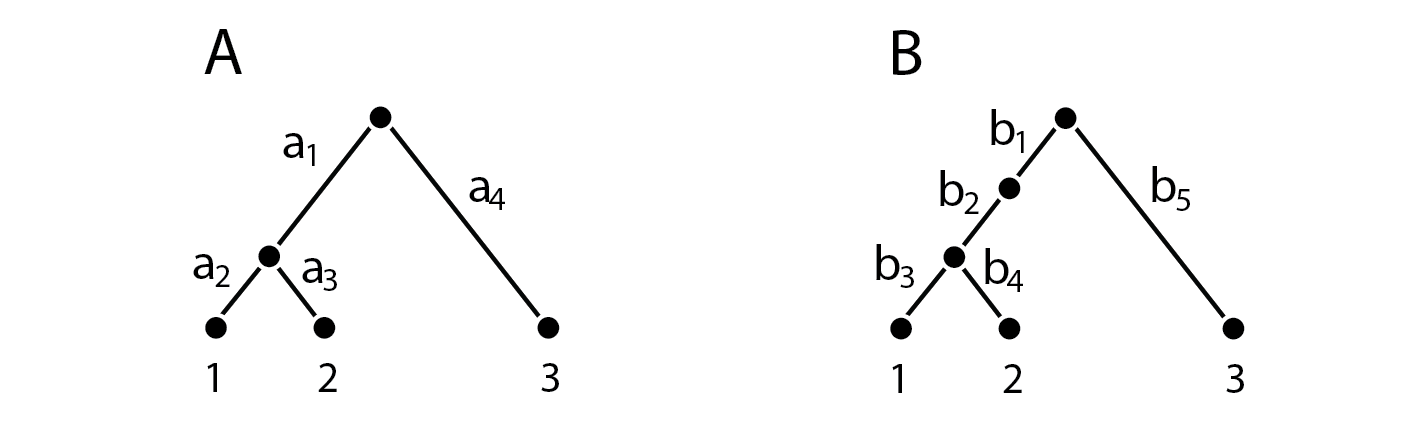}
  \caption[Two similar trees with undefined edge weights]{Trees $A,B\in\gamma^w_S$ with undefined edge weights.}
  \label{fig:RFLe}
\end{figure}

One interesting thing to note is the limitations of $d_{RFL}$, since can give multiple results for the same input and it isn't symmetric. Consider the trees $A,B\in\gamma^w_S$ in Figure \ref{fig:RFLe}.

It should be trivial to understand that:
\begin{itemize}
  \item The edge sets $E'_A$ and $E'_B$ equal $E_A$ and $E_B$ respectively;
  \item There are two possible functions $h_{(A,B)}:E_A\longrightarrow E_B$: one that maps $a_1$ into $b_1$ and other that maps $a_1$ into $b_2$;
  \item $h_{(B,A)}:E_B\longrightarrow E_A$ maps $b_1$ and $b_2$ in $a_1$.
\end{itemize}

Given this, one should also realize that we have two different values for $d_{RFL}(A,B)$ depending on the chosen $h_{(A,B)}$ function. These are results of several problems in \cite{bib:12}: Theorem 4 proves the existence of $h$ matching function between identical trees, however, there's no unicity assured; The authors define (differently to Day, Notation \ref{nota:day}) $T'_1$ and $T'_2$ as the trees generated by collapsing edges in $E_1\backslash E'_1$ and $E_2\backslash E'_2$ respectively as identical, but in the example we just exposed that does not hold; Definition 5 assumes a unique $h$ between $T'_1$ and $T'_2$ but that might not be the case, as we just exemplified.

But the problems don't end here. We ask the reader to check if the symmetry and identity of indiscernibles holds in $d_{RFL}$ for all $A,B\in\gamma^w_S$, which the second can be easily refuted by considering the weight of every edge in $A$ and $B$ of our example as $1$.

From a computational standpoint, the challenge of implementing $d_{RF}$ and $d_{RFL}$ are approximately the same, but not without realizing the relation between the $E'_A$ and $E'_B$ edge sets with $C(A,B)$, given $A,B\in\gamma^w_S$. If we take the definition of the \textit{strict consensus tree} (Definition \ref{dfn:scmsct}) for two trees and ask what's its edge set, we understand that it must consist of a set of edges that must be matched in both trees. \textbf{If we assume that there's an one-to-one correspondence} between the \textbf{connected components} (or \textbf{edges}) from the definition of \textit{partitioning function} (Definition \ref{dfn:partfunc}) and the \textbf{clusters} from the \textit{cluster representation} (which may not be the case, as we shown with Figure \ref{fig:RFLe}) we can prove the existence of unique bijective functions between $E'_A$, $E'_B$ and $E_{C(A,B)}$. This implies that the complexity of computing sets $E'_A$ and $E'_B$ is reducted to the complexity of computing the tree $C(A,B)$ which, by William Day's work, we know it is $O(n)$.

\begin{dfn}
Let $A\in\gamma^w_S$ and $B$ a subtree of $A$. $S\mid_B$ is the subset of $S$ in which its elements are labels of some vertex in $B$. Also, let $v\in V_A$. We define $A(v)$ as the subtree of $A$ that consists of $v$ and all its lineal descendants.\par
\textbf{Also, for the next proof, for any trees $A,B\in\gamma_S$ we'll denote by $A'$ and $B'$ as the trees generated by collapsing all the edges in $A$ and $B$ that belong in the set $E_A\backslash E'_A$ and $E_B\backslash E'_B$ respectively, and $(A)'$ and $(B)'$ as the cluster representation of $A$ and $B$ respectively.}\par
Assume as well that for every edge $uv\in E_X$ for $X\in\gamma^w_S$, $v$ is deeper than $u$.
\end{dfn}

\begin{thm} \label{thm:rfldif}
Let $A,B\in\gamma^w_S$, $f_A,f_B$ the respective partitioning functions and $C(A,B)$ their strict consensus tree. Assuming $(\dag)$ there's a one-to-one correspondence between edges of $A$ and $B$ and their respective clades and $(\dag\dag)$ for all $a\in E_A$ there's no $a'\in E_A$ such that $S\mid _{A(a)}= S\backslash \big( S\mid _{A(a')}\big)$, there's bijective matching functions $h_{(C(A,B);A')}$ and $h_{(C(A,B);B')}$.
\end{thm}

\begin{proof}

Let $xy\in E_{C(A,B)}$. By definition of the strict consensus tree $C(A,B)$
\begin{equation}(S\mid_{C(A,B)(y)})\in (A)' \wedge (S\mid_{C(A,B)(y)})\in (B)' \end{equation}
by the $(\dag)$ property, we have that exists one and only one $a_1a_2\in E_A$ and one and only one $b_1b_2\in E_B$ such that \begin{equation}(S\mid_{C(A,B)(y)})=(S\mid_{A(a_2)})=(S\mid_{B(b_2)})\end{equation} which is equivalent to state that \begin{equation} f_A(a_1a_2)=\{(S\mid_{C(A,B)(y)}),S\backslash (S\mid_{C(A,B)(y)})\}=f_B(b_1b_2).\end{equation}
which implies that $a_1a_2$ and $b_1b_2$ are matched edges, hence $a_1a_2\in E'_A$ and $b_1b_2\in E'_B$. Then, we can establish that our matching functions will be such that $h_{(C(A,B);A')}(xy)=a_1a_2$ and $h_{(C(A,B);B')}(xy)=b_1b_2$.

We'll now prove that $h_{(A;C(A,B))}$ is bijective (the proof for $h_{(C(A,B);B)}$ will be left for the reader).
Let $a_1a_2=a$ and $a'_1a'_2=a'$ be edges of $A$. If $h_{(A;C(A,B))}(a)=h_{(A;C(A,B))}(a')$ then $f_A(a)=f_A(a')$. Since there's a one-to-one correspondence between the edges of $A$ and it is clades $(\dag)$ and it can't be the case that $S\backslash (S\mid_{A(a_2)})= S\mid_{A(a_2')}$ $(\dag\dag)$ we have that $a$ must be equal to $a'$, proving the injectivity of $h_{(A;C(A,B))}$. For surjectivity, let $xy\in E_{C(A,B)}$. By definition of strict consensus tree, we have that $S\mid _{C(A,B)(y)}\in (A)'$. We have that, by $(\dag)$, $\exists a_1a_2\in E_A$ such that $(S\mid_{A(a_2)})=(S\mid_{C(A,B)(y)})$. This means that $f_A(a_1a_2)=f_{C(A,B)}(xy)$ hence $h_{(A;C(A,B))}(a_1a_2)=xy$, proving surjectivity.
\end{proof}

Regarding the \textbf{discriminatory power}, one should understand that, other than the characteristics inherited from the fact that $d_{RF}$ is purely a topological measure and $d_{RFL}$ considers branch length, \textbf{Robinson Foulds} inspired metrics (at least the ones discussed here) are really sensitive to the scalability of $S$ \cite{bib:4}.

\begin{figure}[!htb]
  \centering
  \includegraphics[width=0.61803398875\textwidth]{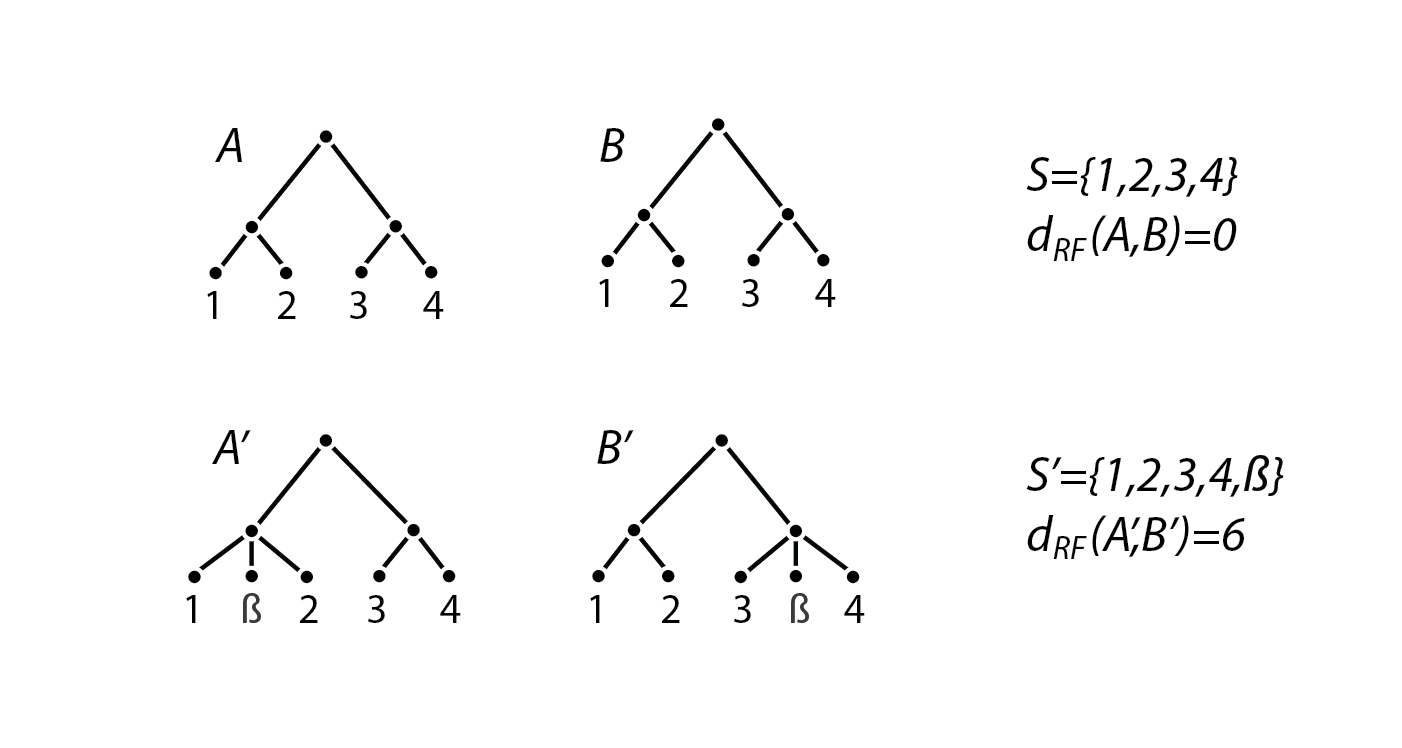}
  \caption[Trees in $\gamma_S$ and $\gamma_{S'}$ and $d_{RF}$ between them.]{Trees in $\gamma_S$ and $\gamma_{S'}$ and the \textit{RF} distance between them.}
  \label{fig:RFOutl}
\end{figure}

For instance, if we have $A,B\in\gamma_S$ and $d_{RF}(A,B)=k$ for some $k$, having $A',B'\in\gamma_{S'}$ such that $A$ and $B$ are \textit{non-trivial} subtrees of $A'$ and $B'$ respectively (meaning: $S\varsubsetneq S'$, $V_A\varsubsetneq V_{A'}$, $V_B\varsubsetneq V_{B'}$, $E_A\varsubsetneq E_{A'}$, $E_B\varsubsetneq E_{B'}$), there's no direct relation between $d_{RF}(A',B')=k$ whatsoever, since all clades that were shared between $A$ and $B$ can now be different. The same applies for $d_{RFL}$. To illustrate this, consider the Figure \ref{fig:RFOutl} for $d_{RF}$.

So, would be reasonable to assume that if one wants to adopt an iterative method for his problem on size of $|S|$, \textit{RF} based metrics wouldn't be a good approach.

Not only the scalability of $S$ is a problem, but moving a single leaf could lead to great discrepancies in the distance value. All around, \textit{RF} metrics end up being a fairly unstable metric to work with, although serving its purpose for distinguishing trees \cite{bib:1}. All due to the fact that it is a metric that works with shared clades: if a leaf is replaced on the tree, all the clades which it belongs to will necessarily be different. Another consequence of this fact is that \textit{RF} will overperform in close to resolved trees other than to unresolved ones \cite{bib:1,bib:16} (being an unresolved tree a tree which the internal nodes have mostly degree greater than $3$).


\subsection{Quartets, Triplets and Triplets Length }
\label{subsection:trip}

The methods that we're about to introduce are recent compared to previous ones and results on complexity differ according to the features of the considered data structures. First approach was made in 1985 by George F. Estabrook, F. R. McMorris and Christopher A. Meacham with the publication of the \textit{Quartets distance} in \cite{bib:16} and wasn't until 11 years later when Douglas E. Critchlow, Dennis K. Pearl and Chunlin Qian provided a formal definition for the \textit{Triplets distance} in \cite{bib:14} which is heavily inspired by the former. In 2014 Mary K. Kuhner and Jon Yamato made a study to compare practical performance of a variety of different metrics \cite{bib:1} and for that matter thought it was interesting to consider a metric that would take the topology analysis properties of the latter but consider branch length as well.

\begin{figure}[!htb]
  \centering
  \includegraphics[width=0.61803398875\textwidth]{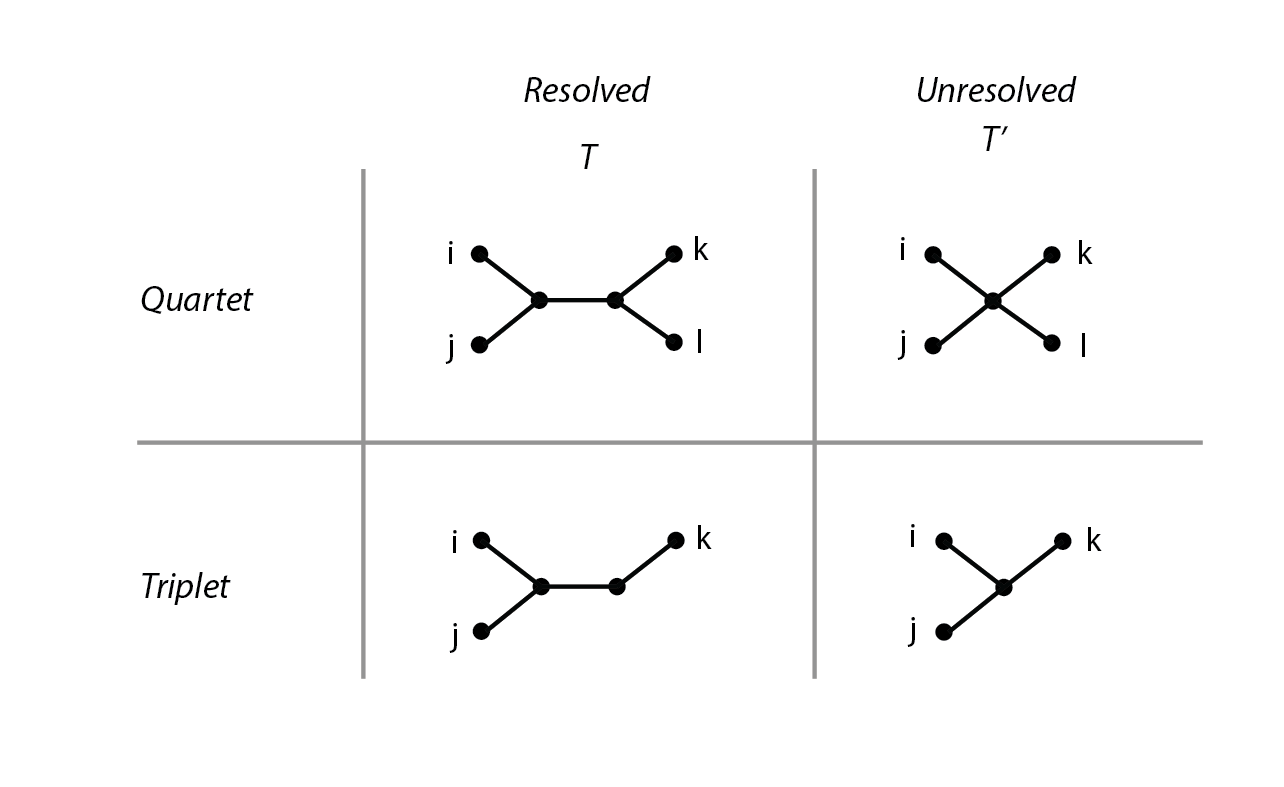}
  \caption[Resolved and unresolved quartets and triplets.]{Resolved and unresolved quartets and triplets.}
  \label{fig:QTTopo}
\end{figure}

The initial thought behind Estabrook, \textit{et al.} \textit{Quartet Distance} was how phylogenetic tree agreement behaves with respect to the topologic aspect of the branching alone, disregarding direction. However, and as stated previously, the problem differs regarding the structure we're applying the distance: \textit{binary} trees only lead to \textbf{\textit{resolved quartets/triplets}} while \textit{non-binary} can lead also to \textbf{\textit{unresolved quartets/triplets}}. These quartets/triplets can be consulted in Figure \ref{fig:QTTopo}.

\begin{dfn}\label{dfn:quart1}
{\textbf{Quartet Distance} (informal)}\\
Let $A,B\in \gamma_S$, $V_X$ the vertex set for any tree $X\in \gamma_S$, $T$ the quartet depicted in Figure \ref{fig:QTTopo} for the resolved case and $\bar{d}(a,b)$ the usual edge distance between vertices $a,b\in V_X$ in the tree $X$. The \textbf{Quartet distance} $d_Q$ consists in:
\begin{itemize}
 \item Consider every subset $S'$ of size $4$ from the set of leaves $S$;
 \item Build maps $\sigma_A$ and $\sigma_B$ (in case they exist) from the vertex set of the subtrees of $A$ and $B$ generated by considering only edges connecting leaves from $S'$ (that we will designate as $A\mid _{S'}$ and $B\mid _{S'}$) to the vertex set of $T$ such that, given $X\in\{A,B\}$, for every $v_1,v_2,v_3\in V_{X\mid_{S'}}$: \begin{equation}\bar{d}(v_1,v_2)\leq \bar{d}(v_1,v_3) \Rightarrow \bar{d}(\sigma_X(v_1),\sigma_X(v_2)) \leq \bar{d}(\sigma_X(v_1),\sigma_X(v_3))\end{equation}
 
 \item Build partitions $T^p_{A\mid_{S'}}$ and $T^p_{B\mid_{S'}}$ for the labels $S'$ such that, for all $X\in\{A, B\}$, $v_1,v_2\in V_{X\mid_{S'}}:\ v_1, v_2\ \mathrm{labeled\ vertices}$: \begin{equation}(\bar{d}(\sigma_X(v_1),\sigma_X(v_2))=2)\Rightarrow (\{label(v_1),label(v_2)\}\in T^p_{X\mid_{S'}})\end{equation}
 \item If $T^p_{A\mid_{S'}}\neq T^p_{B\mid_{S'}}$ or exactly one of the maps $\sigma_A$ and $\sigma_B$ does not exist, account $1$ for the quartet distance $d_Q$. 
\end{itemize}
\end{dfn}

If exactly one of the mappings $\sigma_A$ or $\sigma_B$ does not exist, it means one of the quartets is unresolved in tree $A$ or $B$, so they necessarily differ. If both mappings $\sigma_A$ and $\sigma_B$ don't exist, it means that in both trees, $A$ and $B$, the quartet is unresolved, hence, they agree.

However, in article \cite{bib:11} for \textit{Triplet distance} is presented an informal definition that we find more suitable to understand the concept behind these two metrics (\textit{Triplets} and \textit{Quartet distances}), however, this falls short by semantic reasons.

\begin{dfn}{\textbf{Quartet and Triplet distance} (informal)}\\
Let $A,B\in\gamma_S$. Consider $S'$ as every subset of $S$ of size $k$ and the indicator function defined as
\begin{equation} I_{S'}=
\begin{cases} 
      1 & \mathrm{if\ labels\ on\ }S'\mathrm{\ have\ different\ }\mathrm{subtrees\ in\ }A\mathrm{\ and\ }B \\
      0 & \mathrm{otherwise} 
\end{cases}\end{equation}

Then, the \textbf{Triplet distance} $d_{Trip}(A,B)$ (for $k=3$) and \textbf{Quartet distance} $d_{Q}(A,B)$ (for $k=4$) are given by \begin{equation} \sum_{S'\subset S\ :\ |S'|=k} I_{S'}\end{equation}

\end{dfn}

The problem with this definition is that the \textit{different subtrees} referred in indicator $I_{S'}$ isn't the straightforward notion of \textit{different}. Actually, to achieve the comparison between subtrees that Critchlow, \textit{et al.} (1996) (from \cite{bib:14}) are referring, one would need to erase every label information on $A$ and $B$ other than $S'$ and then consider the trees $A'$ and $B'$ generated with the smallest amount of contractions of Bourque $\alpha$ from $A$ and $B$ with the same topology as $T$ or $T'$ (depending on which one requires least contractions) from Figure \ref{fig:QTTopo} and at most $1$ label for each leaf, $0$ labels for internal nodes. \textbf{To obtain the \textit{Triplet distance} from the Definition \ref{dfn:quart1} (of the \textit{Quartet distance}) we need to consider subsets $S'$ of size $3$ instead of size $4$ and consider $T$ from Figure \ref{fig:QTTopo} for triplets instead of quartets}.

In the article by Kuhner \textit{et al.} \cite{bib:1} a new metric was considered to maintain $d_{Trip}$'s topologic discriminatory power and also account for weight information. The main idea was to account the path weight between the tips with labels from $S'$, but do it when the topology of the \textit{``subtrees"} is \textbf{equal} rather than different. Considering the definition from Critchlow, \textit{et al.} (1996) \cite{bib:14} as a starting point, we define the \textit{Triplet Length distance} below:

\begin{dfn}{\textbf{Triplet Length distance} (informal)}\\
Let $A,B\in\gamma_S$. Consider $S'=\{i,j,k\}$ as every subset of $S$ of size $3$, $v_{(i,X)}\in V_X$ the singular vertex in $X$ with label $i$ and the indicator functions $I_{S',1}$ and $I_{S',2}$ defined as
\begin{equation} I_{S',1}=
\begin{cases} 
      0 & \mathrm{if\ labels\ on\ }S'\mathrm{\ have\ different subtrees\ in\ }A\mathrm{\ and\ }B \\
      |\bar{d_l}(v_{(i,A)},v_{(j,A)})-\bar{d_l}(v_{(i,B)},v_{(j,B)})| & \mathrm{otherwise} 
\end{cases}\end{equation}

\begin{equation} I_{S',2}=
\begin{cases} 
      0 & \mathrm{if\ labels\ on\ }S'\mathrm{\ have\ different subtrees\ in\ }A\mathrm{\ and\ }B \\
      |\bar{d_l}(v_{(i,A)},v_{(k,A)})-\bar{d_l}(v_{(i,B)},v_{(k,B)})| & \mathrm{otherwise} 
\end{cases}\end{equation}

Where $\bar{d_l}(v,u)$ is the weight of the path between $u$ and $v$. Then, the \textbf{Triplet Length distance} is defined as \begin{equation} d_{TripL}(A,B)=\sum_{S'\subset S\ :\ |S'|=3} (I_{S',1}+I_{S',2})\end{equation}

\end{dfn}

This metric is fairly recent, considered for the purposes of the study in \cite{bib:1} and its relevance is underwhelming (as we will state and can be checked on the results of the article).

When comes to \textbf{implementation}, an easy way to structure all the possible situations when analyzing the \textit{quartets} or \textit{triplets} is depicted in Table \ref{tab:QTpos}.

\begin{table}[h]
\centering

\label{tab:QTpos}
\begin{tabular}{l|l|l}
                          & Resolved      & Unresolved           \\ \hline
\multirow{2}{*}{Resolved} & $A$: Agree    & \multirow{2}{*}{$C$} \\ 
                          & $B$: Disagree &                      \\ \hline
Unresolved                & $D$           & $E$                 
\end{tabular}
\caption[Categorization of the different types of Quartets and Triplets.]{Categorization of the different types of Quartets/Triplets, necessary for the computation of the metrics.}

\end{table}

And actually, as Brodal, \textit{et al.} specifies in \cite{bib:13}, a paper focused on efficient algorithms to compute Triplet and Quartet distance, the lines and rows of this table can be calculated in $O(n)$ time through \textit{dynamic programming}. Since the quartet and triplet distance consists only in adding up $B+C+D$, then the main idea is to find a way of computing $E$ and $A$, and that's the focus of \cite{bib:13}.

The conclusion is that the algorithm for finding $A$ and $E$ differs in complexity depending on the structure: For rooted trees and triplets, $A$ and $E$ can be computed in time $O(nlog(n))$; unrooted trees and quartets, $A$ can be computed in $O(nlog(n))$ and $B$ in $O(dnlog(n))$ where $d$ is the maximum vertex degree of any node in the two trees.

As for the \textbf{discriminatory power} is interesting to understand how these metrics relate to the scalability of $S$, where \textit{RF} metrics underperforms. Actually, if we consider $A_0,B_0\in\gamma_{S_0}$ and a chain of non-trivial \textit{supertrees} $A_i,B_i\in\gamma_{S_i}$ for $i\in\mathbb{N}$ where $S_i\varsubsetneq S_{i+1}$ and $A_i$ and $B_i$ are non-trivial subtrees of $A_{i+1}$ and $B_{i+1}$ respectively, it is reasonable to understand that $d_{Q}(A_i,B_i)\leq d_{Q}(A_{i+1}, B_{i+1})$ and $d_{Trip}(A_i,B_i)\leq d_{Trip}(A_{i+1}, B_{i+1})$. This is due to the fact that these metrics value relations between subsets of size $k$ ($k=4\vee k=3$), and any change done to leaves will only affect the part of the sum related to quartets/triplets where that leaf is contained.

However, regarding its practical performance on the main field of application (phylogenetics) \cite{bib:1}, \textit{Quartet} based metrics didn't perform well, and according to Kuhner, \textit{et al.} is due to these distances being more sensitive to the bottommost branchings of the tree, once a large portion of these branches are contained in these branchings. This last conclusion might be too specific for the dataset in the reference but, if that's the case, this will lead us to believe that \textit{Quartet} based metrics will overperform in unresolved trees over resolved ones (being an unresolved tree a tree which the internal nodes have mostly degree greater than $3$).

\subsection{Geodesic distance}
\label{subsection:geo}

The most recent metric that brought original insight for the classification problem was the product of Louis J. Billera, Susan P. Holmes and Karen Vogtmann. Since the classical problem of phylogeny is to find a tree which is more consistent with the \textit{taxonomical} data, knowing how much the calculated tree is correct becomes also a statistical problem: would a small change in the data will result in a change of choice in the resulting tree (as we saw this is a limitation of \textit{Robinson Foulds metric}). The fact that this can be considered as a problem in the estimation process lead various authors to suggest to partition the space of trees into regions, and that's what Billera, \textit{et al.} specifies in \cite{bib:15}. The \textbf{Geodesic distance} is a distance built over the space of trees.

We are first left with the question of how many minimal (with no edges $ab,bc\in E$ such that $degree(b)=2$ and $b$ is not the root) non-identical (Definition \ref{dfn:ide}) binary trees (Definition \ref{dfn:basic}) exist. This will be a key factor for the space we will later formalize.

\begin{thm}
  The number of minimal non-identical binary trees with $n$ leaves is $(2n-3)!!$ (where $!!$ stands for the \textbf{double factorial}).
\end{thm}

Most literature points to \cite{bib:35} for a proof, but this can be instead done in the following way: given trees with $n$ leaves ($S=\{1,2,...,n\}$) and root $r$, if we identify all the possible $n-2$ internal vertices (except the root) as $a_1,a_2,...,a_{n-2}$, the Pr{\"u}fer code (you can find a description in Reference \cite{bib:36}) gives us a bijection between those trees and sequences of size $2n-3$ with one $r$ and two of each $a_i$ (for $1\leq i \leq n-2$). Therefore, we can conclude that the number of sequences is

\begin{equation}
\frac{(2n-3)!}{2^{n-2}}
\end{equation}

Finally, we need to remove from this counting the trees with permutations of labels on the internal vertices $a_1,a_2,...,a_{n-2}$, which are $(n-2)!$. Then we obtain the desired result:

\begin{equation}
\frac{(2n-3)!}{2^{n-2}(n-2)!}=(2n-3)!!.
\end{equation}

That leaves us with the task of primarily formalizing the space where the distance will be built. Take into consideration that an \textbf{internal edge} is any edge that is not connected to a leaf of a tree. Do not forget that when we refer minimal trees we are specifically referring to our context previously explained (that these have no edges $ab,bc\in E$ such that $degree(b)=2$ and $b$ is not the root).

\begin{dfn} \label{dfn:stsn} \textbf{Space of trees with $n$ labels, $\mathcal{T}_n$}\\
Consider $S$ a set of labels and $|S|=n$. For every minimal non-identical (Definition \ref{dfn:ide}) binary tree (Definition \ref{dfn:basic}) $B_i\in\gamma_S$ (there is a total of $(2n-3)!!$ minimal non-identical binary trees \cite{bib:35}) generate an $(n-2)$-dimensional space $\mathcal{T}^o_B$ (that we designate as \textbf{orthant}) such that every component $c_e$ is identified to one (and only one) internal edge $e\in E_B$ and takes real values between $[0 , \infty [$. For all pairs of spaces $\mathcal{T}^o_{B_1}$ and $\mathcal{T}^o_{B_2}$ (with $e_1\in E_{B_1}$ and $e_2\in E_{B_2}$) identify components $c_{e_1}$ and $c_{e_2}$ if and only if the cluster representation of the clades associated with the removal of $e_1$ and $e_2$ from their respective trees match, that is, if $(C_{e_1})'=(C_{e_2})'$ (or, according to Definition \ref{dfn:medg}, $e_1$ and $e_2$ are matched edges). The \textbf{Space of trees with $n$ labels} is the result all the orthants $\mathcal{T}^o_B$ with this identification.
A point $(t_1,t_2,...,t_k)\in\mathcal{T}_n$ specifies a unique tree $A\in \gamma^w_S$ with internal edges $e_i$ such that $w(e_i)=t_i$ for all $t_i\neq 0$. 
\end{dfn}

For better understanding, consider the following two examples:

\begin{ex} \textbf{Space of trees with $3$ labels}, $\mathcal{T}_3$ \\
  The topology of binary trees with $3$ labels is unique, so, if we consider the set $S=\{1,2,3\}$, there are only $3$ minimal non-identical binary trees ($(2\times 3-3)!!=3$), depicted Figure \ref{fig:T3rep}. Each of those will generate a $1$-dimensional space, that will meet by their origin.

\begin{figure}[!htb]
  \centering
  \includegraphics[width=0.8\textwidth]{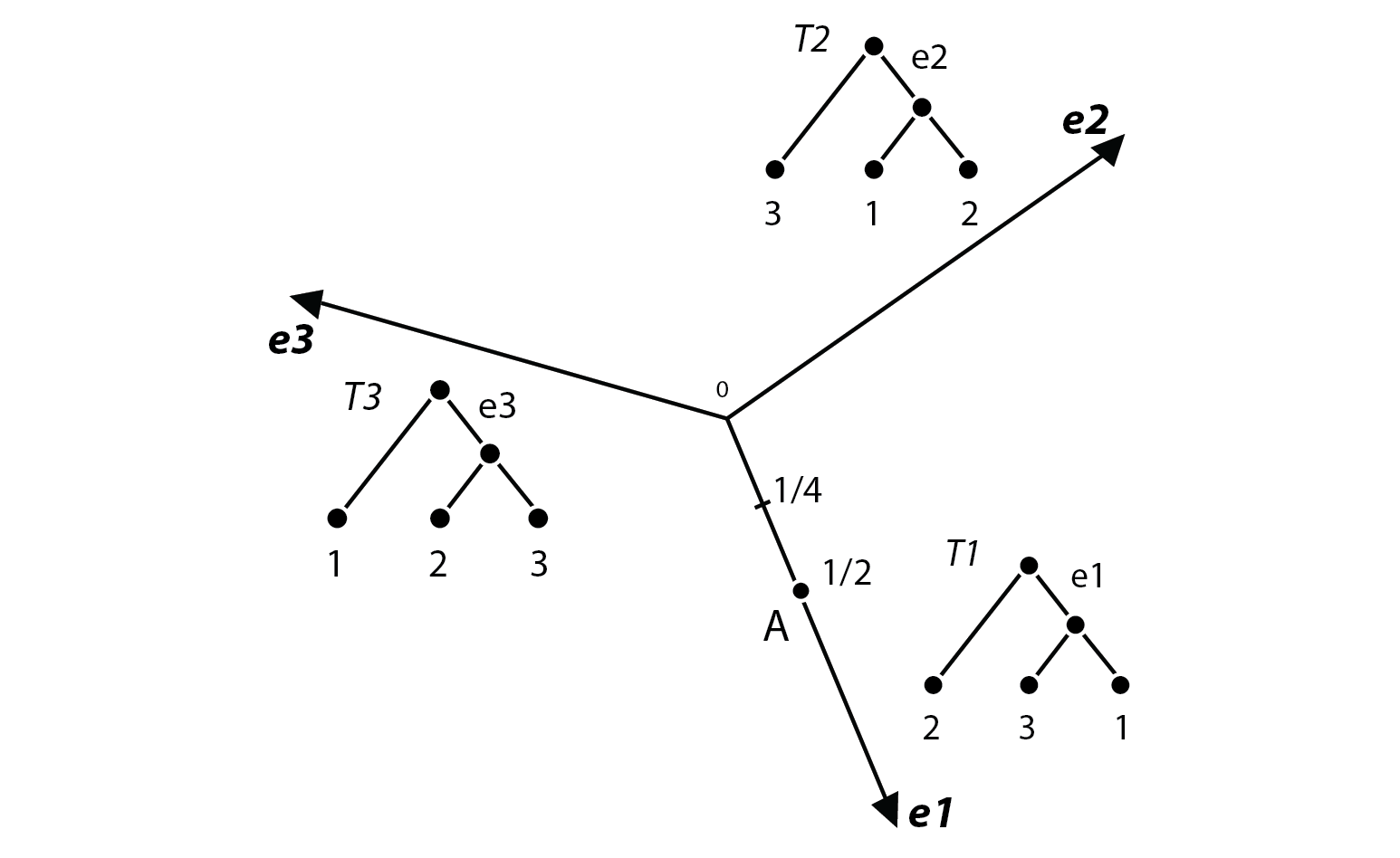}
  \caption[Space of trees with label set $S=\{1,2,3\}$.]{$\mathcal{T}_3$}
  \label{fig:T3rep}
\end{figure}

The tree at point $A$ is a tree whose topology and labeling are equal to those of $T1$, however, $w(e1)=1/2$.

\end{ex}

\begin{ex} \textbf{Space of trees with $4$ labels}, $\mathcal{T}_4$ \\
  Let $S=\{ 1,2,3,4\}$. The dimension of each orthant will be $(n-2)=(4-2)=2$, meaning that each binary tree will have exactly two internal edges. Also, there are $(2\times 4 - 3)!!= 5!! = 5\times 3\times 1 = 15$ different binary trees. Take the next figure as an example of one of its composing orthants $\mathcal{T}^o_B$, for the depicted $B\in\gamma_S$.

\begin{figure}[!htb]
  \centering
  \includegraphics[width=0.5\textwidth]{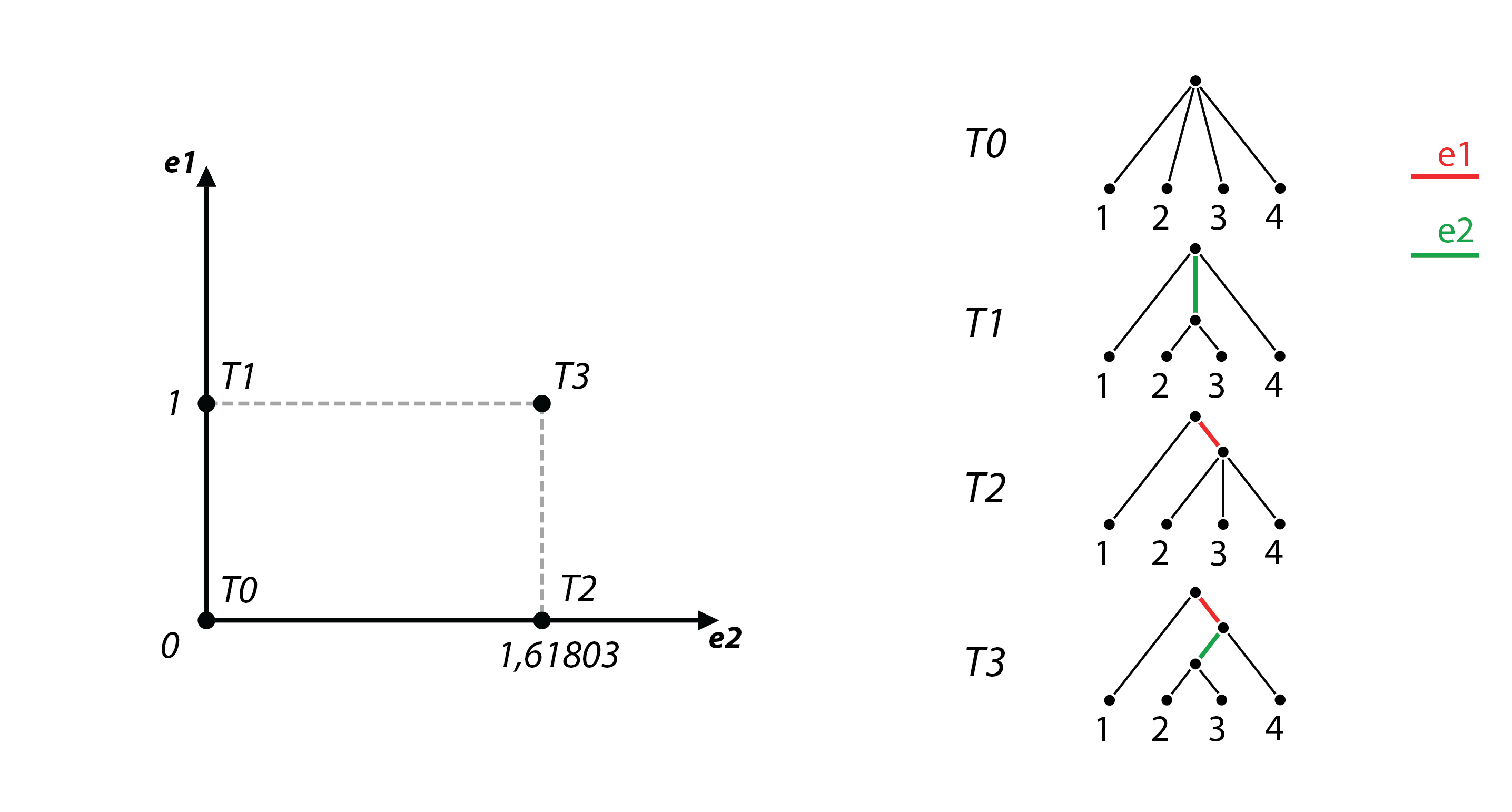}
  \caption[Various trees in orthant of $\mathcal{T}_4$.]{$\mathcal{T}^o_B$, where $B\in\gamma_S$ is the binary tree identical to $T3$. Given $e_1,e_2\in E_{T3}$, we have that $w(e1)=1$ and $w(e2)=1.61803$.}
  \label{fig:T4Orep}
\end{figure}

As a matter of fact, for the topology of trees with $4$ labels, we have five different candidates that can be seen in Figure \ref{fig:T4O5rep}. And also, the identification of edges will be such that the five orthants associated with these trees will be joined by components two by two.

\begin{figure}[!htb]
  \centering
  \includegraphics[width=0.8\textwidth]{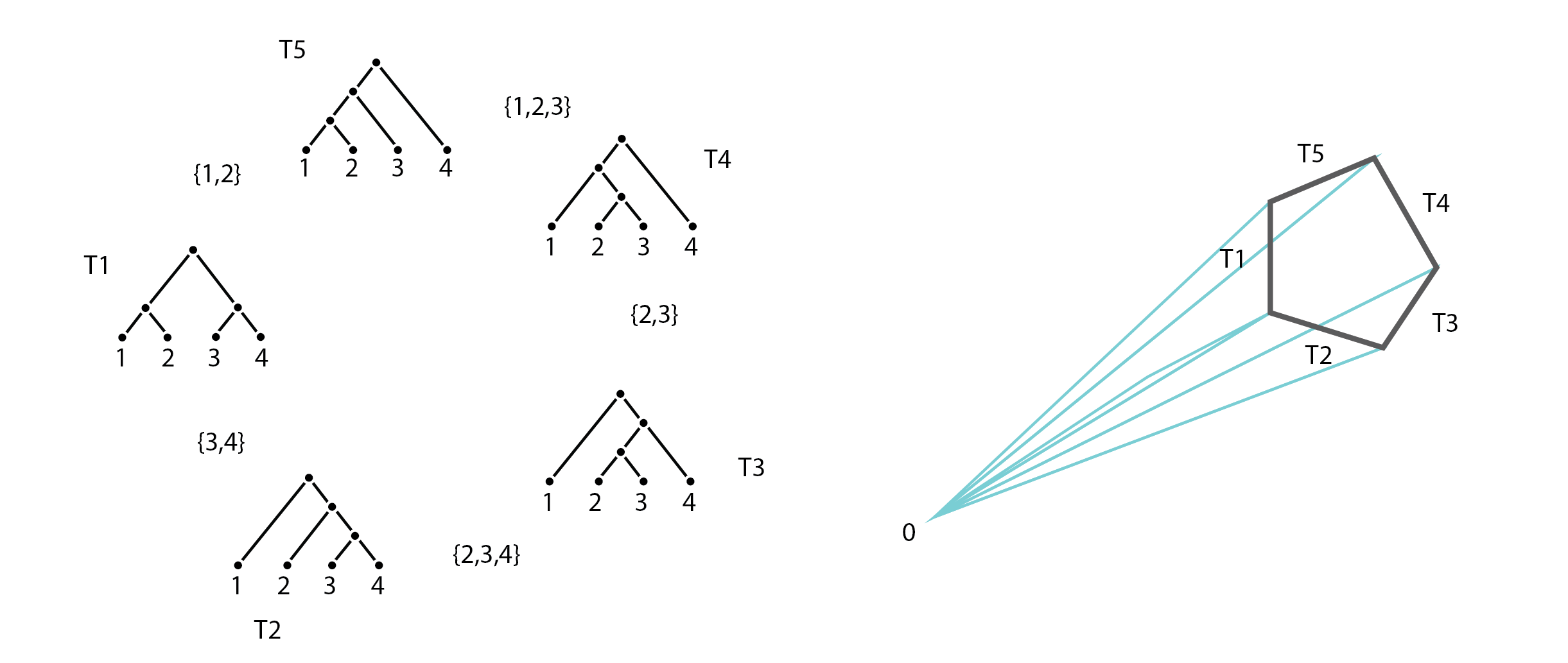}
  \caption[Five different topologies for trees with label set $S=\{1,2,3,4\}$.]{The five different topologies for trees with $4$ labels and the representation of respective orthants after component identification. It is important to note that, since $\mathcal{T}_4$ isn't presentable in $\mathbb{R}^3$ we will loosen our representations a bit for better understanding. Also note that each triangle in this figure it is actually just a region of each orthant, equivalent to the one in Figure \ref{fig:T4Orep}: the five orthants are actually an infinite \textit{`cone"} on the one depicted (with cone point the origin).}
  \label{fig:T4O5rep}
\end{figure}

Since we can permutate the labels on trees, $\mathcal{T}_4$ will actually be composed of several of these spaces, until we cover all $15$ minimal non-identical binary trees. If we consider all $15$ orthants and do the respective identifications we will end up with what can be seen in Figure \ref{fig:T4rep}.

\begin{figure}[!htb]
  \centering
  \includegraphics[width=0.61803398875\textwidth]{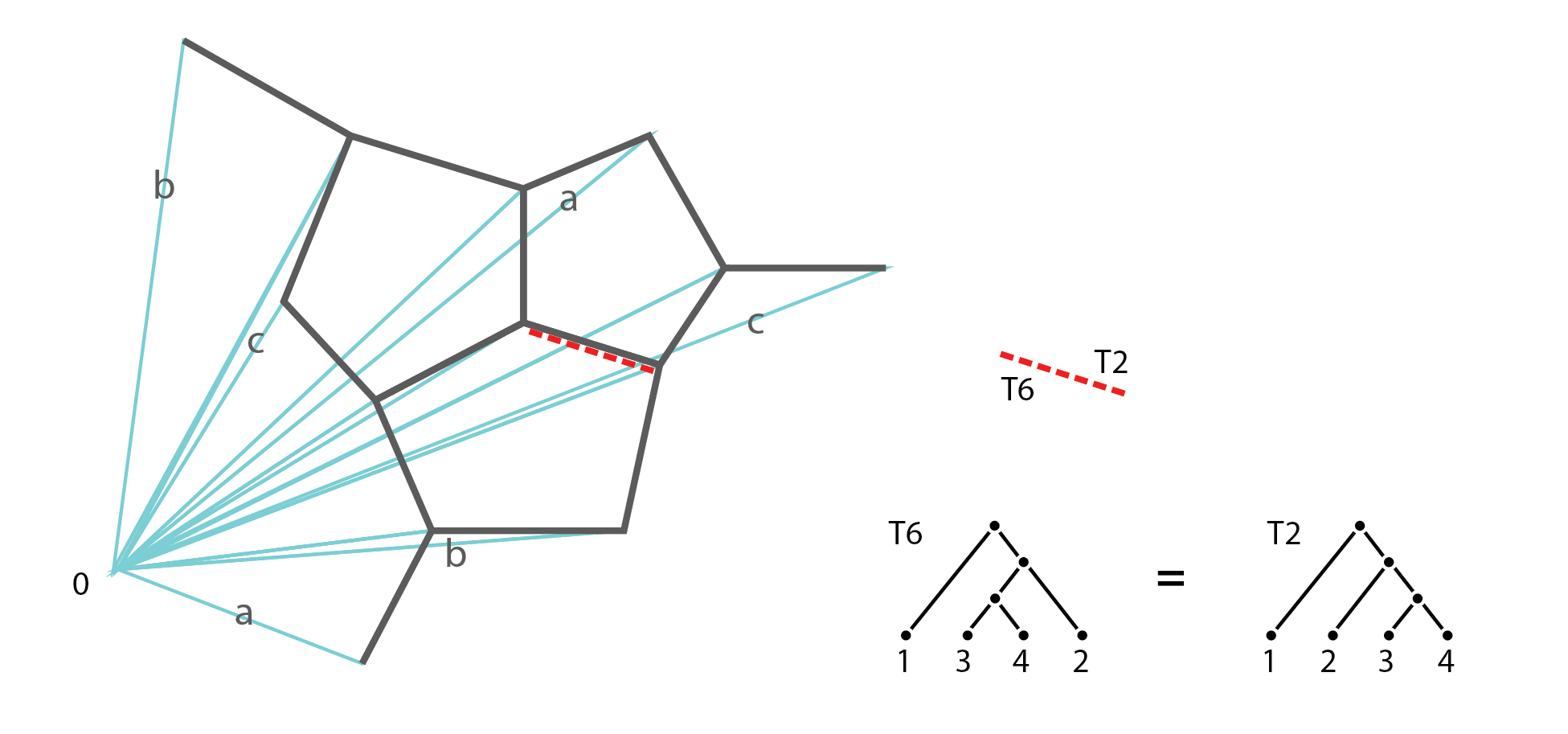}
  \caption[Representation of $\mathcal{T}_4$ with two equal trees.]{Representation of $\mathcal{T}_4$, given the labeled edges, are identified as the same components. The trees $T2$ and $T6$, although belong to different pentagons (that are associated with different label permutations), are actually the same tree.}
  \label{fig:T4rep}
\end{figure}

It is important to understand that, even though the other ``infinite hexagonal cones" are copies of the first considered in Figure \ref{fig:T4O5rep}, when we match their components, trees with supposedly different topologies are actually the same with coinciding orthants, given the label permutation considered. That would lead us to believe that we could identify one $5$-sided polygon for each permutation of the set $S$, and since $|S|=4$ we would have $4!=24$ polygons, instead of $12$ that are identified on Figure \ref{fig:T4Cycles}.

\begin{figure}[!htb]
  \centering
  \includegraphics[width=1\textwidth]{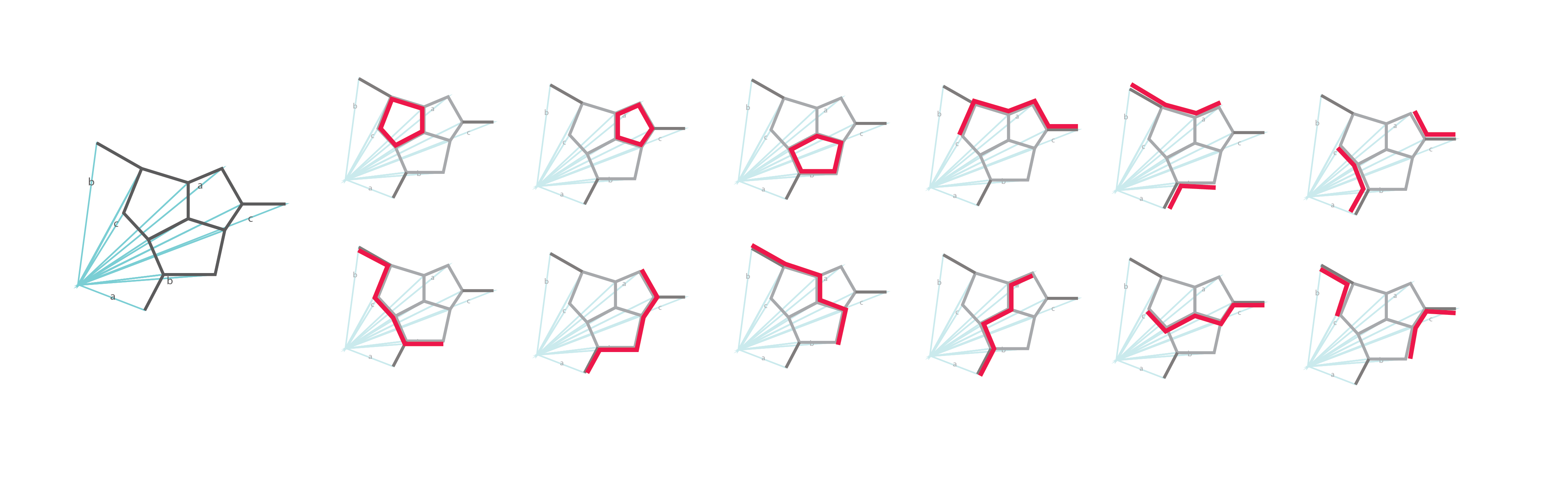}
  \caption[All the $5$-polygons in $\mathcal{T}_4$.]{All the $5$-polygons in $\mathcal{T}_4$. These actually match the $5$-cycles in \textit{Petersen Graph}.}
  \label{fig:T4Cycles}
\end{figure}

However, if we take a good look at Figure \ref{fig:T4O5rep} we notice that the permutation of labels $(2\leftrightarrow 3;\ 1\leftrightarrow 4)$ would lead to the same set of $5$ orthants: not only those trees are identical, but the same tree. So each $5$-side polygon represents, actually, two permutations of labels instead of one.

\end{ex}

Now that we defined this space of trees, we are left to understand what's the metric associated with this space. Its construction leads us to conclude that it already comes equipped with a natural distance function: as a matter of fact, this space is made up of \textit{standard Euclidean} orthants. So, the distance between two points (or trees) in the same orthant will be the usual Euclidean distance. If two points are in different orthants, we can build a path between them that is a sequence of straight segments, each one laying in a single orthant. We can then measure the length of the path by adding up the lengths of the segments. Let us denote this distance in $\mathcal{T}_{|S|}$ space as $d_{\mathcal{T}_{|S|}}(A,B)$.

The existence of this path along orthants is given by the fact that, for all $n$, $\mathcal{T}_n$ is a space with \textit{non-positive curvature} (proof of Lemma 4.1 in \cite{bib:15}) and follows from Gromov, 1987 \cite{bib:27} that all these spaces have a unique shortest path connecting any two points called \textbf{geodesic}, hence the name of the metric.

\begin{dfn} \textbf{Geodesic distance}\\
Let $A,B\in\gamma^w_S$, $\mathcal{T}_{|S|}$ the space of trees with $|S|$ labels and $d_{\mathcal{T}_{|S|}}$ the associated distance function. Then, the \textbf{Geodesic Distance} $d_{Geo}(A,B)=d_{\mathcal{T}_{|S|}}(A,B)$.
\end{dfn}

Although Billera, \textit{et al.} \cite{bib:15} approaches how to calculate this metric (and we recommend the article for more insight), it is far from describing an implementation. Far from previous approaches that lead to exponential time algorithms, Megan Owen and J. Scott Provan (2009) described a polynomial time algorithm to compute this distance in \cite{bib:9}, which we will approach next. This will, however, need some extra notation, defined in \cite{bib:17}:

\begin{dfn}
\label{dfn:split}
Let $T=(V,E,S,w)\in\gamma^w_S$ be a weighted rooted tree where every leaf and root are labeled and have no duplicate labels (there are no two distinct vertices $v_1, v_2\in V$ such that $label(v_1)=label(v_2)$). Let $V'\subseteq V$ be the set of labeled vertices and $X=\bigcup_{v\in V'} \{label(v)\}$ (note that it is always the case that $S\subseteq X$). Given a set of labels $L$, a \textit{$L$-split} is a two set structure $A|B$ such that $\{A,B\}$ is a partition of $L$ of two non-empty sets and, in our work, if we omit the set and refer only \textit{split} we are referring to $X$-splits of the tree at hand. It is straightforward to understand that each edge $e\in E$ induces a partition of the set $X$, building the sets $A$ and $B$ from the labeled vertices contained in each of the connected components generated from removing $e$ from $T$. We will denote the $X$-split induced by $e\in E$ as $\sigma_e = \sigma^X_e=X_e|\bar{X_e}$, $\Sigma$ will denote an arbitrary collection of $X$-splits and $\Sigma(T)=\bigcup_{e\in E}\{X_e|\bar{X_e}\}$. These are similarly established to any subtree $T'\subseteq T$ with label set $X'\subseteq X$ of vertices of degree at most $2$.
\end{dfn}



\begin{dfn} \label{dfn:cscss} \textbf{(Compatible $X$-Splits and Compatible $X$-Split Sets)}
Let $\Sigma$ be any non-empty collection of $X$-splits and $A|B,C|D\in\Sigma$. We say that $X$-splits $A|B$ and $C|D$ are \textbf{compatible} if at least one of the sets $A\cap C$, $A\cap D$, $B\cap C$, $B\cap D$ is empty. Additionally, we say that $\Sigma$ is a \textbf{compatible $X$-split set} if every pair of $X$-splits is compatible.
\end{dfn}


\begin{dfn} \label{dfn:ces} \textbf{(Compatible Edge Sets)}\\
Let $T_1,T_2 \in \gamma^w_S$ and $A \subseteq E_{T_1}$ and $B \subseteq E_{T_2}$. 
\begin{itemize}
  \item A set $C$ is a \textbf{compatible edge set} if, for all $c_1,c_2\in C$, the $X$-splits $X_{c_1}|\bar{X_{c_1}}$ and $X_{c_2}|\bar{X_{c_2}}$ are compatible;
  \item We say that $A$ and $B$ are \textbf{compatible edge sets} if for every $a\in A$ and $b\in B$ the $X$-splits $X_a|\bar{X_a}$ and $X_b|\bar{X_b}$ are \textbf{compatible}.
\end{itemize}
\end{dfn}

The following theorem is important for the existence of the space of trees as it is, as a matter of fact, is what stops it from collapsing. It can be found on \textit{Phylogenetics} by Charles Semple and Mike Steel \cite{bib:17}, but the statement was originally Peter Buneman's work in 1971.

\begin{thm} \label{thm:set} (\textbf{Split Equivalence Theorem}, \cite[Theorem 3.1.4]{bib:17})\\
Let $\Sigma$ be a set of $X$-splits. $\Sigma$ uniquely defines a tree if and only if $\Sigma$ is a \textbf{compatible $X$-split set}.
\end{thm}

To prove Theorem \ref{thm:set} we will need to prove some other results first:

\begin{dfn} \label{dfn:mrs}
 Let $A\in\gamma^w_S$ and $S'$ a set of labels of the tree $A$ that may or may not be contained in $S$. $A(S')$ is the minimal rooted subtree of $A$ that connects the vertices labeled by all the elements of $S'$ (meaning, the smallest subtree of $A$ in terms of vertices and edge count, that satisfies the exposed condition). Furthermore, we denote by $A|S'$ the tree generated from $A(S')$ with every non-root vertex of degree 2 suppressed. This is equally established for trees in $\gamma_S$.
\end{dfn}

\begin{lmm} \label{lmm:set1}
Let $T=(V,E,S,w)$ be a tree and let $\sigma_1,\sigma_2\in\Sigma(T)$ such that $\sigma_1\neq\sigma_2$. Then, $X$ can be partitioned into three sets $X_1, X_2$ and $X_3$ such that $\sigma_1 = X_1|(X_2\cup X_3)$ and $\sigma_2 =(X_1\cup X_2)|X_3$. Furthermore, the intersection of the vertex sets of the minimal subtrees of $T$ induced by $T(X_1)$ and $T(X_3)$ is empty.
\end{lmm}
\begin{proof}
Let $e_1=u_1v_1$ and $e_2=u_2v_2$ be unique edges of $T$ such that $\sigma_1=X_{e_1}|\bar{X_{e_1}}$ and $\sigma_2=X_{e_2}|\bar{X_{e_2}}$. Evidently, there's a path $\mathcal{P}$ in $T$ such that $e_1$ and $e_2$ are first and last edges, respectively, that are traversed by $\mathcal{P}$. Without loss of generality, we may assume that $u_1$ and $u_2$ are initial and terminal vertices of $\mathcal{P}$, respectively. Observe that $u_1$ and $u_2$ are distinct but $v_1$ and $v_2$ might not be distinct. Let $V_1,V_2$ and $V_3$ denote the vertex sets of the components of $T\backslash\{e_1,e_2\}$ containing $u_1,v_1$ and $u_2$, respectively. By choosing, for each $i\in\{1,2,3\}$, $X_i=label(V_i)$ we get the desired result.
\end{proof}

\begin{figure}[!htb]
  \centering
  \includegraphics[width=1\textwidth]{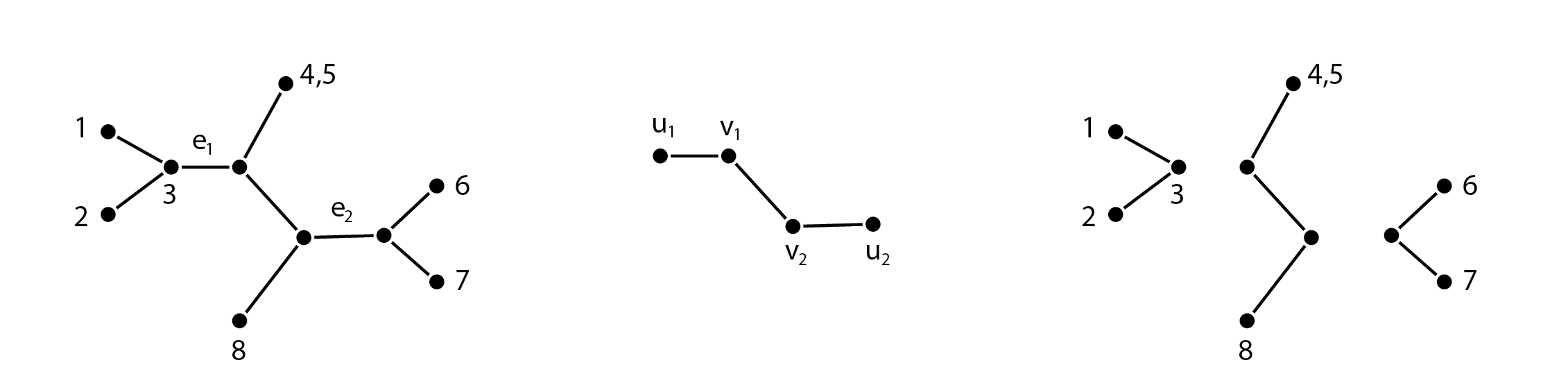}
  \caption[Minimal subtrees induced by path $\mathcal{P}$.]{\textit{(Left to right: Tree $T$; Path $\mathcal{P}$; Minimal subtrees induced by $T(X_1), T(X_2)$ and $T(X_3)$, respectively)} The splits $\sigma_1=\{1,2,3\}|\{4,5,6,7,8\}$ and $\sigma_2=\{6,7\}|\{1,2,3,4,5,8\}$ are $X_{e_1}|\bar{X_{e_1}}$ and $X_{e_2}|\bar{X_{e_2}}$ respectively, and can be built from the subsets of $X$ as follow: $\sigma_1=X_1|(X_2\cup X_3)$ and $\sigma_2=X_3|(X_1\cup X_2)$.}
  \label{fig:SeqEx}
\end{figure}

You can check an illustration of the previous lemma on Figure \ref{fig:SeqEx}.

The next lemma is a general property of trees. For it, we need to define what is a colouring of the vertex set induced by a function $f$: Let $T=(V,E,S,w)\in\gamma^w_S$ and $f:Y\longrightarrow V$ where $Y$ is a finite set. Colour the elements of $Y$ either red or green. Let $v\in f(Y) \subseteq V$. If all elements of $f^{-1}(v)$ have the same colour, $v$ has that particular colour. If not, assign $v$ red \textbf{and} green. A subgraph of $T$ is monochromatic if all its vertices have the same colour.

\begin{lmm}\label{lmm:colour}
Let $T=(V,E,S,w)\in\gamma^w_S$ and let $f$ be a mapping from a finite set $Y$ into $V$. Consider the colouring of $V$ induced by $f$. Suppose that, for each edge $e\in E$, exactly one of the components of $(V,E\backslash\{e\},S,w)$ (we will refer this tree as $T\backslash v$ for short, for some $v\in V$) is monochromatic. Then, there exists a unique vertex $v\in V$ for which each component of $(V\backslash\{v\},E,S,w)$ (we will refer this tree as $T\backslash e$ for short, for some $e\in E$) is monochromatic.
\end{lmm}
\begin{proof}
We first show that there exists at least one such vertex. For each edge $e\in E$, assign an orientation from the end of $e$ that is incident with the monochromatic component of $T\backslash e$ to the other end of $e$. Then, there exists in $V$ a vertex $v$ of \textit{out-degree} zero (mind Definition \ref{dfn:iodr}); otherwise, we would have a directed path of infinite length.
We now show that there can be at most one such vertex. Suppose that distinct vertices $v$ and $v'$ both have the claimed property. Select an edge $e$ in the path connecting $v$ and $v'$. Then, exactly one of the two components in $T\backslash e$ is not monochromatic. Without loss of generality, assume that $v$ is in that component. This contradicts the assumption that each component of $T\backslash v'$ is monochromatic.
\end{proof}

\begin{lmm}\label{lmm:set2}
Let $A|B$ be an $X$-split. Suppose that $T=(V,E,S,w)\in\gamma^w_S$ is a tree such that $A|B$ is not a split of $T$, but $A|B$ is compatible with each $X$-split of $T$. Then, there exists a unique vertex $v$ of $T$ such that, for each component $(V',E')$ of $T\backslash v$, either $label(V')\subseteq A$ or $label(V')\subseteq B$.
\end{lmm}
\begin{proof}
Let $X$ be the set of labels of vertices of degree 2 or less. Colour the elements in $X$ that belong to $A$ with red and the ones that belong to $B$ with green, and consider the corresponding colouring of the vertices of $T$ induced by $label^{-1}:V\longrightarrow X$. Then, for each edge $e\in E$, exactly one of the components of $T\backslash e$ is monochromatic under the colouring of the vertices of $T$ by $label^{-1}$.
Applying \textbf{Lemma \ref{lmm:colour}} with $f=label^{-1}$ and $Y=X$, there exists a unique vertex $v$ of $T$ for which each component of $T\backslash v$ is monochromatic. This implies that $A|B$ satisfies the condition of the lemma.
\end{proof}

We now follow with the proof of the \textit{Split Equivalence Theorem}:

\begin{proof}
$(\Rightarrow)$
First, suppose that $\Sigma$ is equal to the set of $X$-splits induced by a tree, and let $\sigma_1$ and $\sigma_2$ be distinct elements of $\Sigma$. By Lemma \ref{lmm:set1}, there is a partition of $X$ into three sets $X_1$, $X_2$ and $X_3$ such that $\sigma_1 = X_1 | (X_2\cup X_3)$ and $\sigma_2=(X_1\cup X_2)|X_3$. Since $X_1\cap X_3 =\emptyset$, the $X$-splits of $\sigma_1$ and $\sigma_2$ are compatible.\\
$(\Leftarrow)$
Suppose that $\Sigma$ is a pairwise compatible collection of $X$-splits. We use induction on the cardinality of $\Sigma$ to simultaneously prove that $\Sigma = \Sigma(T)$ for some tree $T$ and, up to isomorphism, the choice of $T$ is unique. If $|\Sigma|=0$, then it is clear that, up to isomorphism, there is a unique tree $T$ such that $\Sigma =\Sigma(T)$, namely the tree consisting of a single vertex labeled $X$.
Now, suppose that $|\Sigma|=k+1$, where $k\geq 0$, and that the existence and uniqueness properties hold for $|\Sigma|=k$. Let $A|B\in\Sigma$. Since $\Sigma\backslash\{A|B\}$ is pairwise compatible, it follows by our induction assumption that there is, up to isomorphism, a unique tree $T'=(V',E',S,w)\in\gamma^w_S$ with $\Sigma\backslash\{A|B\}=\Sigma(T')$. By Lemma \ref{lmm:set2}, exists a unique vertex $v'$ of $T'$ such that, for each component, $\mathcal{C}_i=(V_i,E_i), i\in I$ of $T'\backslash v'$, either $label(V_i)\subseteq A$ or $label(V_i)\subseteq B$ for every $i\in I$.
Let $T=(V,E,S,w)\in\gamma^w_S$ be the tree obtained from $T'$ by replacing $v'$ with two new adjacent vertices $v_A$ and $v_B$, and attaching the subtrees that were incident with $v'$ to the new vertices in such a way that the subtrees consisting of the vertices in $label'^{-1}(A)$ and $label'^{-1}(B)$ ($label'$ being the label function of $T'$) are attached to $v_A$ and $v_B$ respectively. Let us define the map $label^{-1}:X\longrightarrow V$ as follows:

\begin{equation} label^{-1}(x)=
\begin{cases}
label'^{-1}(x) & \mathrm{if\ } label'^{-1}(x)\neq v'\\
v_A & \mathrm{if\ } label'^{-1}(x)=v' \mathrm{\ and\ } x\in A\\
v_B & \mathrm{if\ } label'^{-1}(x)=v' \mathrm{\ and\ } x\in B
\end{cases}\end{equation}

It is easily checked that $label$ is the label function of T, and that we have $\Sigma=\Sigma(T)$. Moreover, as $T'$ is the unique tree for which $\Sigma\backslash\{A|B\}=\Sigma(\mathcal{T})$, it is easily seen that, up to isomorphism, $\mathcal{T}$ is the only such tree satisfying $\Sigma=\Sigma(\mathcal{T})$. 
\end{proof}

\begin{dfn} \label{dfn:psg} \textbf{(Path Space and Path Space Geodesic)}\\
Let $T_A,T_B\in\gamma^w_S: T_A, T_B$ non-identical and $\mathcal{A}=(A_1,...,A_k)$ and $\mathcal{B}=(B_1,...,B_k)$ partitions of $E_{T_A}$ and $E_{T_B}$ such that the pair $(\mathcal{A},\mathcal{B})$ satisfies:
\begin{itemize}
  \item[] \textbf{(P1)} For each $i>j$, $A_i$ and $B_j$ are compatible sets.
\end{itemize}
Then, for all $1\leq i \leq k$, $B_1\cup ...\cup B_i\cup A_{i+1}\cup ... \cup A_k$ is a compatible set, hence, from the \textbf{splits equivalence theorem}, uniquely defines a binary $|S|$-tree $T_i$ and an associated orthant $\mathcal{T}^o_{T_i}$. The connected space $\mathcal{P}=\bigcup^k_{i=1} \mathcal{T}^o_{T_i}$ is the \textbf{path space} with support $(\mathcal{A},\mathcal{B})$ and the shortest path from $T_A$ to $T_B$ contained in $\mathcal{P}$ the \textbf{path space geodesic} for $\mathcal{P}$.
\end{dfn}

\begin{thm} (From Billera \textit{et al.} \cite[Proposition 4.1]{bib:15})\\
For trees $T_1,T_2\in\gamma^w_S$ with disjoint edge sets, the geodesic between $T_1$ and $T_2$ is a path space geodesic for some path space between $T_1$ and $T_2$.
\end{thm}

For a set of edges $A$ we use the notation $||A||=\sqrt{\sum_{e\in A} w(e)^2}$ to denote the norm of the vector whose components are lengths (weights) of the edges in $A$.

\begin{dfn} \label{dfn:pps} \textbf{(Proper Path Space and Proper Path)}\\
Let $T_1,T_2\in\gamma^w_S$ and $\Gamma$ the geodesic in $\mathcal{T}_{|S|}$ between $T_1$ and $T_2$. Then, $\Gamma$ can be represented as a path space geodesic with support $\mathcal{A}=(A_1,...,A_k)$ of $E_{T_1}$ and $\mathcal{B}=(B_1,...,B_k)$ of $E_{T_2}$ which satisfy \textbf{P1} and the following additional property:
\begin{itemize}
  \item[] \textbf{(P2)} $\frac{||A_1||}{||B_1||} \leq \frac{||A_2||}{||B_2||} \leq ... \leq \frac{||A_k||}{||B_k||}$
\end{itemize}
We call a path space satisfying conditions \textbf{P1} and \textbf{P2} a \textbf{proper path space} and the associated path space geodesic a \textbf{proper path}.
\end{dfn}

\begin{thm} \label{thm:p3} (From Owen \textit{et al.} \cite[Theorem 2.5]{bib:9})\\
Given $T_1,T_2\in\gamma^w_S$ a \textbf{proper path} $\Gamma$ between $T_1$ and $T_2$ with support $(\mathcal{A},\mathcal{B})$ is a geodesic if and only if $(\mathcal{A},\mathcal{B})$ satisfy the property:
\begin{itemize}
  \item[] \textbf{(P3)} For each support pair $(A_i,B_i)$ there is no non-trivial partitions $C_1\cup C_2$ for $A_i$ and $D_1\cup D_2$ for $B_i$ such that $C_2$ is compatible with $D_1$ and $\frac{||C_1||}{||D_1||}<\frac{||C_2||}{||D_2||}.$
\end{itemize}
\end{thm}

One intuitive path between any two trees that will be useful for the algorithm implementation as the starting point is the \textbf{\textit{cone path}}. This is the path that connects the two trees through two straight line segments through the origin of our space $\mathcal{T}_n$. The \textit{cone path} will function as our starting point with support $(\mathcal{A}^0,\mathcal{B}^0)$ that vacuously satisfies (P1) and (P2). The algorithm goes as follows:

\begin{alg}\textbf{(Geodesic Algorithm, GTP)}
\label{alg:GTP}
\begin{itemize}
  \item[]\textbf{Input}: $T_1,T_2\in\gamma^w_S$;
  \item[]\textbf{Output}: The path space geodesic between $T_1$ and $T_2$.
  \item[]\textbf{Initialize}: $\Gamma^0$ = cone path between $T_1$ and $T_2$ and support $(\mathcal{A}^0,\mathcal{B}^0)=((E_{T_1}),(E_{T_2}))$.
  \item[]\textbf{Step}: At stage $l$, we have proper path $\Gamma^l$ with support $(\mathcal{A}^l,\mathcal{B}^l)$ satisfying conditions (P1) and (P2).
    \begin{itemize}
        \item[] \textbf{if} $(\mathcal{A}^l,\mathcal{B}^l)$ satisfies (P3),
        \item[] \textbf{then} path $\Gamma^l$ is the path space geodesic,
        \item[] \textbf{else} chose any minimum weight cover $C_1\cup D_2$, $C_1\subset A_i$ and $D_2 \subset B_i$ with complements $C_2$ and $D_1$, respectively, having weight $\frac{||C_1||}{||A_i||}+\frac{||D_2||}{||B_i||}<1$. Replace $A_i$ and $B_i$ in $\mathcal{A}^l$ and $\mathcal{B}^l$ by the ordered pairs $(C_1,C_2)$ and $(D_1,D_2)$, respectively, to form a new support $(\mathcal{A}^{l+1},\mathcal{B}^{l+1})$ with associated path $\Gamma^{l+1}$.
    \end{itemize}
\end{itemize}
\end{alg}

Would be reasonable for the reader to ask how can we assure that $(\mathcal{A}^l,\mathcal{B}^l)$ satisfy (P2) with multiple iterations of \textbf{step}, since (P1) is assured by the construction of $(\mathcal{A}^{l+1},\mathcal{B}^{l+1})$. Owen \textit{et al.} assure that condition on the Lemma 3.4 of \cite{bib:9}.

Once we have the final support, we can compute the geodesic distance the following way:

\begin{thm} \label{thm:gdl} \textbf{Geodesic Distance}\\
Let $T_1,T_2\in\gamma^w_S$ such that $T_1$ and $T_2$ have no common edges and $(\mathcal{A}^l,\mathcal{B}^l)=((A_1,B_1),...,(A_k,B_k))$ the resulting support of running the \textit{geodesic algorithm} for $T_1$ and $T_2$. The \textbf{geodesic distance} between $T_1$ and $T_2$ is given by:
\begin{equation}d_{Geo}(T_1,T_2)=||(||A_1||,...,||A_k||)+(||B_1||,...,||B_k||)||\end{equation}
\end{thm}

The biggest slice of the algorithm complexity lays on checking if a specific support satisfies (P3). This is solved through an equivalent problem that is called the \textit{Extension Problem}, which we will specify later, however, its complexity is $O(n^3)$ \cite[Lemma 3.3]{bib:9}. Adding to that, we need to account for the unsuccessful tries of checking (P3) which are at most $n-3$ (for the maximum $n-2$ possible iterations, that correspond to the maximum number of internal edges that are the higher bound for $|\mathcal{A}|$ and $|\mathcal{B}|$, minus the iteration where the solution is found), hence the complexity of our algorithm is $O(n^4)$.

A later article by Megan Owen provides extra results that aim to optimize the geodesic distance. In [t1], among others, we can find a linear time algorithm to calculate the geodesic distance between two trees that the path space between them is known and an algebraic equivalence when the trees for which the distance is being calculated share at least one split. For the later, we will now formalize some concepts that are needed for better representation and understanding of this equivalence.

\begin{thm} \label{thm:orig1}
Let $T\in\gamma^w_S$, $e\in E_T$ and $\sigma^X_e=C|D\in\Sigma (T)$. The subset of edges of $E_T$ that induce $C$-splits is:
\begin{equation} E^C_T=\{d\in E_T : (X_d\subsetneq C) \vee (\bar{X_d}\subsetneq C)\}\end{equation}
\end{thm}

\begin{proof}
  Let $d\in E_T$ such that $\sigma^C_d=C_d|\bar{C_d}$ is a $C$-split. Removing $d$ from $T$ will break it into two connected components (since $T\in\gamma^w_S$). Assume, without loss of generality, that $C_d$ is the set of labels of the vertices with a degree at most $2$ from the connected component which \textbf{does not} contain $e$. By definition of $X$-split, we have that $\sigma^X_d=C_d|(X\backslash C_d)$. Since $\{C_d, \bar{C_d}\}$ is a non-empty set partition of $C$, we have that $C_d\subsetneq C$, so $d\in E^C_T$.

  Conversely, let $d\in E^C_T$. Assume, without loss of generality, that $X_d\subsetneq C$ (implying that $\bar{X_d}\nsubseteq C$). By definition of $X$-split, we have that $\{X_d,\bar{X_d}\}$ is a partition of $X$. Then:
  \begin{equation} X_d \cup \bar{X_d} = X \Leftrightarrow (X_d \cup \bar{X_d})\cap C = X\cap C \Leftrightarrow (X_d \cap C) \cup (\bar{X_d} \cap C) = C\end{equation}

  where $(X_d \cap C)$ and $(\bar{X_d} \cap C)$ are all the respective labels of $X_d$ and $\bar{X_d}$ that are in $C$, concluding that (since $X_d\cap\bar{X_d}=\emptyset$ and $X_d\subsetneq C$) $\{(X_d \cap C),(\bar{X_d}\cap C)\}$ is a non-empty set partition of $C$, hence $(X_d \cup C)|(\bar{X_d}\cup C)$ is a $C$-split.
\end{proof}

For the next definition we will need to slightly adapt the operation $\alpha$ (contraction of Bourque, Definition \ref{dfn:alphaop}) to $\alpha^w:\gamma^w_S\times E \longrightarrow \gamma^w_S$ and, for $T\in\gamma^w_S$ and $J=\{e_1,e_2,...,e_j\}\subseteq E_T$, we abbreviate $\alpha^w(\alpha^w(...\alpha^w(T,e_j)...,e_2),e_1)$ as $\alpha^w(T,J)$.

\begin{dfn} \label{dfn:TC}
Let $T\in\gamma^w_S$, $e\in E_T$, $\sigma_e=C|D\in\Sigma(T)$ and $\alpha^w$ the adapted contraction of Bourque for weighted trees. We define the tree of internal edges of $C$ as: 
\begin{equation} T^C=\alpha^w(T,E^D_T\cup\{e\})\end{equation}
\end{dfn}

\begin{thm} \label{thm:gdse}
(Geodesic distance for trees with shared $X$-splits - Megan Owen [t1])\\
If $T_1,T_2\in\gamma^w_S$ are such that at least one pair of edges $e_1\in E_{T_1}$ and $e_2\in E_{T_2}$ satisfy $\sigma_{e_1}=\sigma_{e_2}=C|D\in (\Sigma(T_1)\cap\Sigma(T_2))$, then:
\begin{equation} d_{Geo}(T_1,T_2)=\sqrt{d_{Geo}(T^C_1,T^C_2)^2+d_{Geo}(T^D_1,T^D_2)^2+(w(e_1)-w(e_2))^2}\end{equation}
\end{thm}

The \textit{Geodesic distance} and the way it is formalized brings a new dimension to visualization. Since there's a continuous path between any two trees, one can technically visualize how the tree deforms along that path. Regarding its \textbf{discriminatory power}, it can be seen as holding emphasis according to shared internal edges and its length's between the trees, rather than internal path length (such as the case with \textit{Quartet based metrics}). One could also see it as closer to \textit{Robinson Foulds Length} since each internal edge has a one-on-one correspondence with clades: traversing $\mathcal{T}_n$ can be boiled down to contractions and decontractions of Bourque ($\alpha$ and $\alpha^{-1}$ operations, respectively). To which degree these two metrics relate, is a topic probably worth discussing.


\section{Other Approches}
\label{section:oapproach}

We will now cover some other metrics that are less used. Some of them might be interesting or promising approaches to the problem at hand, but by some reason are disregarded or unused by the scientific community, such as really expensive computations (as the case of the \textit{Hybridization Number}). We will not, however, cover these distances with the same detail as the ones in the previous section.

\subsection{Maximum Agreement Subtree, Align}
\label{subsection:mast}

The \textbf{Maximum Agreement Subtree} is a concept initially formalized in 1980 by Gordon A.D., and it was covered in \cite{bib:1} when its practical performance was analyzed.

\begin{dfn} (\textbf{Maximum Agreement Subtree distance})\\
Let $A,B\in\gamma_S$, $|S|=n$, $T$ the maximum subtree of $A$ and $B$ and $t$ the number of leaves of $T$. So, the Maximum Agreement Subtree Distance \begin{equation} d_{MAST}(A,B)=n-t\end{equation}
\end{dfn}

\begin{figure}[!htb]
  \centering
  \includegraphics[width=0.61803398875\textwidth]{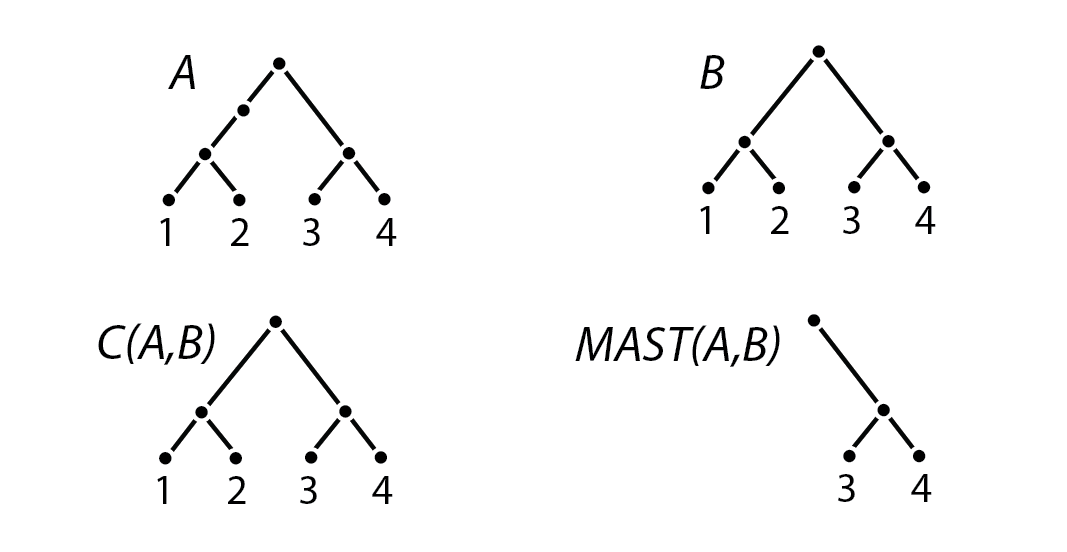}
  \caption[Example to distinguish the \textit{Maximum Agreement Subtree} and \textit{Stric Consensus Tree}.]{A simple counterexample to illustrate the difference between a \textit{Maximum Agreement Subtree} and a \textit{Strict Consensus Tree}}
  \label{fig:MAST}
\end{figure}

We could go further on this definition by going through the specifics of what is a ``subtree" (establishing an isomorphism between subsets of $V_A$, $V_B$ and $E_A$, $E_B$ that preserves labeling). One could also be lead to believe that a relation between the \textit{Maximum Agreement Subtree} and the \textit{Strict Consensus Tree}, however, we provide in Figure \ref{fig:MAST} a simple counterexample.

Regarding complexity, the article \cite{bib:18} provides a study of the complexity of computing \textit{MAST}, which the conclusion is that can be solved in $O(n^{O(d)})$ time where $n$ stands for the size of $S$ and $d$ the maximum vertex degree in both trees, concluding that this is a polynomial-time problem. However, other papers \cite{bib:1,bib:19} adopt other implementations with other complexities associated: as a matter of fact, imposing some constraints in the data structures can also lead to $O(nlog(n))$ complexity \cite{bib:19}.

Regarding its \textbf{discriminatory power}, based on its definition one can conclude that this metric is very sensitive to small variations of the same tree since it only considers what's strictly equal to both trees and disregards everything else. So it should be safe to assume that its power lies close to the one of \textit{Robinson Foulds}, and should be considered in cases where one desires to analyze how identical to trees are, instead of how similar.

\vspace{10mm}

The first time \textbf{Align} distance was formalized was in 2006 by Nye, \textit{et al.} in \cite{bib:8}. The original motivation behind it was to build an alignment between the two trees at hand (analogous to sequence alignment) building a match between edges according to their topological characteristic. The Align distance is a topologic measure.

As seen previously, in every tree $A\in\gamma_S$ there is an associated partitioning function (Definition \ref{dfn:partfunc}) that delineates partitions of the set $S$ from the set of edges of $A$, $E_A$. Consider $A,B\in\gamma_S$, $f_A$ and $f_B$ the respective partitioning functions, $e_A\in E_A$, $e_B\in E_B$. Consider as well that, for every $X\in\gamma_S$, partitioning function $f_X$ and edge $e_X\in E_X$, $f_X(e_X)=\{P^X_{(e_X,0)},P^X_{(e_X,1)}\}$. We now define parameters $a_{(r,s)}$ as \begin{equation} a_{(r,s)} = \frac{|P^A_{(e_A,r)}\cap P^B_{(e_B,s)}|}{|P^A_{(e_A,r)}\cup P^B_{(e_B,s)}|} \end{equation} that represents the proportion of elements shared by $P^A_{(e_A,r)}$ and $P^B_{(e_B,s)}$. The score $s(e_A,e_B)$ is then defined by \begin{equation} s(e_A,e_B)=max\{min\{a_{(0,0)},a_{(1,1)}\},min\{a_{(0,1)},a_{(1,0)}\}\}\end{equation}

The Align distance is then defined by:

\begin{dfn} \textbf{(Align distance)}\\
Let $S$ be a set of labels of size $n$, $A,B\in\gamma_S$ such that $|E_A|=|E_B|$. The \textbf{Align distance} is given by \begin{equation} d_{Align}(A,B)=\sum_{e_A\in E_A} s(e_A,f(e_A))\end{equation} where $f:E_A\longrightarrow E_B$ is a bijective function that maximizes the sum. 
\end{dfn}

Finding the bijection $f$ is actually what bounds the time complexity of the implementation, the Munkres Algorithm has $O(n^3)$ time complexity (Munkres, J. \cite{bib:20}; Bourgeois \textit{et al.}\cite{bib:21}).

There is, as stated in \cite{bib:1}, a strong emphasis on the shared clades on the distance function, making it related to \textit{Robinson Foulds} when issuing its \textbf{discriminatory power}. But since it uses information from clades that are almost the same the same way it uses from the ones that are the same, the conclusion is that it lies on the opposite side of \textit{MAST} when compared to \textit{RF}.

\subsection{Cophenetic correlation coefficient, Node, Similarity based on Probability}
\label{subsection:coph}

The \textbf{Cophenetic Correlation Coefficient} is known as ``the first effective numerical method known" by most \cite{bib:7}, was originally described by Sokal and Rohlf in 1962 in \cite{bib:22} and its motivation was to measure how faithful a dendrogram preserves the pairwise distances between the original unmodeled data points. Given the architecture of the method, one can also use it to measure how close the trees at hand are regarding pairwise distance between their leaves.

Another interesting fact is that on \cite{bib:22}, Sokal and Rohlf also stated that \textit{``One of the initial schemes which occurred to us (...) was to try to compare different dendrograms with the same set of leaves by counting the number of breaks and rearrangements necessary to convert one dendrogram into another"} (adapted) and that this would later inspire the original idea of the \textit{Robinson Foulds distance}.

\begin{figure}[!htb]
  \centering
  \includegraphics[width=1\textwidth]{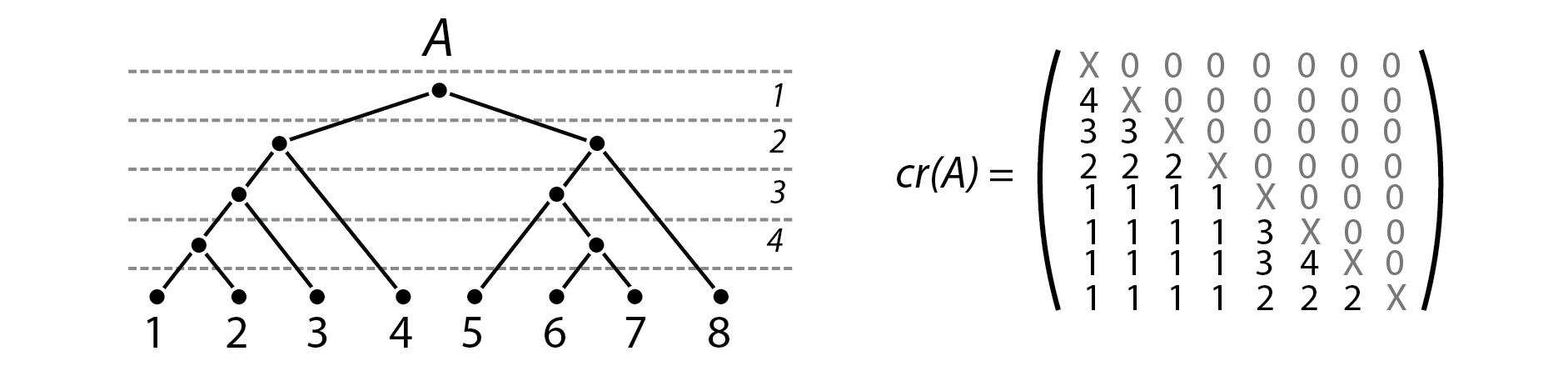}
  \caption[Dendrogram with matrix of cophenic relations.]{A dendrogram with internal vertices split for class values and its respective cophenetic relations. It could be the case that vertices with different depths are in the same class value.}
  \label{fig:CCC}
\end{figure}

The first step to calculate the \textit{cophenetic correlation coefficient} for a dendrogram and a corresponding set of data is dividing the internal nodes into suitable \textbf{class values}. These are distributed across the depth of the tree, such that if one node is deeper or as deep as another one, its class value will be greater or equal than the class value of the compared vertex (check Figure \ref{fig:CCC} as an example). 

This process is left to the scientist, keeping in mind that the number of class of values must be picked taking in consideration the number of leaves of the dendrogram (for less than 10 leaves one should stick with 4 or fewer class values, for 100 leaves one should choose at least 10 class values \cite{bib:22}). Then, the cophenetic relation between two leaves is the class value of the least deep vertex on the path from one leaf to another. The cophenetic relation for a dendrogram is a matrix of size $n \times n$ (where $n$ is the size of the data sample) that comprises all cophenetic relations between its set of leaves. We will denote, for a dendrogram $A\in\gamma_S$, the cophenetic relation of $A$ as $cr(A)$, and the cophenetic relation between leaves $v_i, v_j\in V_A$ is stored on the entry $cr(A)_{ij}$.

\begin{dfn}\textbf{Cophenetic correlation coefficient}
Let $S=\{X_i\}_{i\in [n]}$ a set of data, $\bar{d}$ a distance function for the data, and $A\in\gamma_S$ a dendrogram estimated from the data. The \textbf{cophenetic correlation coefficient} is given by

\begin{equation} c(A) = \frac
{\sum^n_{j=1}\sum^j_{i=1} (cr(A)_{ij}-\bar{cr}_A)(\bar{d}(X_i,X_j)-\bar{X})}
{\sqrt{[\sum^n_{j=1}\sum^j_{i=1} (cr(A)_{ij}-\bar{cr}_A)^2][\sum^n_{j=1}\sum^j_{i=1} (\bar{d}(X_i,X_j)-\bar{X})^2]}}
\end{equation}

where $\bar{X}$ is the average distance for all possible data pairs from $\{X_i\}_{i\in[n]}$ and $\bar{cr}_A$ the average cophenetic relation between all pairs of vertices for the dendrogram $A$.\\
If we instead want to calculate the cophenetic correlation coefficient between two dendrograms $A,B\in\gamma_S$ we should instead consider

\begin{equation} d_{CCC}(A,B) = \frac
{\sum^n_{j=1}\sum^j_{i=1} (cr(A)_{ij}-\bar{cr}_A)(cr(B)_{ij}-\bar{cr}_B)}
{\sqrt{[\sum^n_{j=1}\sum^j_{i=1} (cr(A)_{ij}-\bar{cr}_A)^2][\sum^n_{j=1}\sum^j_{i=1} (cr(B)_{ij}-\bar{cr}_B)^2]}}
\end{equation}

where $\bar{cr}_A$ and $\bar{cr}_B$ are the average cophenetic relation between all pairs of vertices for the dendrogram $A$ and $B$ respectively.

\end{dfn}

This coefficient is nothing less than the \textit{Pearson correlation coefficient}, in the latter case of the definition (for two matrices) also referred to as the product-moment correlation coefficient between $A$ and $B$.


Regarding its \textbf{complexity}, one should note that after the computations done to obtain $cr(A)$, nothing more is left other than make a calculation that can be done in linear time. To calculate every entry of $cr(A)$ we can do a calculation that will take $O(log(n))$ time. Every vertex $v$ of our dendrogram $A$ induces a partition of its labels dictated by the leaves of the trees in the forest generated by removing $v$ from $A$. The cophenetic relations between the leaves on these partitions is given by the class value of the vertex $v$ removed. From this point on we can apply this method recursively to the neighborhood of the removed vertex $v$ and by the end of the computation $cr(A)$ will be calculated.

Considering the tree $A$ in the Figure \ref{fig:CCC}, removing the root vertex would give us the partition of $S$ $\{P^1_1,P^1_2\}=\{\{1,2,3,4\},\{5,6,7,8\}\}$, concluding that $cr(A)_{ij}$ equals the class value of the root vertex, which is $1$, for all $i\in P^1_1$ and $j\in P^1_2$. From this point on, we would apply the same procedure at the child vertices of the root vertex (that lay in class value $2$), which would give partitions $\{\{1,2,3\},\{4\}\}$ and $\{\{5,6,7\},\{8\}\}$.

This was, as stated, the first effective numerical method to compare classifications, there are limitations and concerns as far its \textbf{discriminative power}. Williams \textit{et al.}, in its article about a variant of this metric which we'll approach next, refers how this metric has drawbacks in regards of how the class values are not a property from the classification itself, but something instead defined by the scientist \cite{bib:7}. This alone makes the \textit{Cophenetic correlation coefficient} is obsolete.

\vspace{10mm}

The \textbf{Node distance} was formalized by W. T. Williams and H. T. Clifford in 1971 \cite{bib:7} and is a variant of the cophenetic correlation coefficient pairwise heuristic, looking forward to improve on the limitation in discriminatory power (as pointed out in \cite{bib:7}).

\begin{dfn} \textbf{(Node distance)}\\
Let $S$ be a set of labels such that $|S|=n$, $A,B\in\gamma_S$ and, for all tree $X\in\gamma_S$ and $v_i,v_j\in V_X$, $d^X(v_i,v_j)$ the distance of the path from $v_i$ to $v_j$ in the tree $X$. Consider as well a function $l_A$ and $l_B$ that, given a label $s\in S$, $l_A(s)$ and $l_B(s)$ returns the vertices with label $s$ from $V_A$ and $V_B$ respectively (the $label$ ``inverse" function). The \textit{Node distance} function is given by

\begin{equation} d_N(A,B)=\frac{2\sum_{s_1,s_2\in S}|d^A(l_A(s_1),l_A(s_2))-d^B(l_B(s_1),l_B(s_2))|^k}{n(n-1)}, k=1 \end{equation}

\end{dfn}

Article \cite{bib:1} also refers that a similar metric as proposed by Penny, et all in 1982 which follows the \textit{Node distance} outlines but for $k=2$ instead, by the name of \textit{Path Difference metric}.

The \textbf{complexity} of the \textit{Node distance}, just as the \textit{Cophenetic Correlation Distance}, is determined before the actual calculation, this time by determining the distance between every leaf of the two subject trees. The path length can be obtained through a topological sorting algorithm, of complexity $\Theta (|V|+|E|)$ for some tree $T=(V,E)$. Since this operation will be done once for each labeled vertex, the time complexity of \textit{Node} will be $O(n^2)$.

As for the \textbf{discriminatory power} goes, Penny, \textit{et al.} refers to \textit{Path Difference metric} in \cite{bib:26}: describes it as ``sensitive to the tree distribution" (since its formulation wasn't accounting for $n(n-1)$ division, which is done here) and points out that another useful application would be ``when the topic of interest is the relative position of subsets of nodes, rather than the comparison of trees themselves" (adapted).

\vspace{10mm}

The \textbf{Similarity based on probability} is a metric firstly defined in \cite{bib:23} by Hein \textit{et al.} and like most metrics defined in this work, it was studied in \cite{bib:1}. This metric is rather unique compared to the others given its probabilistic approach, which suits the book where its integrated. The following definition was adapted from Chapter 7 of \cite{bib:23}.

\begin{dfn} \textbf{(Similarity based on probability)}\\
We start by defining an indicator function $I$ and measure $M$, where $r_X\in V_X$ is a vertex such that $label_X(r_X)='root'$ and $a^X_u\in V_X$ is the ancestor of $u\in V_X$ in the tree $X$ 
\begin{equation} I_{\{u,v\}}=
\begin{cases}
1 & if\ f_A(u)=f_B(v)\\
0 & otherwise
\end{cases};\end{equation}\\ \begin{equation}
M_{XY}=\frac{\sum_{u\in V_X\backslash\{r_X\};v\in V_Y\backslash\{r_Y\}}I_{\{u=v\}}w(a^X_u u)w(a^Y_v v)}{l_X l_Y}
\end{equation}
For all $X,Y\in\gamma^w_S$ (with $f_X$ and $f_Y$ as the respective partitioning functions - Definition \ref{dfn:partfunc} ), $u\in V_X$, $v\in V_Y$, $l_X=\sum_{e\in E_X} w(e)$ and $l_Y=\sum_{e\in E_Y} w(e)$. We now define the \textbf{similarity based on probability} function $S$, for two weighted trees $A,B\in\gamma^w_S$, as

\begin{equation} S(A,B)=\frac{M_{AB}}{M_{AA}}\end{equation}
\end{dfn}

The meaning behind this similarity measure $M_{AB}$ is, as described in \cite{bib:1}, the \textit{``probability that a point chosen randomly in A will be on a branch leading to the same set of tips as a point chosen randomly in B"}, which is afterward normalized by $M_{AA}$. This leads to a non-symmetry scenario (a requirement for a distance), which is solved in \cite{bib:1} in the following way:

\begin{dfn}\textbf{(Similarity based on probability distance)}\\
Let $A,B\in\gamma^w_S$ and $S$ the similarity based on the probability function. We define the \textbf{Similarity based on probability distance} $d_{Sim}$ as
\begin{equation} d_{Sim}(A,B)=1-\frac{S(A,B)+S(B,A)}{2}\end{equation}
\end{dfn}

The \textbf{time complexity} of this algorithm lies on the indicator function since every other steps are merely calculations which can be done in linear time. To obtain the partitioning described the the functions $f_X$ and $f_Y$ one should apply a breadth-first search which complexity is $O(|V|+|E|)$, and since this must be done for every for every vertex on each tree (storing results not to repeat calculations), the time complexity of the indicator function is $O(|V|^2)$.

Regarding its \textbf{discriminatory power}, article \cite{bib:1} reported that its performance was underwhelming for trees with five tips and with branch length zero, but excluding these cases behaved similarly to other branch length metrics. It also states that for problems where branch proportion are important but their absolute value isn't, this should be the selected metric.

\subsection{Hybridization Number}
\label{subsection:hyb}

This fairly recent concept was brought up around mid of the first decade of the two thousand, and the main concept behind it is the assumption that evolution does not need to be described by a tree structure since cross-breeding can be an event behind a species existence. Cross-breeds are often called \textit{hybrids}, hence the name of this concept. 

That leads us to expand the standard data structure we've worked until this point since now we can have two distinct paths from the root to a leaf. Following the main motivation for this concept, \textit{directed acyclic graphs} (or DAGs) suit the discussed problem. The article followed in our research was \cite{bib:6}, since it gives a good introduction to the subject, even if its main purpose is presenting results regarding the complexity of the problem that we will approach right after defining some key concepts. Since in \textbf{directed graphs} the edges $(a,b)$ and $(b,a)$ are different, we need to adapt some concepts such as the degree of a vertex:

\begin{dfn} \label{dfn:iodr} \textbf{(In and Out-degree, Split and Reticulation Nodes)}\\
Let $T=(V,E)$ be a directed graph (a graph where, for $u,v\in V$, the edges $uv$ and $vu$ are different). The \textbf{in-degree} of a vertex $u$, denoted as $d^-(u)$, is determined by $|E_{\cdot u}|$ where $E_{\cdot u}=\{xu\in E: x\in V\}$. Similarly, the \textbf{out-degree} of $u$, denoted as $d^+(u)$, is determined by $|E_{u\cdot}|$ where $E_{u\cdot}=\{ux\in E: x\in V\}$.\\
A \textbf{split node} of a directed graph $T=(V,E)$ is a $u\in V$ such that its in-degree is at most 1 and its out-degree at least 2. A \textbf{reticulation node} of a directed graph $T$ is a $v\in V$ such that its in-degree at least 2 (reticulations for short).
\end{dfn}

\begin{dfn} \textbf{(Hybridization Number)}\\
For a graph $T=(V,E)$, the \textbf{hybridization number} $H_T$ is given by \begin{equation} H_T=\sum_{v\in V} (d^-(v)-1)\end{equation}
\end{dfn}

\begin{dfn} \textbf{(Hybridization Problem; Hybridization Distance)}\\
 Given a forest $\mathcal{F}=\{T_0,T_1,...,T_k\}$, where $T_i\in\gamma_S$ for every $0\leq i \leq k$, the \textbf{hybridization problem} consists in finding a graph (which we refer to as \textbf{\textit{hybridization network}}) $N$ such that:
\begin{itemize}
 \item[] $(1)$ For every $T_i$ there is an injective map $h_i:V_{T_i}\rightarrow V_N$ that preserves vertex adjacency (that is, if $uv\in E_{T_i}$ then $h_i(u)h_i(y)\in E_N$);
 \item[] $(2)$ $H_N$ is minimum amongst all trees that satisfy $(1)$.
\end{itemize}

Assuming this problem is solved by $\mathcal{P}$, we can now define a distance $d_H$ between two trees $A,B\in\gamma_S$ as \begin{equation} d_H(A,B)=H_{\mathcal{P}(\{A,B\})}\end{equation} were $\{A,B\}$ is the forest for $\mathcal{P}$. We name this distance the \textbf{hybridization distance}.
\end{dfn}

As stated in \cite{bib:6}, \textit{``The holy grail for this problem is to develop algorithms that can cope with many input trees and non-binary input trees"}, since there's no actual efficient way to compute such metric. We are talking about a metric in which research is still being done given its good interpretation value on phylogenetics, and even though it is being formalized for input sets with the arbitrary number of trees, computing the problem for two specific trees it is a problem considered to lay in \textbf{NP-Hard} and \textbf{APX-Hard}.

As far as its \textbf{discriminatory power} goes it is interesting to note that this metric values not only the shared clades between two trees but the clades in which its cluster representation are not disjoint. Also, the interpretative value for phylogeny is fairly relevant in this case, however, it is still relatively early to come with practical conclusions regarding its discriminatory power since testing is not yet a viable task.

\subsection{Subtree Prune and Regraft}
\label{subsection:SPR}

The idea of \textbf{Subtree Prune and Regraft} distance goes back to Sokal and Rohlf idea of identifying how many operations are two trees apart from each other. This operation, which is named \textit{prune and regraft}, has a far more relevant interpretation in phylogeny when compared to the $\alpha$ operation from where Robinson and Foulds started drafting, and actually, that's the main reason behind the intense research and study made around this operation. As for now (and just like the hybridization number), only looks promising since computing this operation is far from a trivial task from a complexity standpoint.

Our main resource for this subject was \cite{bib:2}, a paper on optimization of the \textit{Subtree Prune and Regraft}, or \textit{SPR}, from 2016. This is a very complete article that compiles not only a good background for understanding the intricacies of the subject at hand, as a handful of important results towards a practical and usable distance formalization. As for us now, let us go over the definition of the \textit{SPR} operation and remarkable results.

\begin{dfn} \textbf{(Rooted Subtree Prune and Regraft operation - rSPR)}\\
Let $A\in\gamma_S$, $\rho\in V_A$ the root of $A$, $u\in V_A\backslash\{\rho\}$, $E_{C_u}\subseteq E_A$ the set of edges of the clade defined by $u$, the vertex $v\in V_A$ as the ancestor of $u$, $Adj_v=\{e_1,e_2,...,e_k\}\subseteq E_A$ the set of edges connecting $v$ to its neighboors, and $xy\in E_A\backslash (E_{C_u}\cup Adj_v)$. The \textbf{rooted Subtree Prune and Regraft} operation is a function $uSPR:\gamma_S\times V\times E\longrightarrow \gamma_S$ such that $uSPR(A,u,xy)=(V,E,S)$ generated by the following procedure:

\begin{itemize}
  \item[] $(1)$ $V_0=V_A\cup\{v'\}$; $E_0=(E_A\backslash\{vu,xy\})\cup\{xv',v'y,v'u\}$;
  \item[] $(2)$ Remove $vu$ from $Adj_v$ and relabel its elements;
  \item[] $(3)$ $(V,E,S)=\alpha(\alpha(...\alpha((V_0,E_0,S),e_{k-1})...,e_2),e_1)$;
  \item[] $(4)$ if $v'$ is adjacent to $\rho$ then $label(\rho)='NULL'$ and $label(v')='root'$.
\end{itemize}
\end{dfn}

The \textbf{unrooted Subtree Prune and Regraft} is also defined as the previous operation in data structures without root, leaving aside all requirements and steps that involving it, and also disregarding the requirement that $v$ is an ancestor of $u$.

\begin{dfn} \textbf{(Subtree Prune and Regraft distance)}\\
  Let $A,B\in\gamma_S$. The Subtree Prune and Regraft distance, denoted by $d_{SPR}$, equals the number of SPR operations required to transform $A$ in $B$.
\end{dfn}

A lot of work has been put recently on researching about this operation. Allen and Steel (2001) proved a theorem about a distance defined the same way as this previous one, but for a more general operation that relates to SPR, the \textit{tree-bisection-reconnection} (or TBR for short) \cite{bib:24}. Ultimately, that lead to Bordewich and Semple (2005) proving the same conclusion in regards of our SPR operation \cite{bib:25}. We'll expose that result after exposing the concept of \textit{maximum agreement forest}. This was established for \textit{rooted binary trees}, but for other data structures should work similarly. Take in consideration Definition \ref{dfn:mrs}:

\begin{dfn} \textbf{(Maximum Agreement Forest)}\\
Let $A,B\in\gamma_S$ be binary rooted trees. An \textbf{agreement forest} for $A$ and $B$ is a collection $\mathcal{F}=\{T_\rho,T_1,T_2,...,T_k\}$ where $T_\rho\in\gamma_{S_\rho},T_1\in\gamma_{S_1},T_2\in\gamma_{S_2},...,T_k\in\gamma_{S_k}$ and $T_1,T_2,...,T_k$ are binary and the following properties are satisfied:
\begin{itemize}
  \item[] $(1)$ $S_\rho,S_1,S_2,...,S_k$ partition $S\cup\{\rho\}$ and, in particular, $\rho\in S_\rho$;
  \item[] $(2)$ For every $i\in\{\rho,1,2,...,k\}$, there is bijective maps that preserve labeling between $T_i$, $A|S_i$ and $B|S_i$;
  \item[] $(3)$ The trees in $\{A(S_i):i\in\{\rho,1,2,...,k\}\}$ and $\{B(S_i):i\in\{\rho,1,2,...,k\}\}$ are vertex-disjoint subtrees (trees which their vertex set are disjoint) of $A$ and $B$, respectively.
\end{itemize}

The \textit{agreement forest} in which the $k$ is minimized is called \textbf{maximum agreement forest} and that $k$ is denoted by $m(A,B)$.
\end{dfn}

For unrooted trees, the previous definition holds without the requirement of $S_\rho$ and $T_{S_\rho}$.

\begin{thm} (Bordewich and Semple, 2005)\\
Let $A,B\in\gamma_S$. Then, the \textbf{Subtree Prune and Regraft distance} is given by
\begin{equation} d_{SPR}(A,B)=m(A,B)-1\end{equation}
\end{thm}

Regarding \textbf{complexity}, there is, until the time of publication of \cite{bib:2}, that being 2015, no solid idea on the complexity of the \textit{SPR} distance, however, it is stated that is conjectured that another distance, by the name of \textit{replug}, that captures a lot of \textit{SPR} distance properties, is \textbf{NP-Hard}, which leads to believe that \textit{SPR} falls under the same category. However, and how we stated before, this is still an ongoing topic of discussion due to its application relevance.

It is interesting to understand, given the metric formulation, how \textit{SPR} shares some of \textit{Robinson Foulds} characteristics regarding its \textbf{discriminatory power}, but it is also interesting to try to understand how promising it might be, given that the limitation of small variations on the trees from the \textit{RF} distance is solved with a simple vertex replug. Just like the \textit{Hybridization Number}, the interpretative value for phylogeny is fairly relevant in this case, however, it is still relatively early to come with practical conclusions regarding its discriminatory power since testing is not yet a viable task.





\cleardoublepage


\chapter{An analysis of the Geodesic Distance}
\label{chapter:GeodesicAnalysis}

After an extensive overview of distances for tree-like structures, we now focus our attention to the Geodesic Distance. In this chapter, we will give further specification of this distance, discuss and analyze one implementation of this distance in detail.

\section{Motivation, Tools and Methodology}
\label{section:MTMethod}

Before we dive into this chapter's objective, we will now clarify it, expose the available tools and discuss the methodology such as present the reasoning behind it.

As exposed in the thesis objectives, one of the purposes of this thesis is providing a more accessible, detailed and complete guide to the implementation of the Geodesic Distance. As we saw in the previous chapter, documentation for the Geodesic Distance is scarce even though is one of the most promising distances. 

We will use the programming language of \textbf{\textit{Wolfram Mathematica version 11.3.0.0}}, even though we see no reasons for our code to be incompatible in previous versions. There is also no reason for this code to be adapted to other coding languages, the decision of using \textit{Mathematica} relied in already acquired coding dexterity, and no coding specific tools were used to ease the code performance. \textit{Mathematica} is a high-level coding language and actually, as we will see further in this work, proved to be a drawback in analysing the complexity of the code given the \textit{closed source} nature of the software, that lead us having no insight in a lot of low-level prebuilt (but essential) functions such as set intersection or finding an element in a set. 

One of the first concerns regarding this implementation challenge was how to represent graphs in \textit{Mathematica} (or any programming language). Representing graphs in code is, such as any way to represent a graph other than a sketch, troubling. So the first problem we needed to approach was how can we save ourselves time and be more practical with specifying \textit{input}, taking in consideration the set of graphs we're dealing with, which are trees in the tree space $\mathcal{T}_n$. 

Next we have the implementation of the \textbf{\textit{GTP Algorithm}} itself. Here we have several things to take in mind, such as having a detailed sketch and a deep understanding of the inner architecture of the algorithm. Article \cite{bib:9} has a good overview of the algorithm, but some details and proofs are sometimes left open or incomplete, even considering the references it points out.

Finally, we follow with code commenting and time complexity analysis.

\textbf{As noted in the beginning of the next subsection, in this chapter, the set $S$ refers to the set of labels $\{0,1,2,...,n\}$ where $0$ is reserved for the root label.}

\section{\textit{SplitToTree} - Data structure for tree-like graphs}
\label{section:S2Tree}

As pointed out, representing graphs in a practical, clean and organized way is troubling, mainly because it is really hard (and still an on-going debate in graph theory) to comprise a small set of parameters to uniquely define a graph.

However, given the trees in $\gamma^w_S$, $S$ is a set of labels of size $n$, set that we will define, without loss of generality, as $\{0,1,2,...,n\}$ (where the label $0$ is reserved for the root and the remaining elements of the set $S$ are labels of leaves), the internal edges $E'\subset E$ must be minimal for the set of partitions the edges $e\in E'$ induce in $S$ upon their removal and every tree has an associated weight function $w$ defined on their edges (recall Notation \ref{dfn:wlt}), Theorem \ref{thm:set}, the \textit{\textbf{Split Equivalence Theorem}} uniquely defines trees based on the set of splits they are associated with. This means that from a set of distinct compatible splits with size at most $n-2$ (from the \textit{\textbf{Space of trees of size $n$}}, Definition \ref{dfn:stsn}) we can identify the topology of the associated tree. We also need a way to embed the weight of the internal edges in this data structure.

\subsection{Defining an order}
\label{subsection:order}

Given the specification of the space of trees of size $n$ available (Definition \ref{dfn:stsn}), seemed only natural to use vectors in this space to represent the trees we want to work with. More than that, from the \textit{GTP Algorithm}, it seemed like an advantage to represent trees in this manner, given that, as an example of a simple advantage, to calculate the length of the cone path between two trees $T_1,T_2\in\gamma^w_S$ as $||T_1||+||T_2||$, being $||T_1||$ the norm of the vector that represents $T_1$ in $\mathcal{T}_{|S-1|}$.

Orthants in $\mathcal{T}_n$ are identified by a split that is a size two partition that some internal edge induces in the set of labels $S=\{0,1,2,...,n\}$. Establishing an order for these splits will make possible for us to represent trees in $\gamma^w_S$ as a vector of size $(2n-3)!!$, therefore, let us designate the set of all valid splits as $\Sigma(S)=\bigcup_{T\in\gamma^w_S} \Sigma(T)$.

\begin{dfn}
Let $S=\{0,1,2,...,n\}$, the set
$
U=\{  X|Y: X\subseteq S,
 2\leq |X| \leq n-1 ,
 Y = S\backslash X \} 
$ and $\sim$ an equivalence relation such that $A|B \sim C|D$ if and only if $(A=C\wedge B=D)\vee (A=D \wedge B=C)$. The set of valid splits is the quotient of the set $U$ from the relation $\sim$, $\Sigma(S)=U/\sim$.
\end{dfn}

There should be no problem prooving that $\sim$ is indeed an equivalence relation. Next, let's consider the following theorem:

\begin{thm}
\label{thm:fbij}
Let $S=\{0,1,2,...,n\}$, $\sigma(S)=K \cup L$ where $K=\{X: X\subsetneq S, 2\leq |X| \leq \lfloor n/2 \rfloor \}$ and $L=\emptyset$ if $n$ is even and $L=\{\{0\}\cup X: X\subset S \mathrm{\ such\ that\ } |X|=\lfloor n/2 \rfloor\}$ if $n$ is odd and $f:\sigma(S)\rightarrow \Sigma(S)$ such that $f(X)=X|(S\backslash X)$. $f$ is a bijective function.
\end{thm}

\begin{proof}
We start by proving that for every $X|Y\in\Sigma(S)$ there exists an $A\in\sigma(S)$ such that $X|Y=f(A)$. Assume, without loss of generality, $|X|<|Y|$. Let $A=X$, since the set $K$ consists of  partition elements smaller or equal than $\lfloor n/2\rfloor$ (recall that $|X|+|Y|=n+1$). Hence, $f(X)=X|(S\backslash X)$ which then equals $X|Y$ since $X|Y \in \Sigma(S)$. The symmetric case (that leads to the conclusion that $A=Y$) it is not different since $X|Y\sim Y|X$. Finally, assume that $|X|=|Y|$. From this, we can infer two things: first, $|X|=n/2=|Y|$ and also $|S|=n+1$ is even, meaning that $n$ is odd, therefore $L\neq\emptyset$. Assume, without loss of generality, that $0\in X$, hence $X\in L \Rightarrow X\in\sigma(S)$. Let $A=X$. We have that $f(X)=X|(S\backslash X) \Leftrightarrow f(X)=X|Y$. This proves that $f$ is surjective.

Let us assume that $A\in\sigma(S)$ and $A\notin L$. Then, $|A|<\lfloor n/2\rfloor$ and let $A|(S\backslash A)\in\Sigma(S)$. By definition of $f$, we have that $f(A)=A|(S\backslash A)\in\Sigma(S)$. Now, assume that there exists a $B\in\sigma(S)$ such that $B\neq A$ and $f(A)\sim(B)$. Since $A|(S\backslash A)\sim B|(S\backslash B)$ holds, $((A=B)\wedge ((S\backslash A)=(S\backslash B)))\vee((A=(S\backslash B))\wedge ((S\backslash A)=B))$ also holds. Since $(A\neq B)$ we have that $A=(S\backslash B)$ necessarily, hence $|A|=n+1-|B|\Leftrightarrow |B|=n+1-|A| \Leftrightarrow B=n+1-|A|\geq n+1-\lfloor n/2 \rfloor > \lfloor n/2 \rfloor \Rightarrow B\notin \sigma(S)$. It is only left for us to assume the case in which $A\in L$. In this case, $n$ is necessarily odd, $|A|=(n+1)/2$ and $0\in A$. Lets assume that there is a $B\in\sigma(S)$ such that $B\neq A$ and $f(A)=f(B)$. Then, we have again that $A=(S\backslash B)$ holds. From this we can conclude that $0\notin B \Rightarrow B\notin L$ and since $|A|=n+1-|B|\Leftrightarrow |B|=n+1-(n+1/2)=(n+1/2)>\lfloor n/2 \rfloor$, $B\notin K$. Reaching the contradiction that $B\notin \sigma(S)$, that we assumed as true. This lead us to conclude that $f$ is injective, therefore, bijective.
\end{proof}

With this theorem we just proved that $f$ is a one-to-one correspondence between $\sigma(S)$ and $\Sigma(S)$, meaning that finding an order in $\sigma(S)$ will provide an order in $\Sigma(S)$. Let us also assume that the sets of $\sigma(S)$ are ordered properly in an increasing way (meaning that if $X=\{x_1,x_2,...,x_k\}\in\sigma(S)$, then $x_i<x_j$ if and only if $i<j$).

\begin{dfn}
\label{dfn:lex}
\textbf{(Lexicografical order for $\sigma(S)$)}\\
Let $S=\{0,1,2,...,n\}$, $X,Y\in\sigma(S)$, $X=\{x_1,x_2,...,x_i\}$ and $Y=\{y_1,y_2,...,y_j\}$ such that $i,j\in {1,2,...,(2n-3)!!}$. We define the \textbf{lexicographical order} for $\sigma(S)$ as a relation such that $X\leq Y$ if and only if:
\begin{itemize}
	\item $i\leq j$, or,
	\item There exists a $k\leq i$ such that $x_k\leq y_k$ and for all $k'<k$ we have that $x_{k'}=y_{k'}$.
\end{itemize}
\end{dfn}

\begin{rmk}
$(\sigma(S),\leq)$, where $\leq$ is the lexicographical order for $\sigma(S)$, is a totally ordered set.
\end{rmk}

After establishing this order, we have now a way how to reference splits by designating which position in the ordered set they occupy. More than that, we can also reference a set $X$ of splits by a binary vector $\vec{v}$ of size $(2n-3)!!$:

\begin{rmk}
Let $S$ be a set of labels and $\sigma(S)={s_1,s_2,...,s_{(2n-3)!!}}$ the respective ordered set of splits. There is a bijective function $h:\{0,1\}^{(2n-3)!!}\longrightarrow \mathcal{P}(\sigma(S))$ given by \begin{equation} h((v_1,v_2,...,v_{(2n-3)!!}))=\bigcup_{1\leq i \leq (2n-3)!!} \{g(v_i)\}\end{equation} and $g:\{0,1\}\longrightarrow \sigma(S)\cup\{\emptyset\}$ a function that returns nothing if $v_i=0$ and returns the $i$th element of $\sigma(S)$ if $v_i\neq 0$.
\end{rmk}

\begin{ex}
Let $S=\{0,1,2,3\}$. Then, $\sigma(S)=\{\{0,1\},\{0,2\},\{0,3\}\}$ is the ordered set of splits. Let $X=\{\{0,1\},\{0,3\}\}$. $X$ can be referenced by $\vec{v}=(1,0,1)$ given that the first element of $\sigma(S)$ is $\{0,1\}$ and \{0,3\} the third element of $\sigma(S)$
\end{ex}

So, as long as $f(h(\vec{v}))$ is a set of compatible splits, $\vec{v}$ uniquely determines a tree in $\gamma_S$. Meaning that if we want to represent trees in $\gamma^w_S$ we can, instead of using $0$ and $1$ on the vector, we can put on the $i$th entry of vector $\vec{u}$ the weight of the $i$th split of $S$ (considering once again the order from Definition \ref{dfn:lex}).

With this in mind, we implemented a function able to return an adjacency list for a tree from its respective space of trees vector. We called this function \textit{SplitToTree} and it demonstrated to be a really practical tool to represent and input trees of $\gamma^w_S$ without the hassle of writing every edge of the tree, given that we have access to the ordered set $\Sigma(S)$. As an example, a tree with $7$ labels and $10$ edges can be represented by $4$ easily identifiable splits.

We will now cover some details of the function \textit{SplitToTree}. Take also note that this is not necessary for our main goal but extremely helpful for visualization, so we did not took efficiency that much into consideration when writing this code.

\subsection{Implementation}
\label{subsection:S2TImp}

The implemented algorithm can be accessed in Appendix A.1.

This algorithm relies in two main programs: \texttt{SMerge} and \texttt{RSize}. While \texttt{SMerge} merges all splits to a \textit{super split}, which will be a tree in Newick format, \texttt{RSize} rebuilds the tree from its leaves, applying a recursive program through the layers of lists in the \textit{super split} that will output an adjacency list and respective edge weights, the data structure for graphs in \textit{Mathematica}. 

Before these two functions are applied, we also need to determine which of the splits is appropriate to start building the \textit{super split}. This will be the split with the subset that contains label $0$ as the smallest between all splits $\Sigma(T)$ where $T$ is the input tree.

After the starting split $\sigma=X_1|X_2$ is chosen, all remaining splits partition $X_1$ or $X_2$. For $X_1$ (or $X_2$) chooses the split which partitions $X_1$ in two sets $C_1,C_2$ and choose the split which the minimum size element of this partition is also the minimum size element amongst all elements of all induced partitions by all splits. Let's assume that $C_1$ is such an element. Then, if for any split $D_1|D_2$, it is the case that $D_1$ or $D_2$ is contained in $C_2$, then build a partition of $C_2$ with this set and replace $C_2$ by it. Assume $D_1$ is such element. $X_1$ is now partitioned as $\{C_1,D_1,D_2\cap C_2\}$. Apply the last described step to $D_1$ recursively until you can't execute it anymore. You'll have that $X_1=\{Y_1,Y_2,...,Y_n\}$ for some $n$. Apply the method described in this paragraph for each $Y_i$ and remove the split that generated this element from the set of usable splits.

At the end of the procedure, you'll have multiple sets contained into each other and the elements of $S$ scattered inside these sets. This is when you'll apply \texttt{RSize} to build the structure from the deepest part to the most out. \texttt{RSize} creates a vertex for each label and connects it to a new vertex with identifier $k$ (here we use numbers higher than the highest number of $S$). Replace this list where these labels belong to the vertex identifier $k$ and repeat the process in a higher level. You'll end up with the desired tree.

One thing that the reader should have in mind is that even though this seems like a practical approach to represent trees, there's also a trade-off for memory. To obtain the splits from the vector, the computer performs multiple searches over $\Sigma(S)$ and the size of this set grows much faster than an exponential function. However, there is most likely a way to calculate the $k$th element of $\Sigma(S)$ for a specific $k\in \{1,2,...,(2n-3)!!\}$, with the lexicographical order for $\sigma(S)$, resorting to triangular numbers (the triangular number of $n$ is $T_n=n+(n-1)+(n-2)+...+1)$) and the $K\cup L$ construction specified in Theorem \ref{thm:fbij}. Since this is not under the field of interest of this thesis we did not give it additional thought, but an exploration of what is the less expensive way to compute this task might be an interesting (but probably not so rich) topic.

\section{The \textit{GTP} Algorithm}
\label{section:GTP}

The \textit{GTP} algorithm is the fastest algorithm known to compute the distance between two trees in $\gamma^w_S$, and it was first described in \cite{bib:9} and was presented in this work in Subsection \ref{subsection:geo}, Algorithm \ref{alg:GTP}. In this section, we will discuss the algorithm with further detail and discuss our implementation, with a final time-complexity analysis of our implementation.

\subsection{General Overview}
\label{subsection:Overv}

The \textit{GTP} algorithm receives as an \textit{input} two trees $T_1$ and $T_2$ of $\gamma^w_S$, returns the \textit{path space geodesic} (Definition \ref{dfn:psg}) (or its length) and consists in three parts plus one extra part: \textbf{initialization}, \textbf{guard test}, \textbf{step} (the latter two consisting of the program cycle) and \textbf{common edge handling}.

\begin{itemize}
	\item \textbf{Initialization} consists of constructing the support $(\mathcal{A}^0,\mathcal{B}^0)=((E_{T_1}),(E_{T_2}))$ and initializing all the variables we will need for execution.

	\item The \textbf{guard test} is where we verify if the support at hand $(\mathcal{A}^l,\mathcal{B}^l)$ satisfies (P3), which is equivalent to check if the \textbf{extension problem} has no solution for each pair $(A_i,B_i)$ of this same support.

	\item The \textbf{step} consists of the construction of a new support $(\mathcal{A}^{l+1},\mathcal{B}^{l+1})$ satisfying certain properties.

	\item \textbf{Common edge handling} replaces the last two steps for trees $T_1, T_2$ that share common edges. For this step, we start by identifying a common edge $e\in E_{T_1}\cap E_{T_2}$ and the split $\sigma_e=C|D$, and compute the geodesic distance between $T^C_1$ and $T^C_2$ and the geodesic distance between $T^D_1$ and $T^D_2$ to determine the geodesic distance between $T_1$ and $T_2$. 
\end{itemize}

Let $T_1,T_2\in\gamma^w_S$ and $\vec{t_1},\vec{t_2}$ be their respective vectors according to the previous section.

In the \textit{initialization}, given that we already decided the data structure to represent trees, determining the initial support requires simply storing the positions in which $\vec{t_1}$ and $\vec{t_2}$ differs from $0$ in two lists (which will be \texttt{As} and \texttt{Bs}). We will store each component of the support in these sets, and call them together when needed.

The \textit{guard} consists in verifying the \textit{extension problem} for $(\mathcal{A}^l,\mathcal{B}^l)$ where $l$ hands out to the number of iterations of the program main cycle. After the following definition, we will finally formulate the \textit{extension problem}.

\begin{dfn} \label{dfn:igraph} \textbf{Incompatibility Graph}\\
Let $T_1,T_2\in\gamma^w_S$ and $E_1,E_2$ subsets of $E_{T_1}$ and $E_{T_2}$ respectively. We define the \textbf{incompatibility graph} $G(E_1,E_2)=(V,E)$ such that $V=E_1\cup E_2$ and \begin{equation} E=\{uv:u\in E_1, v\in E_2,\mathrm{\ } \sigma_u \mathrm{\ and\ } \sigma_v \mathrm{\ are\ incompatible\ splits}\}.\end{equation}
\end{dfn}

The initial statement of the \textit{Extension Problem} derives from condition (P3), and the definition presented is a reformulation of the original one. We will guide the reader through this reformulation in the next subsection.

\begin{dfn} \label{dfn:extp} \textbf{Extension Problem} (reformulation)\\
	Let $T_1,T_2\in\gamma^w_S$, $E_1,E_2$ subsets of $E_{T_1}$ and $E_{T_2}$ respectively and $G(E_1,E_2)$ the incompatibility graph of $E_1$ and $E_2$. The \textbf{Extension Problem} lifts the following question:
	\begin{itemize}
		\item Is there any partition $C_1\cup C_2$ of $E_1$ and $D_1\cup D_2$ of $E_2$ such that:
		\begin{itemize}
		 	\item $C_1\cup D_2$ is a minimum weight cover for $G(E_1,E_2)$;
		 	\item $\frac{||C_1||^2}{||E_1||^2} + \frac{||D_2||^2}{||E_2||^2} \geq 1$
		\end{itemize}
	\end{itemize}
\end{dfn}

\begin{lmm} (Lemma 3.2, article \cite{bib:9})\\
	A proper path $\Gamma$ with support pair $(\mathcal{A}^l,\mathcal{B}^l)$ is a geodesic if and only if the \textit{extension problem} has no solution for any support pair $(A_i,B_i)$ of $(\mathcal{A}^l,\mathcal{B}^l)$.
\end{lmm}

To understand the previous Lemma, it is important to take in consideration the following topics:

\begin{itemize}
	\item At any stage of execution of the \textit{GTP algorithm}, condition (P1) is satisfied given the construction of the support $(\mathcal{A}_l,\mathcal{B}_l)$;
	\item At any stage of execution of the \textit{GTP algorithm}, condition (P2) is satisfied given Lemma 3.4 in \cite{bib:9};
	\item The \textit{extension problem} (Definition \ref{dfn:extp}) is a reformulation of (P3), so if there's no solution for it, (P3) is satisfied. If (P1), (P2) and (P3) are satisfied by a proper path, then, by Theorem \ref{thm:p3} (same as proved Theorem 2.5 in \cite{bib:9}), this path is a geodesic.
\end{itemize}

\textit{Step} is where the new support to be validated is constructed from the weight cover $C_1\cup D_2$. Assuming that the \textit{extension problem} had a solution for the pair $A^l_i,B^l_i$ (such that $A^l_i$ is the $i$th element of $\mathcal{A}^l$ and $B^l_i$ the $i$th element of $\mathcal{B}^l$), we construct $(\mathcal{A}^{l+1},\mathcal{B}^{l+1})$ in the following way:

\begin{itemize}
	\item $As^{l+1}_k=As^l_k$, $Bs^{l+1}_k=Bs^l_k$ for $k<i$;
	\item $As^{l+1}_i=C_1$, $Bs^{l+1}_i=D_1$, $As^{l+1}_{i+1}=C_2$, $Bs^{l+1}_{i+1}=D_2$;
	\item $As^{l+1}_k=As^l_{k-1}$, $Bs^{l+1}_k=Bs^l_{k-1}$ for $i+1<k<r+1$ with $r$ equal to the number of sets in $\mathcal{A}^l$ or $\mathcal{B}^l$.
\end{itemize}

The conditions exposed in Algorithm \ref{alg:GTP} are necessarily assured, it is enough to take in consideration that $\frac{||C_1||^2}{||A_i||^2}+\frac{||D_2||^2}{||B_i||^2}$ is indeed the weight of the minimum weight cover $C_1\cup D_2$, which is exactly what determines if the \textit{extension problem} has solution.

\subsection{The Extension Problem}
\label{subsection:ExtP}

Let us start by restating property (P3) from Theorem \ref{thm:p3}:

Given $T_1,T_2\in\gamma^w_S$ a \textbf{proper path} $\Gamma$ between $T_1$ and $T_2$ with support $(\mathcal{A},\mathcal{B})$ is a geodesic if and only if $(\mathcal{A},\mathcal{B})$ satisfy the property:
\begin{itemize}
  \item[] \textbf{(P3)} For each support pair $(A_i,B_i)$ there is no non-trivial partitions $C_1\cup C_2$ for $A_i$ and $D_1\cup D_2$ for $B_i$ such that $C_2$ is compatible with $D_1$ and $\frac{||C_1||}{||D_1||}<\frac{||C_2||}{||D_2||}$.
\end{itemize}

\begin{dfn} \label{dfn:mwis} \textbf{Maximum weight independent set}\\
Let $G=(V,E)$ be a graph. An \textbf{independent set} $I\subseteq V$ is a set of vertices in which for all $u,v\in V$, $uv,vu\notin E$. A \textbf{maximal independent set} is an independent set with maximal cardinality against every other independent set of $G$. A \textbf{maximum weight independent set} is a maximal independent set with minimal weight against every other maximal independent set. 
\end{dfn}

An independent set in $G(A,B)$ is a set of vertices that the splits their edges correspond to are a compatible split set. Also, since scaling the edge lenghts of $A$ and $B$ will lead to the same inequality $\frac{||C_1||}{||D_1||}<\frac{||C_2||}{||D_2||}$ we will scale $A$ and $B$ such that $||A||=||B||=1$ and square both sides of the equation (meaning that the vertex with associated edge $e$ in $G(A,B)$ will have weight $|e|^2/||A||^2$). This leads us to:
$$\frac{1-||C_2||^2}{||D_1||^2}<\frac{||C_2||^2}{1-||D_1||^2} \Leftrightarrow (1-||C_2||^2)(1-||D_1||^2)<||C_2||^2||D_2||^2\Leftrightarrow ||C_2||^2+||D_1||^2 > 1 \Leftrightarrow$$
\begin{equation}\Leftrightarrow \frac{\Sigma_{e\in C_2} e}{||A||^2} + \frac{\Sigma_{f\in D_1} e}{||B||^2} > 1\end{equation}

\begin{dfn}\label{dfn:mwvc}\textbf{Minimum weight vertex cover}\\
Let $G=(V,E)$ be a graph. A vertex cover $C\subseteq V$ is a set of vertices in which for all $uv\in E$, it is the case that $u\in C$ or $v\in C$. A \textbf{minimal vertex cover} is a vertex cover that has minimal cardinality against every other vertex cover of $G$. A \textbf{minimum weight vertex cover} is a minimal vertex cover with minimal weight against every other minimal vertex cover.
\end{dfn}

\begin{thm}
Let $G=(V,E)$ with weight function $w$ defined in $V$. The weight $w_I$ of the maximum weight independent set and the weight $w_C$ of a minimum weight vertex cover satisfy $ (\Sigma_{v\in V} w(v)) - w_I = w_C$.
\end{thm}

\begin{proof}
Let $I$ be a maximum weight independent set. Then, for $e\in E$, $e$ has at least one vertex not in $I$. Hence, $V\backslash I$ is a vertex cover of weight $(\Sigma_{v\in V} w(v)) - w_I$. Suppose $V\backslash I$ is not a minimal vertex cover, then there is $v\in V$ such that $(V\backslash I)-v$ is a vertex cover. This means that all vertices in the neighborhood of $v$, that is, $v$ and its neighborhood are disjoint with $I$. Hence $\{v\}\cup I$ is also an independent cover, which contradicts $I$ maximality. Supposing that $V\backslash I$ hasn't minimal weight also contradicts $I$ maximum weight. You can prove that if $M$ is a minimal weight cover with a symmetric argument.
\end{proof}

This leads us to the formulation of the \textit{extension problem} as we defined it in Definition \ref{dfn:extp}. However, our discussion on how to solve the extension problem isn't over yet. Concluding:

\begin{dfn} \label{dfn:extp2} \textbf{Extension Problem} (final formulation)\\
For all pairs $(A_i,B_i)$ from the support $(\mathcal{A}^l,\mathcal{B}^l)$, we want to check if the minimum weight vertex cover for $G(A_i,B_i)$ with vertex weights

\begin{equation}
w_e=
\begin{cases}
\frac{e^2}{||A_i||^2} & \mathrm{\ if\ } e\in A_i\\
\frac{e^2}{||B_i||^2} & \mathrm{\ if\ } e\in B_i\\
\end{cases}
\end{equation} has weight greater than 1. If that's the case for all the minimum weight covers, the \textit{extension problem} is said to have no solution (hence satisfies (P3)) and we know that we can generate the \textbf{path space geodesic} from the support $(\mathcal{A}^l,\mathcal{B}^l)$.

\end{dfn}

From article \cite{bib:9} it is pointed out that this can be solved as a \textit{\textbf{max-flow}} algorithm, but neither the details of how to adapt the problem to a \textit{max-flow}, neither the references it points out are enough to adapt the problem properly.

\begin{dfn} \label{dfn:mfp} \textbf{Maximum flow problem}\\
Let $N=(V,E)$ be a directed graph with $s,t\in V$, which we will designate by \textbf{source} and \textbf{sink} respectively, $c:E\longrightarrow \mathbb{R}^+$ a capacity function defined on the edges $N$. We designate $(N,c)$ as a \textbf{flow network}. A \textbf{flow function} is a mapping $f:V\times V\longrightarrow \mathbb{R}^+$ such that for every $xy\in E$ we have that $f(xy)\leq c(xy)$, $f(yx) = -f(xy)$ such that $y,x\notin \{s,t\}$ (and $yx$ not necessarily in $E$) and also $\sum_{xy\in E}f(xy)=0$ for any $x\in V\backslash\{s,t\}$. The \textbf{flow} of $N$ is defined as $|f|=\Sigma_{sx\in E} f(sx)$ and the \textbf{maximum flow problem} consists in determining the maximum value of $|f|$.
\end{dfn}

\begin{dfn} \label{dfn:mcp} \textbf{Minimum cut problem}\\
Let $N=(V,E)$ be a directed graph with a weight function $w:E\longrightarrow \mathbb{R}^+$. We define a \textbf{cut} as a partition $\{U_1,U_2\}$ of $V$ such that there is at least one $xy\in E$ such that $x\in U_1$ and $y\in U_2$ and $s\in U_1$ and $t\in U_2$. Let $C\subseteq E$ the subset of edges between $U_1$ and $U_2$. The \textbf{minimum cut problem} consists in finding a \textit{cut} such that the weight of $C$, $w(C)=\Sigma_{e\in C}w(e)$ is minimal amongst the weight of all possible cuts.
\end{dfn}

Both of these problems are actually related by the following theorem:

\begin{thm} \label{thm:mfmct} \textbf{Max-flow min-cut theorem}\\
In a flow network $(N,c)$, the maximum flow $|f|$ equals $w(C)$ where $C$ is the minimum cut of $N$.
\end{thm}

You can find a proof for this important theorem in reference \cite{bib:29}, Section 6.5.

\begin{dfn} \label{dfn:feg} \textbf{Flow equivalent graph}\\
Let $G(A,B)=(A\cup B,E)$ be the incompatibility graph of $A$ and $B$, and $w_e:A\cup B\longrightarrow \mathbb{R}^+$ a weight function defined in the vertices of $G(A,B)$. We define the \textbf{flow equivalent graph} $G_f(A,B)=(V_f,E_f)$, a directed graph for our problem, as:
$$ V_f=A\cup B\cup \{s,v\}$$
$$ E'=\{ab: a\in A, b\in B, ab\in E \mathrm{\ or\ } ba\in E\}$$
$$ E_f=E'\cup\{sv\}_{v\in A}\cup\{vt\}_{v\in B} $$
and the weight function $w_f:E_f\longrightarrow \mathbb{R}^+$ defined as:
\begin{equation} w_f(xy) =
\begin{cases}
w_e(y) & \mathrm{\ if\ } x=s\\
w_e(x) & \mathrm{\ if\ } y=t\\
\infty & \mathrm{otherwise}\\
\end{cases}
\end{equation}

\end{dfn}

\begin{thm}
\label{thm:detapir}
Let $G(A,B)=(A\cup B,E)$ be the incompatibility graph of $A$ and $B$, and $G_f=(V_f,E_f)$ the flow equivalent graph for $G(A,B)$. A minimum cut in $G_f$ induces a minimum weight vertex cover in $G(A,B)$ and vice versa.
\end{thm}

\begin{proof}
Let $S$ be a vertex cover of $G(A,B)$ and $\delta(C)=\{xy\in E_f: x\in C \wedge y\notin C\}$. Consider the cut $(U_1,U_2)$ such that $U_1=(A\backslash S)\cup(B\cap S)$ and $U_2=(A\cup B)\backslash U_1$. Any edge $ij\in E_f$ with $w_f(ij)=\infty$ is not in $\delta(U_1)$ since $(i\in U_1)$ is equivalent to $(i\notin S)$, and $(j\notin U_1)$ is equivalent to $(j\notin S)$. On the other hand, an edge $si\in\delta(U_1)$ if and only if $i\in S$, and $jt\in\delta(U_1)$ if and only if $j\in S$. Also:

$$w(S)= \Sigma_{v\in S} w(v) = (\Sigma_{v\in (S\cap A)}w(v))+(\Sigma_{u\in (S\cap B)}w(u))=$$
\begin{equation} (\Sigma_{v\in (S\cap A)}w_f(sv))+(\Sigma_{u\in (S\cap B)}w_f(ut))= \Sigma_{e\in \delta(U_1)} w_f(e)=w_f(\delta(U_1))\end{equation}
\end{proof}

\begin{figure}[!htb]
  \centering
  \includegraphics[width=1\textwidth]{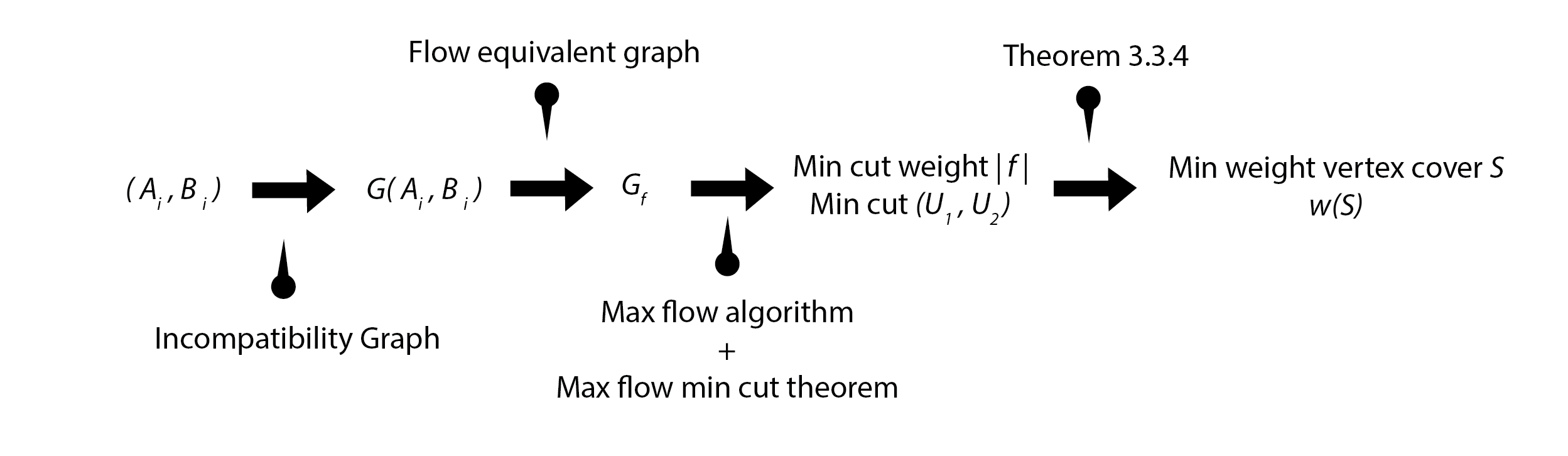}
  \caption[Understanding computation of the minimum weight vertex cover.]{A diagram for the path from the support set until its respective minimum weight vertex cover and its weight}
  \label{fig:extpdiag}
\end{figure}

We are now into condition to delineate a path from the \textit{extension problem} to its solution. In Figure \ref{fig:extpdiag} it is presented a schema for clearer understanding.

The first step from the \textit{extension problem} is to generate the \textit{flow equivalent graph} $G_f$ for $G(A_i,B_i)$ (Definition \ref{dfn:feg}). Then, we follow to solve the \textit{max-flow problem} (Definition \ref{dfn:mfp}) for $G_f$ using edge weight as their capacities in the \textit{max-flow problem}, meaning $c=w_f$, which will give us not only the \textit{maximum flow} (equivalent to the \textit{minimum cut weight}), but also a minimum cut (which is held from the \textit{max-flow min-cut theorem}, Theorem \ref{thm:mfmct}) for the \textit{minimum cut problem} (Definition \ref{dfn:mcp}) for $G_f$. Having the weight of the \textit{minimum weight cut} (and the actual \textit{minimum weight cut} when we want to generate the vertex cover) for $G_f$ we know the weight of the \textit{minimum vertex cover} (or generate this cover) for $G(A_i,B_i)$ using Theorem \ref{thm:detapir}.

This lead us to a final problem, which is how to compute the \textit{maximum flow} for the $G_f$ . This can be done in a multitude of ways, such as a \textit{push-relabel algorithm} \cite{bib:29}, \textit{Dinic's algorithm} \cite{bib:31} or \textit{Orlin's algorithm} \cite{bib:30}. However, taking time complexity and our bipartite graph $G(A,B)$ in consideration, we decided to set for a variation of \textit{Ford-Fulkerson's}, the \textbf{Edmonds-Karp's algorithm} \cite{bib:28}, which has time complexity of $O(|V||E|^2)$, however, it is important to note that algorithms with lower time complexity exist: \textit{Orlin's algorithm} computes the \textit{max-flow} in $O(|V||E|)$; There's a \textit{Goldberg and Rao} blockflow algorithm that leads to a $O(min(|V|^{2/3},|E|^{1/2})|E|\log(|V|^2/|E|)\log(U))$ complexity for capacities in the range of $\{1,2,3,...,U\}$ \cite{bib:32}; A variation on \textit{Spielman and Teng} algorithm for max-flows, resulting in an almost linear time complexity \cite{bib:33} \cite{bib:34}.

\subsection{Edmonds-Karp algorithm}
\label{subsection:EKAlg}

The \textbf{Edmonds-Karp algorithm} is a variant of the \textit{Ford-Fulkerson algorithm} that improves time complexity of $O(|V||f|)$ for $O(|V||E|^2)$ (with $|f|$ being the maximum flow of the analyzed network). A specification of this algorithm can be found in \textit{Introduction to Algorithms}, reference \cite{bib:28}. 

Briefly, the algorithm consists in a cycle that searches for the shortest path $P$ in the \textit{residual graph} of $G_f$ from $s$ to $t$ and pushes flow through $P$ equal to the minimum \textit{residual capacity} of all edges in $P$. \textit{Residual capacity} of an edge $xy\in E$ is given by $r(xy)=c(xy)-f(xy)$ where $f(xy)$ is the cumulative flow pushed through the edge $xy$ during this process. The cycle runs until there's no path from $s$ to $t$.

\begin{dfn} \label{dfn:resg}
The \textbf{residual graph} of $G_f=(V_f,E_f)$ with associated flow function $f$ after the execution of \textit{Edmonds-Karp} is a directed graph $R=(V_r,E_r)$ such that:
\begin{equation} V_r=V_f;\mathrm{\ \ \ } E_r=\{xy: x,y\in V_f \mathrm{\ and\ } r(xy)>0\}. \end{equation}
\end{dfn}

However, the networks in which we'll apply this algorithm have a simplified topology, they're bipartite graphs with vertex sets $A$ and $B$ in which one of the vertex subsets is either connected to the sink or end points of the source, lets assume, without loss of generality, that these are $B$ and $A$ respectively. More than that, edges from $A$ to $B$ have capacity $\infty$, so we do not need to consider these for calculating the minimum residual capacity along the shortest path. And talking about shortest paths, in the first iteration every path from $s$ to $t$ is a shortest path of $3$ edges. We now present a suggestion for a \textit{Edmonds-Karp} formulation taking all this in consideration:

\begin{alg} \label{alg:EKA}\textbf{Edmonds-Karp} (proposed adaptation)
\begin{itemize}
\item[]\textbf{Input:} $G=(A\cup B, E), w_e$;
\item[]\textbf{Output:} The maximum flow for $G_f$, $\Sigma_{a\in A}f'(a)$.
\item[]\textbf{Initialize:} $i=1$; $f':A\cup B\longrightarrow \mathbb{R}^+$ such that $f'(x)=0$ to be redefined point-to-point during the execution of step.
\item[]\textbf{Step:} Iterate while $i\leq size(E)$
\begin{itemize}
	\item[] Let $a_ib_i\in E$; $\delta=min\{w_e(a_i)-f'(a_i),w_e(b_i)-f'(b_i)\}$;
	\item[] $f'(a_i)=f'(b_i)=\delta$;
\end{itemize}
\end{itemize}
\end{alg}

Keep in mind that these operations are being made over the incompatibility graph $G(A,B)=(A\cup B, E)$ for sake of simplicity (the \textit{Edmonds-Karp} is defined for flow networks and $G(A,B)$ is not even a directed graph): In $G_f$ there is always an edge from $s$ to every vertex $a\in A$ with capacity $c(a)=w_e(a)$, there is an edge $a_ib_i$ of $\infty$ capacity if and only if $a_ib_i\in E$ and there is always an edge from every vertex $b\in B$ to $t$ with capacity $c(b)=w_e(b)$. In this first run, there is at most as many shortest paths from $s$ to $t$ as edges in $E$. 

There is, however, the possibility of after all the paths of length $3$ being saturated, more paths of length $2n+1$ exist. After traversing an edge of length $\infty$, there might be an edge in the residual graph pointing back to some $a\in A$, and that might lead to a new shortest path in the graph (check Figure \ref{fig:ekex} as an example).

So after Algorithm \ref{alg:EKA} is executed, we need to search for other shortest paths. Here, the strategy we adopted was implementing a \textit{Breadth First Search} (BFS for short) that will build, one vertex at the time, all the possible paths from $s$ until they reach $t$, and in that case, the BFS will stop and the path is defined as the shortest, or discarded if they reach a previsited vertex or have no adjacent vertices.

Further, we need to generate the minimum-cut of $G_f$, which is done by seeing which vertices of $G_f$ are reachable from $s$ on the \textit{residual graph} after the execution of \textit{Edmonds-Karp}. This is also another BFS, under different conditions.

Take into consideration (keeping in mind Definition \ref{dfn:mfp}) that if there's flow $k$ in the arc $xy\in E_f$ then, $yx\in E_r$ and $f(yx)=k$. \textbf{Edges from $s$ or with endpoint $t$ have no return residual edges}.

Once again, we can look at the problem of how to compute the minimum-cut, that we inevitably will need to do compute the vertex cover, by trying to contextualize in our problem. As a matter of fact, after the execution of Algorithm \ref{alg:EKA}, we can determine which vertices are reachable from $s$ by analyzing vertex capacities, i.e. for $v\in A\cup B$ what is the value of $w_e(v)-f'(v)$. If $v\in A$ and $w_e(v)-f'(v)=0$ then $v$ is not reachable. (*) Otherwise not only $v$ is reachable, but also every vertex of $B$ that it is connected to. Then, for every vertex $b_i\in B$ to whom $v$ is connected to, we need to check if the edge $b_i a_j$ is in the residual graph, with $j\in J$ (since everytime there's flow in one way, there must be a returning edge with the same flow value). If there are one or more vertices in $\{a_j\}_{j\in J}$ that weren't already reachable from $s$ (from an $sa_j$ arc), we must repeat the procedure from (*) for all those.

So, unfortunately, we also need to save some information from the return flow. Consider the following revised version of \textit{Edmonds-Karp} bellow

\begin{alg} \label{alg:EKA2}\textbf{Edmonds-Karp} (revised)
\begin{itemize}
\item[]\textbf{Input:} $G=(A\cup B, E), w_e$;
\item[]\textbf{Output:} The maximum flow for $G_f$, $\Sigma_{a\in A}f'(a)$ and the cut set $U_1$.
\item[]\textbf{Initialize:} $i=1$; $f':A\cup B\longrightarrow \mathbb{R}^+$ such that $f'(x)=0$ to be redefined point-to-point during the execution of step.
\item[]\textbf{Step:} Iterate while $i\leq size(E)$
\begin{itemize}
	\item[] Let $a_ib_i\in E$; $\delta=min\{w_e(a_i)-f'(a_i),w_e(b_i)-f'(b_i)\}$;
	\item[] $f'(a_i)=f'(a_i)+\delta$;
	\item[] $f'(b_i)=f'(b_i)+\delta$;
\end{itemize}

\item[]\textbf{Step:} Iterate while there's no more paths $P$ from $s$ to $t$:
\begin{itemize}
	\item[] Let $P=u_1,u_2,...,u_{2n+1}$ be a sequence of vertices such that $u_1=s$ and $u_{2n+1}=t$ and $u_iu_{i+1}$ are valid edges in the residual graph from $G_f$; 
	\item[] $\delta =min\{w_e(u_1)-f'(u_1),f(u_3u_2),f(u_5u_4),...,f(u_{2n}u_{2n-1}),w_e(u_{2n+1})-f'(u_{2n+1})\}$;
	\item[] $f'(u_1)=f'(u_1)+\delta$;
	\item[] $f(u_{i+1}u_i)=f(u_{i+1}u_i)+\delta, \forall 1<i<2n$;
	\item[] $f'(u_{2n+1})=f'(u_{2n+1})+\delta$;
\end{itemize}
\item[] Compute the set $U_1$ of reachable vertices from $s$ in the residual graph of $G_f$.
\end{itemize}
\end{alg}

\subsection{Common Edge Handling}
\label{subsection:GEHandling}

If we're running the \textit{GTP Algorithm} for two trees $T_1$ and $T_2$ such that these trees share an internal edge $e\in T_1\cap T_2$ with corresponding split $\sigma_e=C|D$, the algorithm must execute different operations, distinct from the operations we discussed before. From Theorem \ref{thm:gdse} we know that we need to compute $T^C_1$, $T^C_2$, $T^D_1$ and $T^D_2$. This is done by intersecting the set $X$ with respective splits from the sets $E^X_{T_1}$ and $E^X_{T_2}$, with $X=C$ or $X=D$.

After this operation, we do not compute $T^C$ and $T^D$ per se, as suggested by Definition \ref{dfn:TC}, but in the following manner:

\begin{thm}
Let $T_1,T_2\in\gamma^w_S$, $\sigma_e=C|D\in\Sigma(T_1)\cap\Sigma(T_2)$, $E^X_{T_i}$ for $X\in\{C,D\}$ and $i=1,2$. The tree $T^X_i$ induced by $X\in\{C,D\}$ in $T_i$ can be specified by the split set:
\begin{equation} \Sigma(T^X_i)=
\begin{cases}
\{(X\cap A)|((X-A)\cup \{0\}):A\in E^X_{T_i}\} & \mathrm{\ if\ } 0\notin A\\
\{(X\cap A)|((X-A)):A\in E^X_{T_i}\} & \mathrm{\ otherwise }
\end{cases}
\end{equation}
\end{thm}

After determining the set of splits $\Sigma(T^X_i)$, we can build the split vector for $T^X_i$ after relabing the leaf labels and passing the weights from the original $T_i$. Everything else to determine distance between trees with shared edges (from Theorem \ref{thm:gdse}) proceeds as expected.

\subsection{Implementation}
\label{subsection:GTPImplementation}

The implemented algorithm can be accessed in Appendix A.2. The code is heavily commented, and with the explanation provided in previous sections, the reader should have no trouble understanding the implementation, is mostly straightforward. We advise special care with small technical details such as the variable syntax between each subprogram of the algorithm.

\subsection{Time complexity analysis}
\label{subsection:tcanalysis}

Reference \cite{bib:9} states that the \textit{GTP algorithm} has $O(n^4)$ time complexity. In this subsection we will analyze the time complexity of our implementation and compare it to the statement, pointing out places of possible improvement and discussing the limitations of the chosen programming language of \textit{Wolfram Mathematica} and how they set us back.

\textit{Wolfram Mathematica 11.3} was chosen given the already acquired code dexterity and at the time of the decision, prebuilt functions seemed like an advantage for code performance. This last reason is debatable given that \textit{Mathematica} is considered a fairly high-level language. Later, we've found out that all \textit{Mathematica} prebuilt functions have a non-user accessible implementation, given that it is \textit{closed source} software (\textit{Wolfram Mathematica} is proprietary code). Given this, when analyzing complexity, we discussed the reliability of empirically estimating the complexity of prebuilt functions but ended up considering worst case complexity for the least time-complex known for functions at hand. Meaning that during the complexity analysis of the \textit{GTP algorithm}, every time that we will refer to the complexity of a \textit{Mathematica} prebuilt function $f$, we will consider the worst case time complexity for the best algorithm to compute $f$.

The \textit{GTP algorithm}, as described in Subsection \ref{subsection:Overv} has four main phases, however, the complexity of the algorithm most likely resides in around solving the \textit{extension problem} and \textit{handling trees with common edges}. You can find the implementation in Appendix A.2.

Let $T_1,T_2\in\gamma^w_S$ be the \textit{GTP Algorithm} input.

In regards of the \textit{initialization}, the time complexity of the used functions are:

\begin{itemize}
	\item \texttt{Position}, $O(1)$;
	\item \texttt{Flatten}, $O(n)$ for $n$ size of the list;
	\item \texttt{Table}, $O(n)$ for $n$ size of the created list;
	\item \texttt{Length}, usually $O(n)$ for $n$ size of the argument list, but since \textit{Mathematica} stores it inside the data structure, its complexity is $O(1)$;
	\item \texttt{Part} (under the form of double straight brackets \texttt{[[ ]]}), $O(1)$.
\end{itemize}

As for the \texttt{CanSplitOrder[]}, we should point out that this is specific for our implementation, and we'll go into details later.

This piece of code is executed once, so the overall impact of the function execution will be $O(n)$, for $n$ equal to the size of one of the two input tree vector (which are equal).

Now, the paths for the function split depending if $T_1$ and $T_2$ have shared internal edges. Let us start with the \textit{common edge handling}, which can be found in the latter part of the algorithm.

This section of the code starts with a cycle, but there is no other instructions to do inside of it other than \texttt{i++} when the \texttt{If} does not trigger. Plus, the first execution of \texttt{If} body will set \texttt{i=}$\infty$, which will lead this cycle to break. In the worst case, the \texttt{If} triggers at the end of \texttt{CommonEdges} vector. When it comes to prebuilt functions that didn't appear in \textit{initialization}, the following are present:

\begin{itemize}
	\item \texttt{MemberQ}, $O(1)$;
	\item \texttt{NumberQ}, $O(1)$;
	\item \texttt{SubsetQ}, $O(n)$ for $n$ size of the tested subset;
	\item \texttt{Append} and \texttt{Prepend}, $O(n)$ for $n$ size of the list to pre or append;
	\item \texttt{Sort}, $O(n\log(n))$ for $n$ size of the list;
	\item \texttt{ReplaceAll}, $O(n)$ for $n$ the size of the list to replace elements in.
\end{itemize}

Adding to this, we will approach the complexity of $\mu$\texttt{vb}, and \texttt{CompatibleSplitsListIndexes["o"]} later. However, one might say that the higher complexity resides in the $\mu$\texttt{vb} function by performing \texttt{Sort} operations, which time complexity leads to $O(n^2\log(n))$. However, all \texttt{Sort} operations are for two elements list, making its complexity actually constant. 

Before we direct our attentions to the \textit{extension problem}, we will consider the operations made in \textit{step}, where we rebuild the support. This part of the code only features the prebuilt function \texttt{DeleteCases} under the original function \texttt{SetSub}. \texttt{DeleteCases} has time complexity $O(n)$.

As for the \textit{extension problem}, the code is captured by two functions, \texttt{ExtP} and \texttt{EdmondsKarp}, with \texttt{EdmondsKarp} contained in \texttt{ExtP}. Prebuilt functions in \texttt{ExtP} are:

\begin{itemize}
	\item \texttt{Join}, $O(n)$, for $n$ the size of the lists to be joined;
	\item \texttt{N}, $O(1)$.
\end{itemize}

The \textit{extension problem} can be executed by the main \textit{GTP algorithm} cycle, in worst case, the same amount of times as many internal edges there are, meaning $|S|-2$ times.

And finally, in regards to \texttt{EdmondsKarp}, the list of previously unmentioned functions are:

\begin{itemize}
	\item \texttt{First}, $O(1)$;
	\item \texttt{Last}, $O(n)$, for $n$ the size of the input list;
	\item \texttt{OddQ, EvenQ}, $O(1)$;
	\item \texttt{Abs}, $O(1)$;
	\item \texttt{Last}, $O(n)$, for $n$ the size of the input list.
\end{itemize}

\texttt{EdmondsKarp}, as referred in the Subsection \ref{subsection:EKAlg}, has three main parts, each one of them containing a loop:
For the first one, the computation of the paths of length $3$, it runs exactly $|E|$ times, where $E$ is the set of edges of the incompatibility graph generated by a pair of sets from the support at hand. However, there could be implemented cycle skips to run it in $|V|/2$ steps, taking in consideration that his procedure saturates, in the worst case scenario, half of the edges from and with endpoint $s$ and $t$ respectively. This also means that, in the worst case scenario, only $|V|/2$ vertices have remaining residual capacity to carry flow from $s$ to $t$.

Secondly, the algorithm performs a \texttt{While} loop until there are no more paths from $s$ to $t$. One could say that from the book \textit{Introduction to Algorithms} \cite{bib:28}, Section 26.3, the complexity of this task is the same as solving a maximum matching in a bipartite graph by a Ford Fulkerson algorithm has $O(|V||E|)$ time complexity. However, this is not the case: in Section 26.3 of \cite{bib:28} the problem is specified for unweighted graphs, and there is no way of optimizing for weighted graphs since for unweighted graphs the edge capacity from $sa$ for $a\in A$ (which equals $1$) gives a lower bound for edge minimum weight on the shortest path hence for the number of executions, since all the return edges in the residual graph have positive integer weights. This leave us with the only option of executing a regular \textit{Edmonds Karp} until there are no more paths from $s$ to $t$, which is bounded by $|V||E|^2$ executions, hence, has the worst case time complexity of $O(|V||E|^2)$.

\begin{figure}[!htb]
  \centering
  \includegraphics[width=0.68\textwidth]{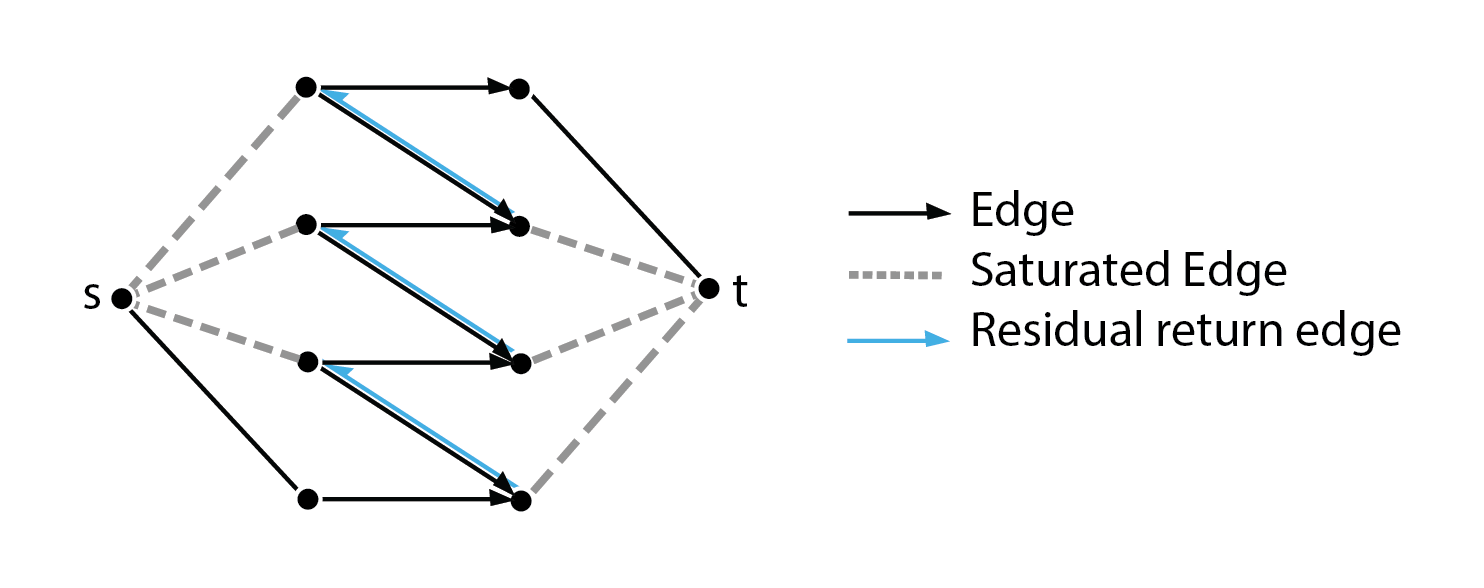}
  \caption[Worst case time complexity for a \textit{BFS} procedure.]{An example to depict a worst-case time complexity for the \textit{BFS} procedure.}
  \label{fig:ekex}
\end{figure}

Finally, the algorithm performs a \textit{BFS} procedure through the final residual graph. The worst case time complexity for this procedure is $O(|V|)$ (for $|V|$ equal to $|A\cup B|$), and the data structure we're using does not help to lower this complexity: it is enough to consider a connected $G(A,B)$ in which every vertex has degree $2$ except vertices for $a,b\in A\cup B$ with degree $1$ and $a$ is such that $sa$ in the residual graph is the only edge from the source $s$. Also, for every edge $a'b'\in G_f$ of $G(A,B)$ there is a return edge on the residual graph (check Figure \ref{fig:ekex} for an example for this case). In this graph, there's only one path that the \textit{BFS} can take, and it must visit every vertex of the graph to collect the cut set.

Putting it all together, the bigger slice of complexity is indeed in the Edmonds Karp algorithm, which is capsuled inside the main program cycle, which the number of runs is, as mentioned before, bounded by the minimum number of internal edges for one of the input trees. Each internal edge in the input trees corresponds to a vertex in the incompatibility graph, \textbf{therefore, we can conclude that the worst case time complexity of the \textit{GTP Algorithm} is $O(|E|^2|V|^2)$}. More specifically, and putting it in terms of the input $T_1$ and $T_2$ instead of vertices and edges of the incompatibility graphs, we have that $|V|$ equals the number internal edges of $T_1$ and $T_2$ summed and $|E|$ equals the number of incompatible splits to which the internal edges of $T_1$ and $T_2$ are associated with. Then, let $t_1$ and $t_2$ be the number of internal edges of $T_1$ and $T_2$ respectively and $k$ the number of incompatible splits between $T_1$ and $T_2$. The complexity of the \textit{GTP Algorithm} is $O(k^2min(t_1,t_2)(t_1+t_2))$.

With the modular implementation we have developed, we can easily swap in and out parts of code as long as they maintain the appropriate structures, however, it is reasonable to think that one disadvantage of this implementation is that we're bounded by \texttt{CompatibleSplitsListIndexes["o"]} (this function builds a list of all compatible splits indexes by comparing them pairwise, so we could later find compatible splits with $O(1)$ time-complexity through look-up hash tables. A small specification can be found in Appendix A.1 but split compatibility can actually be computed on the go, instead of pre-calculated and stored. For comparison of huge data sets of trees with small sets of labels, however, this method is more reliable: calculating and storing compatible splits is not so expensive in time or memory and split compatibility (that requires computing set intersection) does not need to be computed each time its needed to build the incompatibility graph and can be just searched in the table of compatible splits. So it is more a matter of preference considering which dataset it will be explored (and also available memory: after all, the number of splits grows according to a double factorial) rather than anything else. \texttt{CanSplitOrder[]} hands out the \textit{Canonical Split Order} for the current label size $Sz$ (which is calculated in constant time $O(1)$) and $\mu$\texttt{vb} \texttt{While} cycle runs are bounded by the size of the new vector for the new run of the \textit{GTP}, so it's time complexity is linear.

Another interesting topic of discussion would be how can this be optimized using more efficient algorithms for computing \textit{max-flow}. First of all, it is important to realize that we have, in the best case scenario, a linear lower bound time complexity during the support reconstruction cycle. On the other hand, the \textit{max-flow} problem might have lower time complexity, but it is not clear if the algorithms proposed by \textit{Orlin} \cite{bib:30}, \textit{Goldberg and Rao} \cite{bib:32} and finally the ``almost linear time" variation of \textit{Spielman and Teng} \cite{bib:33} \cite{bib:34} can be applied in a network with weighted vertices, since, as we discussed in our work, the \textit{Edmonds-Karp} optimization for flows in bipartite graphs to $O(|V||E|)$ can't be applied in this case. Hypothetically, the \textit{GTP Algorithm} could have a complexity ``almost" of order $O(n^2)$ and this might be an interesting research topic for the future, since it would put it closer to the (by far) most popular \textit{Robinson Foulds distance} and its $O(n)$ time complexity.

The worst case time complexity for the \textit{GTP Algorithm} stated in \cite{bib:9} is $O(n^4)$ which matches our implementation of complexity $O(|V|^2|E|^2)$, which is of order $O(n^4)$. 
Our implementation could also have space for improvement in the best case time complexity (such as analysing the partial flow $|f|(t)$ at each moment $t$ of the Edmonds Karp and quit the computation each time that  $|f|(t)>1$ or rework the relabeling system on the handling of trees with common edges, but unfortunately, for the worst-case time complexity, running \textit{Edmonds Karp} inside of our main program loop, bounds it. 
\cleardoublepage





\chapter{Conclusions}
\label{chapter:conclusions}

\section{Achievements}
\label{section:achievements}

As you witnessed, our work was divided in two core parts: an expositive text regarding thirteen of the most popular or promising methods to compare tree-like structures; specification of details for the \textit{Geodesic Distance}, implementation of the \textit{GTP Algorithm} and an worst-case time complexity analysis. We will discuss conclusions for these two topics separatly.

\subsection{Distance synopsis}
\label{subsection:syn}

After all the formalization and property lifting of all the metrics in Chapter \ref{chapter:background} we can say that we managed to compile information about all these with the same notation, providing a valuable resource for someone who looks for an introduction for comparative metrics. Some original work was developed (as stated in Introduction) and that helped us gluing some gaps in our understanding of the distances we've approached.

We present below two tables compiling the information regarding the complexity and discriminative power of all the metrics. Since $d_{H}$ and $d_{SPR}$ don't have yet feasible algorithms to be calculated, it's presented in Table \ref{tab:dprob} the conjectured problem hardness of their computation. The worst case time complexity of the geodesic distance is also revised by our work in Chapter \ref{chapter:GeodesicAnalysis}.

\subsection{SplitToTree and GTP Algorithm Implementation}
\label{subsection:GTPAI}

In regards to our achievements in Chapter \ref{chapter:GeodesicAnalysis}, we can say that the representation of trees in an edge weight vector, from Section \ref{section:S2Tree}, made it fairly straightforward to handle trees in the \textit{GTP Algorithm}, and representing trees in this way is extremely practical once we have access to a sketch. Those vectors are generated by, first, checking the splits (induced partitions on the set of labels $S$ for each $e$ internal edge, checking which labels belong to each connected component on the graph upon removal of $e$), check their position according to a lexicographical order, and placing the weight of each $e$ in a vector of zeroes of size $(2n-3)!!$ ($|S|=n+1$).

In regards to our implementation, it matches the worst case time complexity reported by Owen et al in \cite{bib:9}. This comes from the fact that the maximum flow problem is solved with an \textit{Edmonds Karp} algorithm, which has $O(|V||E|^2)$ complexity. The data structure we chose for input led us to some compromise in terms of memory, and for big label sets, it might be a good idea to rethink the way compatible splits are checked. Also, its worth to note that the fact we chose \textit{Wolfram Mathematica} as a coding language was also a drawback when it came to analyze complexity, since it is proprietary code and function specification does not cover complexity, neither we have way to check its implementation.

\begin{table}[!ht]
\centering

\label{tab:dcomp}
\begin{tabular}{|l|l|l|}
\hline
            & Time Complexity                         & Discriminatory Power                                                                                                                                                                                                                                       \\ \hline
$d_{RF}$    & $O(n)$                                 & \begin{tabular}[c]{@{}l@{}}- Really sensible to the scalability of S; \\ - Unstable: moving a single leaf could lead\\ to great discrepancies in the distance\\ value; \\ - Overperforms in close to resolved trees\\ other than to unresolved ones.\end{tabular} \\ \hline
$d_{RFL}$   & $O(n)^{(*)}$                           & \begin{tabular}[c]{@{}l@{}}- Shares discriminatory power from RF; \\ - In some cases is non-symmetric and\\ does not has identity of indiscernibles.\end{tabular}                                                                                              \\ \hline
$d_Q$       & $O(dnlog(n))$ to $O(n)$                 & \multirow{3}{*}{\begin{tabular}[c]{@{}l@{}} - Monotonous with the scalability of S; \\ - More sensible to alterations in the\\ bottommost branches; \end{tabular}}                    \\ \cline{1-2}
$d_{Trip}$  & \multirow{2}{*}{$O(nlog(n))$ to $O(n)$} &                                                                                                                                                                                                                                                            \\ \cline{1-1}
$d_{TripL}$ &                                         &                                                                                                                                                                                                                                                            \\ \hline
$d_{Geo}$   & $O(n^4)$                                & \begin{tabular}[c]{@{}l@{}} - Emphasise shared internal edges and its\\ lengths rather than internal path lengths; \\ - Close to RF discriminatory power.\end{tabular}                                                                                        \\ \hline
$d_{MAST}$  & $O(|S|^{O(d)})$                         & \begin{tabular}[c]{@{}l@{}} - Very sensible to small variations; \\ - Should be used to analyze how identical\\ two trees are rather than similar; \\ - Close to RF discriminatory power.\end{tabular}                                                           \\ \hline
$d_{Align}$ & $O(n^3)$                                & \begin{tabular}[c]{@{}l@{}} - Strong emphasis in shared clades; \\ - Related to RF discriminatory power, but\\ not close: it values clades that are similar\\ but not identical.\end{tabular}                                                                    \\ \hline
$d_{CCC}$   & $O(n)$                                  & \begin{tabular}[c]{@{}l@{}} - First effective numerical method to\\ compare classifications; \\ - Subjective parameters make this metric \\not so precise and obsolete.\end{tabular}                                                                             \\ \hline
$d_{N}$  & $O(n^2)$                                & \begin{tabular}[c]{@{}l@{}} - Sensitive to tree distribution {[}26{]}; \\ - Better used when the relative position of\\ subsets of nodes is more important than\\ actual tree comparison {[}26{]}.\end{tabular}                                                  \\ \hline
$d_{Sim}$   & $O(|V|+|E|)$                            & \begin{tabular}[c]{@{}l@{}} - Excluding some particular cases, it\\ behaves similarly to other branch length\\ metrics {[}1{]}; \\ - Good metric to use when branch weight\\ proportion is important despise their\\ absolute value {[}1{]}.\end{tabular}         \\ \hline
\end{tabular}
\caption[Synopsis and comparison of distance's discriminatory power with defined time complexity.]{Synopsis and comparisson of distance's discriminatory power with defined time complexity. In this table, $d$ stands for the maximum vertex degree in the trees in which the distance is being calculated. In $(*)$ the denoted complexity is for trees in which the mapping function $h_{(A,B)}$ is unique, otherwise, the complexity is undefined.}
\end{table}

\pagebreak

\begin{table}[!ht]
\centering
\begin{tabular}{|l|l|l|}
\hline
          & Problem Hardness & Discriminatory Power                                                                                                                                                                                                                                       \\ \hline
$d_{Hyb}$ & \begin{tabular}[c]{@{}l@{}} APX-HARD,\\ NP-HARD \end{tabular}         & \begin{tabular}[c]{@{}l@{}} - Values not only the shared clades between\\ classifications but the  clades in which\\ their cluster representation is not disjoint; \\ - Interpretative value for phylogeny sets it\\ apart from the other metrics.\end{tabular} \\ \hline
$d_{SPR}$ & NP-HARD$^{(*)}$               & \begin{tabular}[c]{@{}l@{}} - May solve the RF problems in trees with\\ small variations given the replug move; \\ - Being the interpretative value for\\ phylogeny fairly relevant, $d_{SPR}$ is fairly\\ promising.\end{tabular}                              \\ \hline
\end{tabular}
\caption[Synopsis and comparison of distance's discriminatory power with undefined time complexity.]{Synopsis and comparisson of distance's discriminatory power with undefined time complexity. In $(*)$ the denoted problem hardness is estimated.}
\label{tab:dprob}
\end{table}

\section{Future Work}
\label{section:future}

The comparison of classifications, since its genesis in phylogenetics, has come a long way, and its advancements were beneficial in pretty much in everything that benefited from drawing distances between tree-like structures (as an example out of phylogenetics field of study, comparison metrics could be used as a way to compare algorithms by their running tree). 

However, given the field of study that bloomed all these methods (alongside the fact that was still a research topic), rigor and formalization was not the top-most priority, as we can see from our references. Formalizing or attempting to formalize these metrics (as we just did) is a challenge by itself and even though we see our effort and result as mostly satisfying, there's still some cases where the definition isn't as clean as possible (such as the case of \textit{Quartet based metrics} and defining the space of trees $\mathcal{T}_n$ for every $n\in\mathbb{N}$).

There's the discussed problem with \textit{Robinson Foulds Length} that was exposed in the respective subsection: $d_{RFL}$ is formalized on top of the existence of a matching function $h_{(A,B)}$ which might not be unique. That's a problem since if there exists more than one matching function, the distance between two trees might be undefined given the existence of two possible results. More than that, this distance is not symmetric as well, meaning that it might be the case that $d_{RFL}(A,B)\neq d_{RFL}(B,A)$. One interesting researching topic might be studying the viability of correcting these problems while maintaining the metric properties and keeping it close to its initial formulation.

When we approached the \textit{Geodesic distance}'s discriminatory power we brought up that $d_{Geo}$ actually relates to $d_{RF}$ since traversing $\mathcal{T}_n$ actually corresponds to contracting and decontracting edges. Even though they are formalized for different structures (the \textit{Geodesic distance} is formalized for weighted trees, contrary to \textit{Robinson Foulds}), if we put this fact aside we can actually examine examples to see more clearly the relation between these two metrics.

\begin{figure}[!htb]
  \centering
  \includegraphics[width=0.8\textwidth]{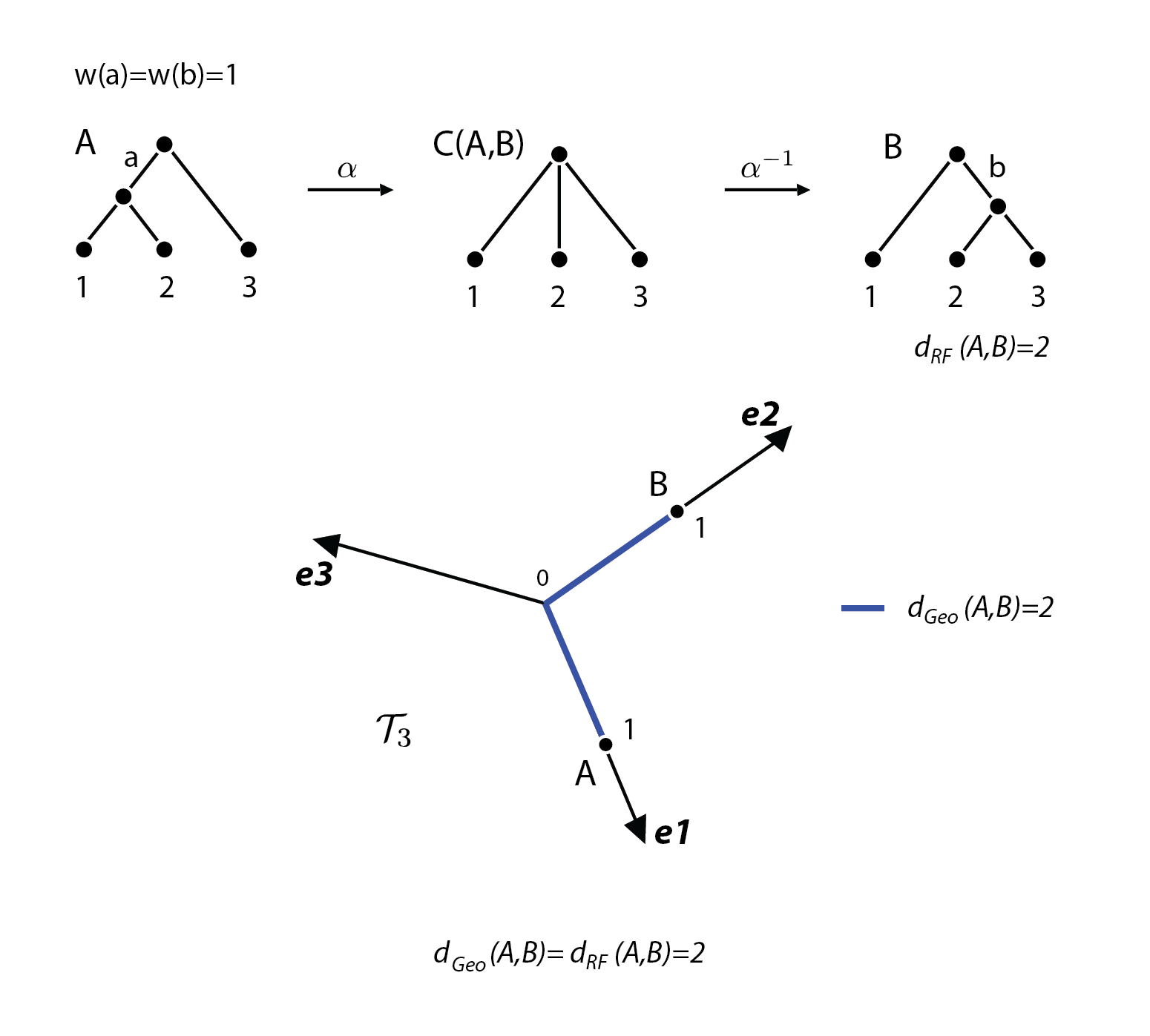}
  \caption[Example where $d_{RF}$ and $d_{Geo}$ coincide.]{Visualization of an example where $d_{RF}$ and $d_{Geo}$ coincide.}
  \label{fig:RFGeoEq}
\end{figure}

As we can see in the Figure \ref{fig:RFGeoEq}, the trees $A$ and $B$ are set apart by two $\alpha$ operations, and the geodesic distance between $A$ and $B$ in $\mathcal{T}_3$ is the cone path between these two trees. If we assume the weight of the internal edges $a$ and $b$, the cone path between $A$ and $B$ is indeed $2$.
However, even when we consider the length of the internal edges as $1$, it's not always the case that $d_{Geo}$ and $d_{RF}$ coincide, since contractions and decontractions can happen \textit{simultaneously} while traversing the space of trees.

\begin{figure}[!htb]
  \centering
  \includegraphics[width=0.8\textwidth]{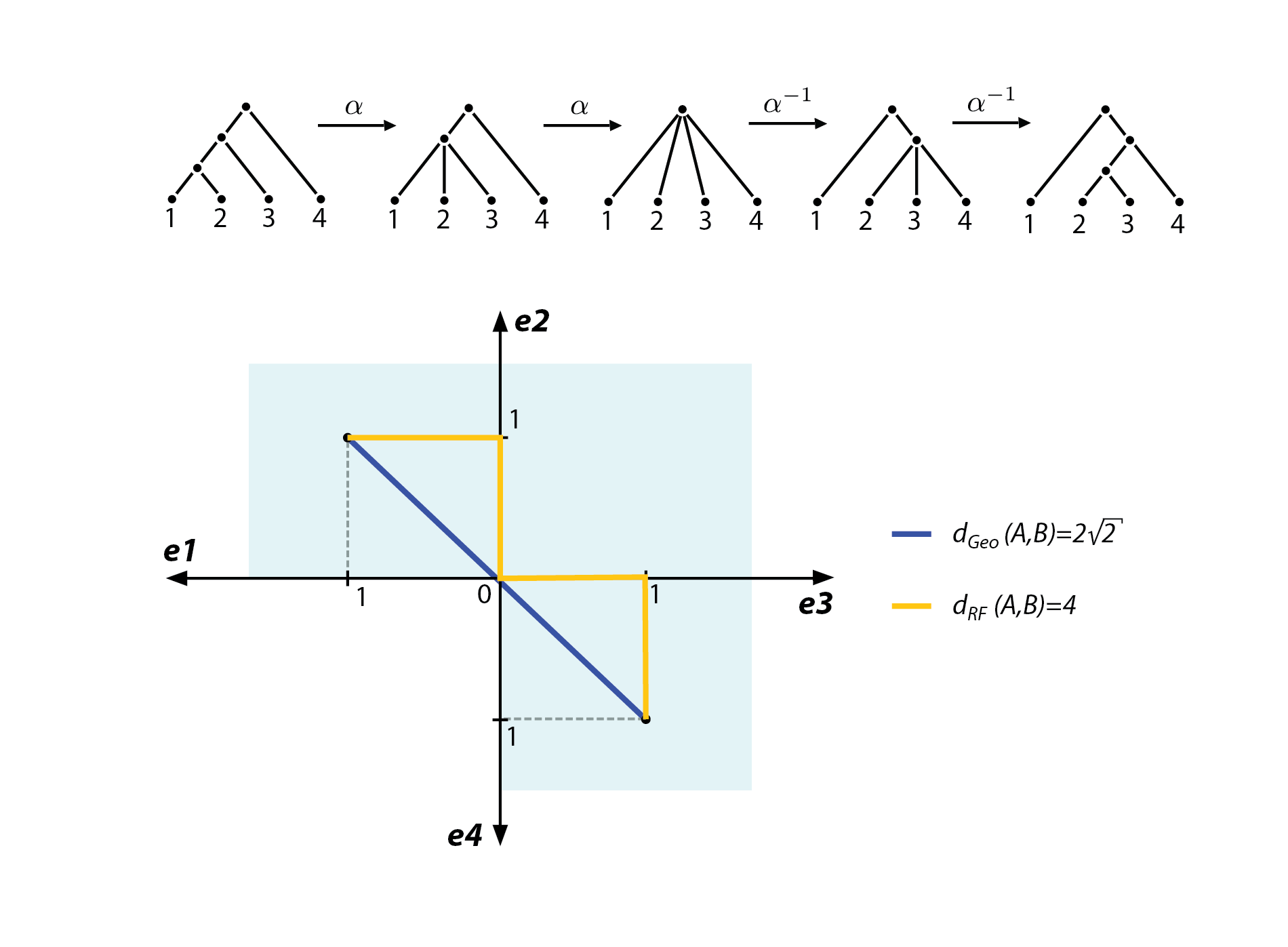}
  \caption[Example where $d_{RF}$ and $d_{Geo}$ do not coincide.]{Visualization of an example where $d_{RF}$ and $d_{Geo}$ don't coincide.}
  \label{fig:RFGeoDif}
\end{figure}

Such is the case in Figure \ref{fig:RFGeoDif}. While in $d_{RF}$ four $\alpha$ operations are done, in $d_{Geo}$ each pair of contractions weight $\sqrt{2}$ for the final result, since they are being done ``at the same time".
This relation between these two metrics could be further studied, as a way to strengthen our understanding between their link: understanding the cases in which the metrics coincide might be a good way to improve the computation time of $d_{Geo}$ for big datasets. The same way we could use $d_{RF}$ to optimize the usage of $d_{Geo}$, a study could be made in how could we use more efficient methods to bypass some cases on the less efficient ones. From a practical point of view, these approaches would be an advantage given the big scope of the data handled on the fields of application, nowadays. Also, this approach does not need to be exact, from a practical point of view an approximation can often be good enough.

Another thing that might be interesting to explore is the relation between the discriminatory power and the efficiency of the metric: it seems that the less efficient metrics are, the greater is their discriminatory power. However, this might also be related to the age of the metrics, since the optimization of the oldest distances were also subject of study for longer than the most recent ones. New distances are born from the limitations of the ones that already exist, so it should be no surprise that they are more powerful from a discriminative standpoint.

When it comes to Chapter \ref{chapter:GeodesicAnalysis}, there's some opportunity for work as well. Assuming the data structure chosen to represent trees $T\in\gamma^w_S$ and its respective order, finding the position of a split in that order is, as expected, as complex as searching a list of size $(2n-3)!!$, for $n=|S|$. As we mentioned in \ref{subsection:S2TImp}, there is surely a way to determine the position of a split $\sigma$ in the established order. 

Also, it would be interesting to study if the \textit{max-flow} algorithms proposed by \textit{Orlin} \cite{bib:30}, \textit{Goldberg and Rao} \cite{bib:32} and finally the ``almost linear time" variation of \textit{Spielman and Teng} \cite{bib:33} \cite{bib:34} can be applied in a network with weighted vertices. As we discussed in Subsection \ref{subsection:tcanalysis}, there is a variation of \textit{Edmonds-Karp} with $O(|V||E|)$ complexity for bipartite graphs, but it is not usable in graphs with weighted vertices (or from source/to sink edge weights). This could, hypothetically, put the \textit{geodesic distance} close to \textit{Robinson Foulds} in terms of worst-case time complexity, with complexity of order ``almost" $O(n^2)$. 
\cleardoublepage


\phantomsection
\addcontentsline{toc}{chapter}{\bibname}

\bibliographystyle{acm} 

\bibliography{refer} 

\nocite{*}

\cleardoublepage

%

\end{document}